\documentclass[prd,onecolumn,showpacs,floatfix,superscriptaddress,
nofootinbib]{revtex4-2}
\usepackage{graphicx}
\usepackage{epsfig}
\usepackage{bm}
\usepackage{amssymb}
\usepackage{float}
\usepackage{amsmath}
\usepackage{dcolumn}
\usepackage{cancel}
\usepackage[colorlinks]{hyperref}
\usepackage[usenames,dvipsnames]{color}
\hypersetup{
     breaklinks=true,
    pdfstartview={FitH},    
    colorlinks=true,       
    linkcolor=blue,          
    citecolor=red,        
    filecolor=magenta,      
    urlcolor=blue,           
    anchorcolor=green,      
    linktocpage=true
}

\usepackage{appendix}

\newcommand{\OmegaGW}{\Omega_{\mathrm{GW}}}

\let\oldsqrt\sqrt
\def\sqrt{\mathpalette\DHLhksqrt}
\def\DHLhksqrt#1#2{%
\setbox0=\hbox{$#1\oldsqrt{#2\,}$}\dimen0=\ht0
\advance\dimen0-0.2\ht0
\setbox2=\hbox{\vrule height\ht0 depth -\dimen0}%
{\box0\lower0.4pt\box2}}

\newcommand{\sss}[1]{{\scriptscriptstyle{#1}}}
\newcommand{\boldmathsymbol}[1]{{\ensuremath{\boldsymbol{#1}}}}

\newcommand{\uPl}{\mathrm{Pl}}

\newcommand{\usssPl}{\sss{\uPl}}

\newcommand{\calH}{\mathcal{H}}

\newcommand{\Mp}{M_\usssPl}

\newcommand{\beq}{\begin{equation}}
\newcommand{\eeq}{\end{equation}}
\newcommand{\bea}{\begin{equation}\begin{aligned}}
\newcommand{\eea}{\end{aligned}\end{equation}}

\newlength{\wsingfig}
\newlength{\wdblefig}
\newlength{\wquadfig}
\newlength{\wtriplefig}
\newcommand{\Eq}[1]{Eq.~(\ref{#1})}

\newcommand{\Fig}[1]{Fig.~{\ref{#1}}}

\newcommand{\Sec}[1]{Sec.~\ref{#1}}

\begin{document}

\leftline{KCL-PH-TH/2024-{\bf 63}}
\vspace{1cm}

\title{Gravitational wave signatures from reheating in Chern-Simons running-vacuum cosmology}

\author{Charalampos~Tzerefos}
\email{chtzeref@phys.uoa.gr}
\affiliation{Department of Physics, National \& Kapodistrian University of Athens, Zografou Campus GR 157 73, Athens, Greece}
\affiliation{National Observatory of Athens, Lofos Nymfon, 11852 Athens, 
Greece}

\author{Theodoros~Papanikolaou}
\email{t.papanikolaou@ssmeridionale.it}
\affiliation{Scuola Superiore Meridionale, Largo S. Marcellino 10, 80138, Napoli, Italy}
\affiliation{Istituto Nazionale di Fisica Nucleare (INFN), Sezione di Napoli, Via Cinthia 21, 80126 Napoli, Italy}
\affiliation{National Observatory of Athens, Lofos Nymfon, 11852 Athens, 
Greece}

\author{Spyros~Basilakos}
\email{svasil@academyofathens.gr}
\affiliation{National Observatory of Athens, Lofos Nymfon, 11852 Athens, 
Greece}
\affiliation{Academy of Athens, Research Center for Astronomy and Applied Mathematics, Soranou Efesiou 4, 11527, Athens, Greece}
\affiliation{School of Sciences, European University Cyprus, Diogenes Street, Engomi, 1516 Nicosia, Cyprus}

\author{Emmanuel~N.~Saridakis}
\email{msaridak@noa.gr}
\affiliation{National Observatory of Athens, Lofos Nymfon, 11852 Athens, 
Greece}
\affiliation{CAS Key Laboratory for Researches in Galaxies and Cosmology, 
Department of Astronomy, University of Science and Technology of China, Hefei, 
Anhui 230026, P.R. China}
\affiliation{Departamento de Matem\'{a}ticas, Universidad Cat\'{o}lica del Norte, Avda.
Angamos 0610, Casilla 1280 Antofagasta, Chile}

\author{Nick~E.~Mavromatos}
\affiliation{Physics Division, School of Applied Mathematical and Physical Sciences, National Technical University of Athens, Zografou Campus, Athens 157 80, Greece}
\affiliation{Theoretical Particle Physics and Cosmology group, Department of Physics, King's College London, London WC2R 2LS, UK}

\begin{abstract}

\vspace{0.2cm}

Within the context of a Chern-Simons running-vacuum-model (RVM) cosmology, one expects an early-matter dominated (eMD) reheating period after RVM inflation driven by the axion field. Treating thus in this work Chern-Simons RVM  cosmology as an effective $f(R)$ gravity theory characterized by logarithmic corrections of the spacetime curvature, we study the gravitational-wave (GW) signal induced by the nearly-scale invariant inflationary adiabatic curvature perturbations during the transition from the eMD era driven by the axion to the late radiation-dominated era. Remarkably, by accounting for the extra GW scalaron polarization present within $f(R)$ gravity theories, we find regions in the parameter space of the theory where one is met with a distinctive induced GW signal with a universal $f^6$ high-frequency scaling compared to the $f^7$ scaling present in general relativity (GR). Interestingly enough, for axion masses $m_a$ higher than 1~GeV and axion gauge couplings $f_a$ above $10^{-3} \Mp$, one can produce induced GW spectra within the sensitivity bands of future GW observatories such as the Einstein Telescope (ET), the Laser Interferometer Space Antenna (LISA), the Big Bang Observer (BBO) and the Square Kilometer Arrays (SKA).
\end{abstract}

\maketitle

\section{Introduction}

Modified $f(R)$ gravities~\cite{Sotiriou:2008rp} have been scrutinized in many works 
in the literature, from the point of view of their astrophysical/cosmological consequences, in an attempt to falsify as many models as possible. Some of these effective gravitational models can be embedded in concrete microscopic frameworks. One of such models is a string-inspired Chern-Simons running vacuum cosmology, termed Stringy Running-Vacuum-Model (StRVM)~\cite{bms,bms2,ms1,ms2},
which can be obtained as a low energy limit of microscopic string theories, with phenomenological relevance.

In previous works we have constrained the model phenomenologically from various points of view, ranging from a potential alleviation of cosmological tensions (Hubble and growth of structure tensions)~\cite{gms}, to 
compatibility with Big-Bang-Nucleonsynthesis (BBN), provided the StRVM is viewed as a provider of the entire Dark Energy observed in the Universe today~\cite{Asimakis:2021yct}. Moreover, interesting constraints may arise~\cite{Papanikolaou:2024rlq} from the requirement of avoidance of overproduction of light primordial black holes, whose dominance in epochs of the Universe before the BBN may lead to early Matter Dominated (eMD)
eras, thus inducing a potentially dominant production of gravitational waves (GWs).

An important feature of the StRVM, which leads to stringent constraints in the above contexts, especially in the cosmological-tension and eMD-era fronts, is the presence of 
$R\, {\rm log}(R/R_0)$ curvature corrections, where $R_0$ the curvature of the era we are interested in imposing the constraint. In the context of StRVM, such corrections may be the result of purely quantum-gravity effects~\cite{ms2,gms}.
It is such corrections that are responsible for  the embedding of the model into the modified $f(R)$ gravity framework,  which in turn leads to the presence of an extra polarization mode of GW, associated with the scalaron degree of freedom of $f(R)$ theories.

In this work, we continue to study the GW phenomenology of the StRVM by exploiting the presence of eMD epochs at the exit from inflation, which are driven by the axion fields that characterise the model~\cite{ms1}. 
In particular, similarly to what happens in the case of GR, the presence of an eMD associated with a sudden transition to the late radiation-dominated (RD) era can give rise to a resonantly enhanced GW signal induced by both adiabatic~\cite{Inomata:2019ivs,Inomata:2019zqy,Inomata:2020lmk} and isocurvature perturbations~\cite{Papanikolaou:2020qtd,Domenech:2020ssp, Papanikolaou:2022chm, Domenech:2021and} due to second order gravitational interactions~\cite{Domenech:2021ztg}. 

However,  in contrast to what happens in GR, in the StRVM framework one is met with the presence of the aforementioned extra GW polarisation mode associated with the scalaron degree of freedom of the $f(R)$ gravity theories~\cite{Sotiriou:2008rp}. In this work, upon considering the gravitational axion-Chern-Simons(CS) StRVM framework and taking into account the quantum-gravity logarithmic-curvature corrections to the Einstein-Hilbert gravitational action, we study GWs induced by scale-invariant inflationary adiabatic curvature perturbations, which are favored by the Planck Collaboration observations~\cite{Planck:2018}. We find a GW spectrum associated with the scalaron degree of freedom which is characterised by a universal $f^6$ frequency scaling in contrast to the $f^7$ high-frequency scaling present in GR. Remarkably, we manage to set constraints on the coefficient $c_2$ of the logarithmic-curvature corrections to the quantum effective gravitational action so that the induced GW signal associated with the scalaron dominates over the GR one, leading to a distinctive GW signature of modifications of gravity of the $f(R)$ type, which is potentially detectable by future GW observatories such as ET, LISA, BBO and SKA.

The paper is structured as follows:
in section \ref{sec:ags}, we review the StRVM model, which incorporates  the quantum-gravity-induced logarithmic curvature corrections, and stringy-in-origin axionic fields.  
The CS condensate, induced by primordial GWs, is held responsible for a specific (monodromy) form of the axion potential. In section \ref{sec:reh}, we discuss reheating features of the StRVM, and suggest the scenario of an axion-driven prolonged reheating process, due to the existence of an early axionic-matter-dominance era (AMDE) that interpolates between inflation and radiation epochs, with a rather sharp transition between AMDE and radiation.
In section \ref{sec:GWsignal}, we discuss the GW signal 
induced in the StRVM, 
as a result of the rapid transition from the AMDE to the radiation-dominated epoch of the Universe, and study the impact of the quantum-gravity logarithmic curvature corrections of the model on the spectral shape of the induced GW signal, as compared with the situation in GR. On making the assumption that the induced GW signal dominates over the GR one,
we also set lower-bound constraints on the coefficient $c_2$ of the logarithmic-curvature corrections to the quantum action. Finally, section \ref{sec:concl} contains our conclusions. Some technical aspects of our approach, associated with the 
concept of the geometric anisotropic stress, which we make use of in our analysis of the induced GW signal in section \ref{sec:GWsignal}, are given in Appendix \ref{app:geometric_anisotropic_stress}.

\section{The Axion-gravitational-Chern-Simons Model \label{sec:ags}}

The variant of the Stringy Running Vacuum Model (StRVM) cosmology to be examined below is a Gravitational Axion-Chern-Simons(CS) theory, including non-perturbative world-sheet instanton effects~\cite{Witten-wsinst,cvetic}, which lead to periodic modulations of the axion potential. The dynamics of the model is  
described by the effective action~\cite{svrcek,kaloper,kaloper2,gms}:\footnote{In this work we use units with $\hbar=c=1$ throughout, and we follow the conventions: $(-, +, +, +)$ for the metric signature, 
$R^\lambda_{\,\,\mu\nu\sigma} = \partial_\nu \Gamma^\lambda_{\, \,\mu\sigma} + \Gamma^\rho_{\,\, \mu\sigma} \, \Gamma^\lambda_{\, \,\rho\nu} - (\nu \leftrightarrow \sigma)$, $\lambda,\mu,\nu,\sigma= 0, \dots 3$
for the Riemann curvature tensor,  with 
$\Gamma^\lambda_{\, \,\mu\sigma} = \Gamma^\lambda_{\, \,\sigma\mu} =
\frac{1}{2}\, g^{\lambda \rho}\,\Big(g_{\rho \mu, \sigma} + g_{\rho\sigma,\mu} - g_{\mu\sigma,\rho} \Big)$
the torsion-free (Riemannian) Christoffel connection, symmetric in its lower indices (the comma denotes ordinary derivative),  $R_{\mu\nu} = R^\lambda_{\, \,\mu\lambda\nu}$ for the Ricci tensor, and $R=g^{\mu\nu} \, R_{\mu\nu}$ for the Ricci scalar.} 
\begin{align}\label{sea4}
S^{\rm eff}_B =
\; \int d^{4}x\, \sqrt{-g}\Big\{ 
c_0 + R\left[c_1+c_2\log\left(\frac{R}{R_0}\right)\right]   
 +  \mathcal{L}_{m} + \dots \Big\}.
\end{align}
In the above expression, $c_0 > 0$ is a constant, which contributes to the current-era dark vacuum-energy density,
while the coefficient 
\begin{align}
\label{c1def}
c_1= \frac{1}{16\pi\rm G} = \frac{1}{2\, \kappa^2} + \tilde c_1 >0\,,
\end{align}
determines an effective (3+1)-dimensional gravitational constant G, including weak quantum-gravity corrections $\tilde c_1$, such that $\tilde c_1/M_{\rm Pl}^2 \ll 1$,
where $M_{\rm Pl}= \kappa^{-1} = \frac{1}{\sqrt{G_N}}$ (with $G_N$ the conventional (3+1)-dimensional Newton's constant) is the reduced Planck mass
$M_{\rm Pl} = 2.4 \cdot 10^{18}$~GeV. The $ R\, {\rm log}(R/R_0)$ terms can arise from quantum-gravity effects, and may survive the current epoch, during which they 
can contribute to the alleviation of cosmological tensions~\cite{gms}.
The quantity $R_0$ denotes the scalar curvature of the expanding universe at the onset of the cosmological era we are interested in, namely the early Matter Dominated (eMD) era, driven by the axion fields (see below). The $\dots$ include 
higher-order curvature corrections, which in modern eras are not dominant. 

The quantity $\mathcal L_m$ denotes the Lagrangian density of matter 
\begin{align}\label{matter}
   \mathcal L_m =  - \frac{1}{2}\, \partial_\mu b \, \partial^\mu b 
- \frac{1}{2}\, \partial_\mu a \, \partial^\mu a  - V(b,a) + \dots ,
\end{align}
where $b(x), a(x)$ are axionic fields arising in string theory:
the field $b(x)$ denotes the Kalb-Ramond (KR), or string-model independent, axion, with coupling $f_b$, which in (3+1)-dimensional spacetimes is dual to the 
totally antisymmetric field strength of the spin-one (KR) field of the massless bosonic string gravitational multiplet~\cite{str1,str2,pol1,pol2,kaloper}; $a(x)$ is a string-model-dependent axion  associated with string compactification~\cite{svrcek}, assuming for simplicity that only one such axion species is dominant, with coupling $f_a$. 
The $\dots$ in \eqref{matter} denote matter and radiation fields  other than axionic. 
The axion potential $V(b,a)$ in its generality reads:\footnote{In the present work we take the potential of axions to be of a  specific form, for the purposes of illustrating the basic results of our analysis. 
Our conclusions, however, are valid for more general potentials, provided their parameters lie in the appropriate range.}
\begin{equation}\label{axeffpot1}
 V(b, \, a)={\Lambda_1}^4\left( -1+ f_a^{-1}\, \tilde \xi_1 \, a(x) \right)\, \cos({f_a}^{-1} a(x))+\frac{1}{f_{a}}\Big(f_b \, {\Lambda_0}^3 + \Lambda_2^4 \Big) \, a(x) + {\Lambda_0}^3\, b(x).
 \end{equation}
where the (energy) scale $\Lambda_0$ is associated with the condensate of the gravitational CS anomalous term in the effective action, arising from condensation of primordial chiral gravitational waves~\cite{stephon,Lyth,Dorlis:2024yqw}. 
The respective contributions to the (linear) effective potential for the $b$ field, have the form 
  \begin{align}\label{effectivepot}
V_{\rm eff}^{\rm linear} (b)= A\, \langle R_{CS}\rangle b(x) \equiv \Lambda_0^3 \, b(x) \,,
\end{align}
where 
\begin{align}\label{cond}
    R_{CS} \equiv \frac{1}{2}R^{\mu}_{\,\,\,\nu\rho\sigma}\widetilde{R}^{\nu\,\,\,\,\rho\sigma}_{\,\,\,\mu} \,,  \qquad  \langle R_{CS}\rangle^{total}_{I}=-\mathcal{N}_I\frac{ A \,  \kappa ^4 \, \mu ^4}{\pi ^2}\, \dot{b}_I \, H_{I}^3\,,
\end{align}
with $H_I$ the (approximately) constant Hubble parameter during inflation. The symbol $\widetilde{\dots} $ denotes the dual of the Riemann tensor, 
$\widetilde{R}_{\alpha\beta\gamma\delta}=\frac{1}{2}R_{\alpha\beta}^{\,\,\,\,\,\,\,\,\rho\sigma}\varepsilon_{\rho\sigma\gamma\delta}$, with $\varepsilon_{\rho\sigma\gamma\delta}$ the gravitationally covariant Levi-Civita tensor. 
The notation $\langle \dots \rangle$ is used to denote the respective condensate. 
The reader should observe that the condensate \eqref{cond}, leading to \eqref{effectivepot}, notably breaks the generic shift symmetry $b(x) \to b(x) + {\rm constant}$ of the original effective gravitational CS theory~\cite{Jackiw,Alexander:2009tp}, and leads formally to a linear axion potential of monodromy type~\cite{silver}, encountered in string/brane compactified models, in a different context though. Nonetheless, the non-linear dependence of the condensate \eqref{cond} on the Hubble parameter $H$ constitutes a rather drastic difference from conventional string theory, given that here the energy density of the cosmic fluid is of running-vacuum-model (RVM) type~\cite{rvm1,rvm2,rvm3,rvmqft1,rvmqft2,rvmqft3,rvmqft4}.
 
In the case of the string-inspired~\cite{kaloper,kaloper2} StRVM~\cite{bms,ms1,ms2}, the coupling $A$ 
is given by  
\begin{equation}\label{Aval}
    A=\sqrt{\frac{2}{3}}\frac{\alpha^\prime}{48\kappa},
\end{equation}
where $\alpha^\prime = M_s^{-2}$ is the string Regge slope, with $M_s$ the string mass scale, which is in general different from $M_{\rm Pl}$. The quantity $\mathcal N_I$ in \eqref{cond} denotes the number of sources of GW at the onset of the RVM inflation~\cite{Mavromatos:2022xdo,Dorlis:2024yqw}, while $\mu$ is an UV cutoff of the graviton modes, which in the string theory context is identified with the string scale $M_s$.\footnote{We remark, for completion, that the computation of the GW-induced CS condensate  \eqref{cond} in \cite{Dorlis:2024yqw}, which went beyond the approximations made in \cite{stephon,Lyth}, yields a reduction of the value found in those works by a factor of 2.} The detailed analysis in \cite{Dorlis:2024yqw}, using dynamical systems, demonstrated that the constancy of the gravitational condensate during RVM inflation 
requires $\mathcal N_I \sim 7 \cdot 10^{16}$ compared with the corresponding number of sources sources ${\mathcal N}_S$ at the preceding stiff-axion-$b$-matter era, which we assume to be of $\mathcal O(1)$ for concreteness. 

The axion coupling $f_b$ is determined from the coupling of the $b$ axion to gauge Chern-Simons terms, which in the models of \cite{bms,ms1,ms2} are assumed absent in the early Universe. Nevertheless, their (topological) coupling with 
$b$ exists in general, and it assumes the form~\cite{kaloper,kaloper2,svrcek}
\begin{align}\label{gaugeCS}
\mathcal S_{\rm CS-gauge} = A\, \int d^4x\, \sqrt{-g} \, b(x)\, {\rm Tr}\Big(\mathbf F_{\mu\nu} \, \widetilde{\mathbf F}^{\mu\nu}\Big)\,,
\end{align}
where Tr is a gauge-group trace, and $\mathbf F_{\mu\nu}$ is the (non-Abelian) gauge-field strength, whose dual is defined as:
\begin{align}\label{Fdual}
\widetilde{\mathbf F}^{\mu\nu} = \frac{1}{2} \varepsilon^{\mu\nu\alpha\beta}\, \mathbf{F}_{\alpha\beta}\,.
\end{align}
In the case of non-Abelian gauge groups, which characterise string-inspired theories, 
the following quantity can be a non-zero integer in the presence of topologically non-trivial, non-perturbative gauge-field configurations (instantons)~\cite{Tong:2005un,Eguchi:1980jx}: 
\begin{align}\label{chern}
\frac{1}{16\pi^2} \, \int d^4x \, \sqrt{-g} \, {\rm Tr}\Big(\mathbf F_{\mu\nu} \, \widetilde{\mathbf F}^{\mu\nu}\Big) = n \, \in \, \mathbb Z\,,
\end{align}
with the convention of positive (negative) integers $n > (<)\, 0$ for (anti)instantons. The axion-$b$ (and, in fact, any axion) coupling is defined as the coefficient of the term \eqref{chern} in the effective action, which has, as a consequence, the presence of periodic modulations of the respective potential-energy density, that is terms of the form $\propto  {\rm cos}\Big(b/f_b)$, breaking the generic shift symmetry $ b \to b + {\rm constant}$ to just periodicity in $b/f_b \to b/f_b + 2\pi$. 
In the string-inspired case of \cite{kaloper,kaloper2}, which the StRVM is based upon~\cite{bms,ms1,ms2}, we therefore have for the $b$(string-model-independent)-axion coupling:
\begin{align}\label{fbdef}
f_b= \frac{1}{16\pi^2 \, A} \stackrel{\eqref{Aval}}{=} 0.09 \, \frac{M_s^2}{M_{\rm Pl}}\,.
\end{align}
The compactification axion couplings  $f_a$ are defined through the appropriate Green-Schwarz anomaly terms~\cite{svrcek}, and their values are highly string-compactification-model dependent. For our purposes, in the context of the StRVM framework~\cite{ms1}, we shall treat $f_a$ as a
phenomenological parameter, different from $f_b$.

To determine $\Lambda_0$, we first recall that, during the RVM inflationary era 
we may parameterise (in our metric signature conventions, which are opposite of those of ref. \cite{bms,ms1,ms2}):
\begin{align}\label{slowrollb} 
\dot b =  \sqrt{2\epsilon}\, H\, M_{\rm Pl}\,,
\end{align}
with the overdot denoting derivative with respect to the Robertson-Walker-frame time, and $\epsilon$ a constant slow-roll parameter, of order $10^{-2}$ to match~\cite{Dorlis:2024yqw} the Planck data~\cite{Planck,Planck:2018}. 
Then we take into account that the detailed dynamical system analysis of \cite{Dorlis:2024yqw} yields
\begin{equation}
    M_s\sim 10^{-1}M_{\rm Pl}<M_{\rm Pl}\,,
    \label{cutoff}
\end{equation}
consistent with the transplanckian censorship hypothesis~\cite{TCH1,TCH2,ms2}, 
and the following lower bound for the magnitude of the value of the axion-$b$-field  $b(0)$ at the onset of the RVM inflation
(in order of magnitude):
\begin{align}\label{b0bound}
|b(0)| \gtrsim \mathcal O(10)\, M_{\rm Pl} \,, \quad b(0) < 0\,.
\end{align}
This implies that, in order of magnitude, the $b$-axion remains approximately constant during the entire duration $\Delta t$ of inflation, which is such that $H_I \Delta t \simeq \mathcal N$, with $\mathcal N = \mathcal O(50-60)$ the number of e-foldings~\cite{Planck:2018}.
Indeed, from \eqref{slowrollb}, we observe that the linear-in-the-axion-$b$-field term in the potential \eqref{effectivepot} 
varies linearly with the cosmic time, so that at the end of inflation $t_{\rm end}$ its value is: 
$$ b(t_{\rm end}) = b(0) + 10^{-1} H_I M_{\rm Pl} \Delta t \simeq b(0) + \mathcal O(5-6) \, M_{\rm Pl} \,, \quad b(0) < 0\,,$$
where $t=0$ is the onset of inflation in the StRVM. From \eqref{b0bound}, then, we observe that $b(t)$ remains of the same order of magnitude during inflation, leading to an approximately constant linear-axion-monodromy potential term \eqref{effectivepot}, which thus drives the de-Sitter inflationary phase. From \eqref{cutoff}, \eqref{fbdef}, we also have that
\begin{align}\label{fbval}
f_b \simeq  9 \times 10^{-4} \, M_{\rm Pl} \sim 10^{-3} \, M_{\rm Pl} = \mathcal O(10^{15}) \, {\rm GeV}\,.
\end{align}

Finally, on using the Planck data constraints on the Hubble rate during inflation~\cite{Planck:2018,Planck}
 \begin{align}\label{HI}
 H_I <  2.5 \times 10^{-5}M_{\rm Pl}\,,
\end{align}
we then obtain from \eqref{cond} the estimate, upon saturating the bound \eqref{b0bound} for concreteness:
\begin{align}
|b(0)|\, \Lambda_0^3 \simeq 7.8 \times 10^{-7}\, M_{\rm Pl}^4\,, 
\end{align}
which yields 
\begin{align}\label{L0}
\Lambda_0 \simeq 4 \times 10^{-3} \, M_{\rm Pl}\,.
\end{align}
The parameters $\tilde \xi_1$, axion coupling $f_a$,  and the world-sheet instanton induced scales $\Lambda_1, \Lambda_2 $ appearing in \eqref{effectivepot}
are treated as phenomenological. In \cite{stamou} the following scale-hierarchy constraint had been imposed
 \begin{align}\label{constr}
 \Big(\frac{f_b}{f_a} + \frac{\Lambda^4_2}{f_a\, \Lambda^3_0} \Big)^{1/3}\, \Lambda_0  \, < \, \Lambda_1 \ll \Lambda_0 ~, 
 \end{align}
which ensures that the dominant effects in the potential come from the gravitational anomaly condensate term of the $b$-axion (see discussion below). 
However, the same is true also for the hierarchy 
\begin{align}\label{constr2}
 \Lambda_1 \ll \, \Big(\frac{f_b}{f_a} + \frac{\Lambda^4_2}{f_a\, \Lambda^3_0} \Big)^{1/3}\, \Lambda_0  \, <  \Lambda_0 ~, 
 \end{align}
which, as we shall see below, leaves more room for the range of axion masses.

We remark that the parameters $\xi_1$ is associated with brane instantons~\cite{silver} in specific string compactifications and, hence, we shall set it to zero in our subsequent analysis for concreteness. The same holds for $\Lambda_2$ which we also set to zero. Therefore, in what follows we shall consider the following axion potential
in the StRVM context:
\begin{equation}\label{axeffpot}
 \boxed{V(b, \, a)= - {\Lambda_1}^4\, \cos({f_a}^{-1} a(x))+\frac{f_b}{f_{a}}\, {\Lambda_0}^3 \, a(x) + {\Lambda_0}^3\, b(x) \,.}
 \end{equation}

Hence, we stress once again that, with the hierarchies \eqref{constr} and \eqref{constr2}, the spirit of \cite{bms,ms1,ms2} regarding the induced RVM inflation from the linear $b$-terms in the axion potential, is maintained, as such terms are dominant. On the other hand, the $a$-dependent terms in $V(a,b)$
are responsible for the prolongation of the duration of inflation. Their corresponding periodic modulation is responsible for features in the profile of GW during the radiation era, after inflation exit.

\section{Reheating in the StRVM and early Axion-Matter-Dominated Era \label{sec:reh}}

A basic feature of the StRVM, which we shall explore in the present work, is the possibility, for a certain region of its parameters, of a 
prolonged reheating period after the exit from inflation. The reader should  recall at this stage that in the StRVM~\cite{bms,ms1,ms2}, inflation is driven by the (dominant) fourth-power of the Hubble parameter, that characterises the vacuum energy density $\rho^{\rm RVM}_{\rm vac}$,  due to the formation of the gravitational anomaly condensate \eqref{cond}:\begin{align}\label{rvmener}
\rho^{\rm RVM}_{\rm vac}  \simeq 3\kappa^{-4} \, \Big[ \nu \, (\kappa H)^2 
+  \alpha\, (\kappa\, H)^4 \Big] + \dots \, ,
\end{align}
where $\nu$ and $\alpha$ are calculable coefficients~\cite{bms,ms1,ms2}.\footnote{The coefficient $\nu $ is found negative due to the contributions of the gravitational anomaly to the stress energy tensor~\cite{bms,ms1,ms2}.} The $\dots$ in \eqref{rvmener} denote the quantum-gravity-induced  $H^n {\rm ln}H, \, n=1,2$ power-logarithmic mixed contributions ({\it cf.} \eqref{sea4}), as well as terms arising  from the quantum fluctuations of the KR $b$ axion and the compactification axion $a$ and their non-perturbative periodic potentials, as appearing in \eqref{axeffpot}, which will play a r\^ole in our subsequent discussion on reheating. 

The form \eqref{rvmener} is of a RVM type Universe, which we know that, after inflation, is  characterised~\cite{Lima1,Lima2} by a {\it prolonged adiabatic period} of reheating, rather than an instantaneous reheating process as in standard cosmology~\cite{kolb}. The radiation particles appear as a result of the decay of the running de Sitter vacuum, which is metastable.  As discussed in \cite{thermal1,thermal2,thermal3}, although the energy density during radiation obeys the usual $T_{\rm rad}^4$ scaling with the cosmic temperature of radiation era $T_{\rm rad}$, the standard 
cooling law of the Universe during radiation $\text{a}_{\rm rad}\sim 1/T_{\rm rad}$ (where $\text{a}_{\rm rad}$ is the scale factor, $\text{a}$, during the radiation dominance) is modified to:
\begin{align}\label{radmod}
T_{\rm rad} =  \Big( \frac{90\, H_I^2 \, 
\kappa^{-2}}{\pi^2 \, g_{\star, \rm rad}} \,\Big)^{1/4} \, \frac{\Big(\text{a}/\text{a}_{\rm eq}\Big)}{\Big[1 + \Big(\text{a}/\text{a}_{\rm eq}\Big)^4\, \Big]^{1/2}} =  \, \sqrt{2} \, T_{\rm eq} \, \frac{\Big(\text{a}/\text{a}_{\rm eq}\Big)}{\Big[1 + \Big(\text{a}/\text{a}_{\rm eq}\Big)^4\, \Big]^{1/2}} \, , 
\end{align}
where the suffix ``eq'' indicates the point at which the running vacuum energy density equals that of radiation, and we have defined $T_{\rm eq} = T_{\rm rad}(\text{a}_{\rm eq}) = \Big(\frac{45\, H_I^2}{2 \, \pi^2\, \kappa^2 \, g_\star}\Big)^{1/4}$, with $H_I$ the RVM inflation scale, and $g_\star$ the degrees of freedom of the created massless modes of the cosmological model at hand. For $\text{a} \gg \text{a}_{\rm eq}$ the Universe reaches a perfect fluid adiabatic phase, during which the temperature decreases in the usual way following a $T^{-1}$ cooling law for the Universe scale factor. 

%

The important feature of \eqref{radmod} is the fact that for $\text{a} \ll \text{a}_{\rm eq}$ one observes a phase of the expanding Universe in which the temperature grows linearly with the cosmic scale factor, attaining a maximal value at vacuum-radiation equality $\text{a}=\text{a}_{\rm eq}$. 
Thus, instead of having the usual instantaneous (highly-non adiabatic) reheating~\cite{kolb}, the RVM Universe exhibits a prolonged non-equilibrium heating period, with non-trivial features, as a result of the decay of the vacuum, which drives progressively the Universe into the radiation phase with important consequences also for its thermal history. In particular, the initial linear-$T$  phase is responsible for (most of) the entropy production of the RVM Universe~\cite{thermal1,thermal2,thermal3}. 


In the standard RVM scenarios~\cite{thermal1,thermal2,thermal3}, despite the long reheating process, inflation is succeeded by the radiation era, with no matter-domination  intervention. 
 However,  in the context of the StRVM~\cite{bms,ms1,ms2},  a more careful look is required before a definite conclusion is reached concerning the absence of an eMD phase preceding radiation dominance. Indeed,  
the presence of string compactification axions in our model, with periodic sinusoidal modulations of the axion potential 
\eqref{axeffpot}, induced by  non-pertrurbative (world-sheet instanton) effects, 
may, under some circumstances, change the above conclusion. In fact, as follows  from \eqref{axeffpot}, these
effects lead to quadratic (mass) terms in the vacuum energy density 
\footnote{We note for completeness at this point that the KR axion $b$ is assumed massless. Inflation is driven in the StRVM by the linear potential of this axion, which arises after condensation of the CS  gravitational anomaly term. As discussed in \cite{dorlis2}, one may consider vsriants of the StRVM in which the $b$ axion potential is also characterised by non-perturbatively-induced periodic modulations, but with the corresponding term of opposite sign as compared with the cosine term in \eqref{axeffpot}. Such a potential term does not contribute a mass for the KR axion, but such modulation terms result in better inflationary phenomenology of the StRVM, as far as a fit of the pertinent slow-roll parameters to the Planck data~\cite{Planck:2018,Planck} is concerned.}
\begin{align}\label{massterms}
V(b, a) \ni \frac{1}{2}\, \frac{\Lambda_1^4}{f_a^2}\, a(t)^2 ,
\end{align}
implying an axion mass of order 
\begin{align}\label{axionmass}
m_a \sim \Lambda_1^2/f_a\,. 
\end{align} 
Depending on the values of the parameters (which, in turn, depend on the details of the underlying microscopic string-theory model), one might encounter a situation, in which at the end of the RVM inflation, the axion field $a(t)$ oscillates slowly around its minimum value, and reheats the Universe, in parallel with the decay of the RVM vacuum itself.  In such a case  therefore, one might have a rather long epoch of an early Axion-Matter Dominated Era (eAMDE), before the radiation era, in the spirit of the case assumed in \cite{carr}. The presence of such an era would lead to a prolonged reheating phase of the Universe. Such a matter-dominated reheating will be succeeded by the (also slow) radiation-era reheating \eqref{radmod}, that characterises the decay of the RVM vacuum. 

Once photons are created by the decay of the RVM vacuum at the  end of the RVM inflation era, as assumed in the scenario of \cite{bms,ms1,ms2}), then the axions couple to the  chiral gauge anomaly, in a similar fashion to the KR axion $b$ \eqref{gaugeCS},  
\begin{align}\label{aFFdual}
S_{\rm CS-axion-a} = \int d^4x \, \sqrt{-g}\, \frac{1}{32 \pi^2 \, f_a\, } \, \varepsilon^{\mu\nu\alpha\beta}\,  a(t)\, F_{\mu\nu}\, F_{\alpha\beta}\,,
\end{align}
with $F_{\mu\nu}$ the Maxwell tensor in case of Abelian chiral anomalies we are interested in here (or the non-Abelian field strength tensor in case of chiral anomalies of non-Abelian gauge groups). 

The massive axions will then decay to massless radiation modes for photon $\gamma$ pairs, for instance, with the corresponding (tree-level) decay width being of order 
\begin{align}\label{widthaxion}
\Gamma_a (a \rightarrow \gamma\, \gamma) \sim \frac{m_a^3}{64\pi \, f_a^2} \sim \frac{\Lambda_1^6}{64\, \pi \, f_a^5}\,,
\end{align}
where we used \eqref{axionmass}. At this stage the reader's attention is called to the fact that, if the reheating of the Universe was attributed exclusively to the  decay \eqref{widthaxion} of the axion field $a(x)$, then 
instant reheating would require 
an upper bound on the width $\Gamma$~\cite{Planck}: 
\begin{align}\label{instrehG}
\Gamma_a \le 10^{14}\, {\rm GeV}\,.
\end{align}

On the other hand, the condition for an eMD, interpolating between inflation and  radiation, reads:
\begin{align}\label{eMDcond}
\Gamma_a \ll H \lesssim \, 10^{14} \, {\rm GeV}\,,
\end{align}
where we took into account that $H$ is of the order of the inflationary scale \eqref{HI}. Thus, in the case of our massive axions from compactification, their decays into photons \eqref{widthaxion} would be compatible with instant reheating. 
The condition 
\eqref{eMDcond} allows for the emergence of an eAMDE, at the end of which an abrupt transition to the radiation era may be plausible, for a range of the parameters of the underlying string theory model.\footnote{There are ambiguities however in string models, due to the fact that the axion does not decay only to standard model massless particles, but also to exotic massless modes, e.g. from hidden sector of the underlying theory. This is a problem that we do not address in the current phenomenological study. Nonetheless, studies  in the string/brane literature~\cite{blumen,Halverson}, have lead to the explicit construction of models in which the decays of the compactification  axions to hidden sector particles are suppressed as compared to those into standard model particles. Thus, for such models, to which we restrict our attention in the remainder of this article, one may draw reliable conclusions on the existence of an eADME based solely on the decays \eqref{widthaxion}. }

Such a condition characterises stringy models in which $\Lambda_1 \ll \Lambda_0$, and not simply $\Lambda_1 < \Lambda_0$ as we considered in \cite{stamou}.
The case $\Lambda_1 \ll \Lambda_0$ is met in stringy models characterised by world-sheet instantons  with large Euclidean actions $\gg 1$. The reader should recall at this stage that 
the scale $\Lambda_1 $ is suppressed by the exponential of the corresponding (Euclidean) non-perturbative instanton action $S_{\rm instanton}$, 
\begin{align}\label{sinst}
\Lambda_1 \sim \zeta \, M_s \, \exp\Big(-S_{\rm instanton}\Big)\,,
\end{align}
where $\zeta $ a numerical factor which depends on the specific model of instanton background considered (but expected in general to be of $\mathcal O(1)$).

If one considers target-space (3+1)-dimensional instantons, that characterise the non-Abelian gauge-group of the corresponding (3+1)-dimensional effective string-inspired theory, arising from string compactification, for energy scales $E \ll M_s$, then it is well-known that there exists a lower bound of $S_{\rm instanton}$~\cite{Tong:2005un}, 
\begin{align}\label{ymcoupling}
S_{\rm instanton} \gtrsim \frac{8\pi^2}{g^2_{\rm YM}}\,,
\end{align}
where $g_{\rm YM}$ is the renormalised Yang-Mills coupling, evaluated at the instanton energy scale. 

Much larger flexibility when evaluating the scale $\Lambda_1$ is offered in the cases where one 
consider world-sheet instanton effects in brane models, in particular in the so-called  type IIA framework of 
intersecting D6-brane models (see refs.~\cite{cvetic,cvetic2}, and in particular \cite{Blumenhagen:2009qh}). In such a case, the Euclidean world-sheet instantons 
are associated with E2-branes wrapped around compactified internal spaces, e.g. Calabi-Yau compact three-folds, while the (Euclidean) internal-space D6-branes wrapping different three cycles, corresponding to a different volume Vol$_{\rm D6}$. This yields a $\Lambda_1$ of order:\footnote{The reader should compare this case with that of an E2-instanton, wrapping the same three cycles as those wrapped by  
an auxiliary spacetime-filling D-brane (DE2) wrapping the same internal cycles as the E2-instanton. In such a case, 
the pertinent  action, replacing \eqref{ymcoupling}, is given by the vacuum open string disc (tree-level) amplitude for the E2 instanton~\cite{Polchinski:1994fq,Blumenhagen:2009qh}:
\begin{align}\label{1instdisc}
\mathcal S^{\rm disc}_{\rm 1-instanton\,E2} = \frac{M_s^3}{g_s}\,{\rm Vol}_{\rm E2}\,,
 \end{align}
where $M_s^3 \, {\rm Vol}_{\rm E2}$ is the volume (in string (measured in $M_s^{-1}$ units) of the internal three-manifold wrapped by the E2 instantons. 
The right-hand-side of \eqref{1instdisc} can be identified with $8\pi^2/g^2_{\rm YM,\rm E2}$, with  $g_{\rm YM,\rm E2}$ the Yang-Mills gauge coupling on the auxilliary DE2 brane. Comparing \eqref{1instdisc} with \eqref{E2D6}, we see clearly the 
richer phenomenology that characterises the latter case, based on Calabi-Yau-like intersecting-D6-brane compactification.} 
\begin{align}\label{E2D6}
\Lambda_1 \sim \zeta^\prime \, \exp\Big(-\frac{8\pi^2}{g^2_{\rm D6}}\,\frac{{\rm Vol}_{\rm E2}}{{\rm Vol}_{D6}}\,\Big) \,,
\end{align}
where again $\zeta^\prime$ is a (model-dependent) numerical factor of $\mathcal O(1)$, and $g_{\rm D6}$ is the gauge coupling of the $U(N)$ gauge theory on the D6 branes. In such models, therefore, we have the freedom to arrange for 
the desired hierarchy of scales $\Lambda_1 \ll \Lambda_0$ in our context, so that there is an intermediate massive-axion-matter-dominated era, by arranging appropriately for the ratio of the two volumes in the exponent of \eqref{E2D6}. 
In such a case, the axion-$\gamma\, \gamma$ decay width \eqref{widthaxion} can be much smaller than the (approximately constant) value of the Hubble parameter  during the RVM inflation, $H_I$, 
thereby fulfilling the criteria~\cite{carr} for the existence of an intermediate-axion-matter dominated era, interpolating between the exit from RVM inflation, and the onset of the radiation epoch.

Such a situation prompts comparison with what happens with the so-called flaton-dominated case in flipped SU(5) superstring models of inflation~\cite{Ellis:2018moe,Basilakos:2023jvp}, where again a two step reheating of the pertinent Universe occurs. In such models, the Hubble parameter $H$ during the flaton-$\Phi$ (of mass $m_\Phi$)  dominated era is computed from the corresponding Friedmann equation, and, in the case of strong reheating, 
which implies that the energy density of the flaton is dominated by incoherent thermal fluctuations, it takes the form:
\begin{align}\label{flaton}
H = \Big(\,\frac{\rho_\Phi}{3 M^2_{\rm Pl}}\,\Big)^{1/2} = \Big(\,\frac{\zeta(3)\, m_\Phi \, T_\Phi^3}{3\pi^2 M_{\rm Pl}^2}\,\Big)^{1/2}\,,
\end{align}
where $\rho_\Phi$ is the energy density of the flaton, which in the matter dominated era is taken proportional to the cube of its temperature, $T_\Phi^{3}$, with $T_\phi = T_{\rm rad} \Big(g(T_{\rm rad})/g_{\rm dec}\Big)^{1/3}$, $T_{\rm rad}$ the temperature of the radiation background, and $g(T)$ the number of degrees of freedom at temperature $T$, with the suffix ``dec'' denoting quantities at decoupling. The above considerations are valid for large mass values of $\Phi$, of order~\cite{Ellis:2018moe} $m_\Phi \sim 10^{20} \, \frac{\rm GeV}{M_{\rm GUT}}$ (with $M_{\rm GUT}$ a typical GUT scale, so $m_\Phi \sim 10^4 \, {\rm GeV}$ for $M_{\rm GUT} \sim 10^{16} \, {\rm GeV}$), 
which drive the vacuum expectation value of the field during inflation to zero, as its mass term will dominate the interactions. The temperature $T_\Phi$ can be estimated from equating 
\begin{align}\label{HGamma}
H \simeq \Gamma_\Phi\,, 
\end{align}
where $\Gamma_\Phi$ te appropriate decay width of the flaton in the fliped SU(5) model~(see ref.~\cite{Ellis:2018moe} and references therein).

In our CS theory, the compactification action couples to the anomalous Chern Simons terms, which during inflation condense, as a result of the condensation of primordial GWs. The readers can easily convince themselves that there is 
no minimization of the axion-$a$ potential in that epoch due to the smallness of $\Lambda_1 \ll \Lambda_0$. 
At the end of the RVM inflationary era, the condensate $\Lambda_0^3 \to 0$ and thus implies that 
there is minimization at $\langle a \rangle \to 0$ at that epoch, since in such eras, the effective axion potential \eqref{axeffpot} is characterised only by the periodic cosine term. The axions $a$ are massive in that phase
with mass given by \eqref{axionmass}, which in view of \eqref{cutoff}, \eqref{sinst} (or \eqref{E2D6} in the case of intersecting brane compactifications~\cite{Blumenhagen:2009qh}), yields
\begin{align}\label{axamassval}
m_a = \Big[10^{-2} \, \zeta^2\, \exp\Big(-2\,S_{\rm instanton}\Big)\Big] \,\frac{M^2_{\rm Pl}}{f_a}\,.
\end{align}
To respect the hierarchy \eqref{constr}, we need to have $f_b \ll f_a$. In view of \eqref{fbval}, which, as we have discussed above,  is valid in the 
context of StRVM, this would require 
\begin{align}\label{faval}
f_a \gtrsim 10^{-1} \, M_{\rm Pl} \gg f_b\,,
\end{align}
where $\gg$ is considered here to mean that the respective quantity is at least two orders of magnitude larger. Notice that this is consistent with 
the constraints from Big-Bang-Nucleosynthesis (BBN) (essentially avoiding overproduction of $^4$He during BBN), which require axion couplings greater than $\mathcal O(10^9)$~GeV, for a range of axion masses (see fig. 4 of ref.~\cite{Conlon:2013isa}, and also the review in \cite{Marsh:2015xka}), which includes masses comparable with that of flaton ($\sim 10^4$~GeV) discussed above. Such a range can be achieved from \eqref{axamassval}, upon appropriate choices (i.e. models) of the instanton action and the axion-$a$ coupling, $f_a$.

Let us now perform some phenomenological analysis of the various models, to see what kind of predictions for the axion $a$ mass range can be made. Below we shall assume string scales satisfying \eqref{cutoff}, which is one consistent model
within RVM inflation~\cite{Dorlis:2024yqw}.\footnote{For other initial conditions of the dynamical system approach to inflation, the analysis of \cite{Dorlis:2024yqw} can lead to other values of the string scale $M_s$. For concreteness, we do not discuss here these other choices, but the reader should have in mind that such, more general, analyses lead to richer phenomenology.}
In the case of the hierarchy \eqref{constr}, which implies $f_a$ of order given in \eqref{faval} (assumed saturation, for concreteness), the scale $\Lambda_1$ satisfies: 
\begin{align}\label{L1constr}
\Lambda_1 \gtrsim   4 \times 10^{-11/3}\, M_{\rm Pl} = \mathcal O(10^{-3})\, M_{\rm Pl} = \mathcal O(10^{15})~{\rm GeV}\,,
\end{align}
which is achieved for instanton actions $S_{\rm instanton} = \mathcal O(5)$. In the case of target space gauge-group instantons with (Euclidean) action saturating the bound \eqref{ymcoupling}, this implies strongly coupled Yang-Mills gauge theories with (renormalized)  fine structure constant $\alpha_{\rm YM}=g_{\rm YM}^2/(2\pi) =\mathcal O(1)$, at the instanton energy scale. Richer choices are allowed for the intersecting-brane models with world-sheet instanton action \eqref{E2D6} in order to achieve a $\Lambda_1$ of order \eqref{L1constr}.  With $\Lambda_1$ of order \eqref{L1constr}, one obtains from \eqref{axamassval} axion masses of order $m_a \simeq 10^{-5}\, M_{\rm Pl} = \mathcal O(10^{13})\,$GeV.

To achieve much lighter axions, as is common in string models, we need to consider the hierarchy \eqref{constr2} within our StRVM framework, in which the  instanton scale $\Lambda_1$ is left as a phenomenological parameter. From \eqref{axionmass} we observe that, with $f_a \simeq \mathcal O(10^{-1}) \, M_{\rm Pl}$, as required by \eqref{constr2}, 
one obtains
\begin{align}\label{L1ma}
\Lambda_1 \sim \sqrt{10^{-1}\, m_a \, M_{\rm Pl}}\,,
\end{align}
and hence to achieve axion masses of the order of the flaton mass in flipped SU(5), for instance, that is $m_a = \mathcal O(10^4)$\, GeV, one needs $\Lambda_1 = \mathcal O(5 \times 10^{10})~{\rm GeV} = 2 \times 10^{-8}\, M_{\rm Pl}$, smaller by almost five orders of magnitude as compared to \eqref{L1constr}, which pertains to the hierarchy \eqref{constr}.
This corresponds to an instanton action of order $\mathcal O(15)$, which in case of gauge target-space instantons corresponds to fine structure constant at the instanton scale of order $\mathcal O(7)$. 

The axion can then form non-relativistic matter, with energy density proportional to the inverse cubic power of the scale factor, as per the flaton case, \eqref{flaton}. The corresponding temperature of the non-relativistic axion intermediately-dominant matter, is then provided by equating the corresponding $H$ as given by \eqref{flaton} (but with the replacement of $\Phi$ by the axion $a$) with the width \eqref{widthaxion}, assuming the dominant decay mode of the axion is that to two photons,\footnote{In specific string models, other modes may be in operation.}
 which in this case gives (using \eqref{L1ma}, and the lower bound of \eqref{faval}, for concreteness)
\begin{align}\label{HGammaaxion}
 \frac{\Big(10^{-1}\, m_a \, M_{\rm Pl}\Big)^3}{64\, \pi \, f_a^5} \simeq \Big(\,\frac{\zeta(3)\, m_a \, T_a^3}{3\pi^2 M_{\rm Pl}^2}\,\Big)^{1/2} \quad \Rightarrow  \quad  \boxed{T_a = \Big[\frac{3 \times 10^4}{(64)^2\, \zeta(3)} \frac{m_a^5}{M_{\rm Pl}^2}\Big]^{1/3}}\,.
\end{align}
For axion masses $m_a = \mathcal O(10^4)$~GeV, this yields (taking into account $\zeta(3) \simeq 1.2$) the axion-matter fluid temperature $T_a \simeq  1.1 \times 10^{-9} \, M_{\rm Pl} \simeq 2.7 \times 10^9 \, {\rm GeV} $, which determines the AMDE in this Universe. This value is sufficiently small compared to $H_I$, which justifies that the specific  AMDE we are discussing above occurs at a temperature which lies past the maximum of the temperature  $T$ in \eqref{radmod}, which occurs when the RVM energy density equals that of radiation. This justifies {\it a posteriori} our assumption on the temperature scaling of the Universe scale factor $a \sim T^{-1}$, which usually characterises radiation era. The situation depends delicately on the range of the axion masses and other parameters of the string-inspired model. In general, the dependence of the scale factor on the temperature may be affected by the relative location of the AMDE in the temperature/scale-factor diagram of the StRVM.  Further exploration along these lines constitute the subject of future works.


\section{Induced gravitational wave signal in the StRVM}\label{sec:GWsignal}

As we demonstrated in the previous section, appropriate parameter choices of our StRVM setup allow for the existence of an early matter-dominated era induced by one of our axion fields (AMDE). Subsequently, there is a rapid transition into a late-radiation domination era (lRD) and the picture of standard cosmology follows. What is interesting and we are going to study in this section, is that this rapid transition from the AMDE into lRD induces the production of a resonantly enhanced gravitational wave signal at second order. On top of that, we will investigate the impact of quantum gravity corrections of the form $ R\ln R$ in our initial action \eqref{sea4} at the level of the induced GW signal. As we shall demonstrate, depending on the strength of these corrections, the latter signal can give rise to a quite different spectral shape as compared to the GR case.

\subsection{ Scalar induced gravitational waves in GR}
We commence our study by summarizing first the formalism for the description of the second order scalar induced gravitational waves (SIGWs) in GR~\cite{Matarrese:1992rp,Matarrese:1993zf,Matarrese:1997ay,Mollerach:2003nq} [see~\cite{Domenech:2021ztg} for a review], that is ignoring at first the  StRVM quantum corrections $ R\ln R$, which are of the ``$f(R)$" type. As we will explicitly show in what follows, the gravitational wave signal from the sudden transition from the AMDE to the lRD is adequately described by this formalism.

Working within the so-called Newtonian gauge\footnote{Here, one should refer to the issue of gauge dependence of GWs emitted at second order in cosmological perturbation theory firstly studied in~\cite{Hwang:2017oxa}. As it was shown in~\cite{Tomikawa:2019tvi,DeLuca:2019ufz,Inomata:2019yww,Domenech:2020xin}, the gauge invariance of the second-order GWs is generically retained when the emission is followed by a phase where the GW source is not active anymore. In our case, although the GW emission takes place during an eMD era driven by the axion field, during which the scalar and the tensor modes are coupled to each other, it is followed by the late RD era, during which the scalar perturbations decay very quickly and decouple from the tensor perturbations~\cite{Inomata:2019yww,Domenech:2020xin}. Thus, the GW signal computed here in the Newtonian gauge during the late RD era is gauge-independent.}, the perturbed metric is written as

\beq 
\mathrm{d}s^2 = a^2(\eta)\left\lbrace-(1+2\Psi)\mathrm{d}\eta^2  + 
\left[(1-2\Psi)\delta_{ij} + 
\frac{h_{ij}}{2}\right]\mathrm{d}x^i\mathrm{d}x^j\right\rbrace \label{ newton1}, \quad i,j =1,2,3\,,
\eeq
where $\Psi$ is the first order Bardeen gravitational potential\footnote{There is only one Bardeen potential instead of two, since in the absence of anisotropic stress, they are equal, which is indeed the case for the periods we investigate here~\cite{Domenech:2021ztg}} and $h_{ij}$ the 
second order tensor perturbation. After substituting this metric into Einstein's field equations in Fourier space, neglecting entropic perturbations and assuming that $c^2_\mathrm{s,tot}\equiv \frac{p^{(1)}_\mathrm{tot}}{\rho^{(1)}_\mathrm{tot}} \simeq w_\mathrm{tot}\equiv \frac{p^{(0)}_\mathrm{tot}}{\rho^{(0)}_\mathrm{tot}}$, one obtains that $\Psi_\boldmathsymbol{k}$ obeys the following dynamics:
\beq\label{eq:Psi_eq}
\Psi_\boldmathsymbol{k}^{\prime\prime} + \frac{6(1+w_\mathrm{tot})}{1+3w_\mathrm{tot}}\frac{1}{\eta}\Psi_\boldmathsymbol{k}^{\prime} + w_\mathrm{tot}k^2\Psi_\boldmathsymbol{k} =0,
\eeq
where $\prime$ denotes differentiation with respect to the conformal time. Regarding now $h_\boldmathsymbol{k}$, we obtain straightforwardly that 
\beq\label{eq:h_k}
h_\boldmathsymbol{k}^{s,\prime\prime} + 
2\mathcal{H}h_\boldmathsymbol{k}^{s,\prime} + k^{2} h^s_\boldmathsymbol{k} = 4 
S^s_\boldmathsymbol{k}, 
\eeq
where $s = (+), (\times)$ stands for the two tensor mode polarisation states in GR, with $\mathcal{H}$ being the conformal Hubble parameter, while
the polarization tensors  $e^{s}_{ij}(k)$ are the standard  
ones~\cite{Espinosa:2018eve}. The source function 
$S^s_\boldmathsymbol{k}$ is given by~\cite{Baumann:2007zm}
\bea
S^s_\boldmathsymbol{k}  = \int\frac{\mathrm{d}^3 
\boldmathsymbol{q}}{(2\pi)^{3/2}}e^s_{ij}(\boldmathsymbol{k})q_iq_j\Big[
2\Psi_\boldmathsymbol{q}\Psi_\boldmathsymbol{k-q} 
+ \frac{4}{3(1\!+\!w)}(\mathcal{H}^{-1}\Psi_\boldmathsymbol{q} 
^{\prime}+\Psi_\boldmathsymbol{q})(\mathcal{H}^{-1}\Psi_\boldmathsymbol{k-q} 
^{\prime}+\Psi_\boldmathsymbol{k-q}) \Big]
\label{eq:Source:def}
\eea
and it is evident that it is quadratically dependent on the first order scalar metric perturbation $\Psi$. Interestingly enough, \Eq{eq:h_k} can be solved analytically through the use of the Green's function formalism, where one can write the solution for $h^s_\boldmathsymbol{k}$ as
\bea
\label{tensor mode function}
a(\eta)h^s_\boldmathsymbol{k} (\eta)  = 4 \int^{\eta}_{\eta_\mathrm{d}}\mathrm{d}\bar{\eta}\,  G^s_\boldmathsymbol{k}(\eta,\bar{\eta})a(\bar{\eta})S^s_\boldmathsymbol{k}(\bar{\eta}),
\eea
with the Green's function  $G^s_{\bm{k}}(\eta,\bar{\eta})$ being the solution of the homogeneous equation 
\beq
\label{Green function equation}
G_\boldmathsymbol{k}^{s,\prime\prime}(\eta,\bar{\eta})  + \left( k^{2} -\frac{a^{\prime\prime}}{a}\right)G^s_\boldmathsymbol{k}(\eta,\bar{\eta}) = \delta\left(\eta-\bar{\eta}\right),
\eeq
with the boundary conditions $\lim_{\eta\to \bar{\eta}}G^s_\boldmathsymbol{k}(\eta,\bar{\eta}) = 0$ and $ \lim_{\eta\to \bar{\eta}}G^{s,\prime}_\boldmathsymbol{k}(\eta,\bar{\eta})=1$.  

One then can recast the second order tensor power spectrum  $\mathcal{P}^{(2)}_{h,s}(\eta,k)$, defined as the equal time correlator of the tensor perturbations, as~\cite{Ananda:2006af,Baumann:2007zm,Kohri:2018awv,Espinosa:2018eve}
\beq\label{eq:P_h_GR}
\mathcal{P}^{(2)}_{h,s}(\eta,k) = 4\int_{0}^{\infty} 
\mathrm{d}v\int_{|1-v|}^{1+v}\mathrm{d}u\! \left[ \frac{4v^2 - 
(1\!+\!v^2\!-\!u^2)^2}{4uv}\right]^{2}
\cdot
I^2(u,v,x)\mathcal{P}_\Psi(kv)\mathcal{P}
_\Psi(ku)\,,
\eeq
where the two auxiliary variables $u$ and $v$ are defined as $u \equiv 
|\boldmathsymbol{k} - \boldmathsymbol{q}|/k$ and $v \equiv q/k$, and the kernel 
function $I(u,v,x)$ can be recast as
\bea
\label{I function}
I(u,v,x) = \int_{x_\mathrm{d}}^{x} \mathrm{d}\bar{x}\, \frac{a(\bar{x})}{a(x)}\, k\, G^s_{k}(x,\bar{x}) F_k(u,v,\bar{x}),
\eea
with $x\equiv k\eta$ and $F_k(u,v,x)$ reading as
\beq
\label{F}
\!\!\!\!\!
F(v,u,x)  \equiv 2T_\Psi(vx)T_\Psi\left(ux\right)  + \frac{4}{3(1+w)}\left[\mathcal{H}^{-1}T_\Psi^{\prime}(vx)+T_\Psi(vx)\right]\left[\mathcal{H}^{-1} T_\Psi^{\prime}\left(ux\right)+T_\Psi\left(ux\right)\right].
\eeq
In the above equation, $T_\Psi$ is the transfer function of the gravitational potential $\Psi$, defined through the equation $\Psi_\boldmathsymbol{k}(\eta) \equiv T_\Psi(\eta) \psi_\boldmathsymbol{k}$, where $\psi_\boldmathsymbol{k}$ is the value of the gravitational potential at some initial time, here considered as the time at the beginning of the eMD era driven by the axion field, and $\Psi_\boldmathsymbol{k}(\eta)$ is the potential $\Psi$ in the late-time limit.

\subsection{Scalar induced gravitational waves including StRVM quantum corrections}

Having introduced above the SIGW formalism in the case of GR, we explore now quantum-gravity effects of StRVM type via their impact on the SIGW signal. Specifically, we may use the action
\eqref{sea4} introduced in \Sec{sec:ags}. We mention here for completeness, that this action can be readily expressed as an $f(R)$ modified gravity theory where the $f(R)$ function reads
\beq
f(R)= c_0+R\left(c_1+c_2\log\left(\frac{R}{R_0}\right)\right), \label{RVM}
\eeq
with $R_0$ being the scalar curvature at an initial time, here considered as the onset of the AMDE era.
The advantage of this formulation is that we can straightforwardly use the formalism developed in the recent work by Zhou et al.~\cite{Zhou:2024doz} in order to extract the relevant SIGW signal, including the quantum gravity $R\ln R$ corrections.

First, we note that in our case, we should be generic in keeping both Bardeen potentials, therefore instead of \eqref{ newton1} we have

\begin{equation}\label{eq:ds}
\begin{aligned}
	\mathrm{d} s^2  =a^2(\eta)\left[-\left(1+2 \Phi^{(1)}\right) \mathrm{d} \eta^2 +\left(\left(1-2 \Psi^{(1)}\right) \delta_{i j}+\frac{1}{2} h_{i j}^{(2)}\right) \mathrm{d} x^i \mathrm{~d} x^j\right],
\end{aligned}
\end{equation}
where the index $(1)$ stands for first-order perturbed quantities while the index $(2)$ stand for the second-order ones. For this theory, \Eq{eq:Psi_eq} for the Bardeen gravitational potentials $\Phi^{(1)}$ and $\Psi^{(1)}$ will be recast as~\cite{Tsujikawa:2007gd}
\beq
2\mathcal{H}\left[3(c^2_\mathrm{s,tot} - w_\mathrm{tot})\mathcal{H}\Phi^{(1)} + \Phi^{(1)\prime}+\left(2+3c^2_\mathrm{s,tot}\Psi^{(1)\prime}\right) \right] + 2\Psi^{(1)\prime\prime}+\Delta\Phi^{(1)} -\Delta\Psi^{(1)} - 2c^2_\mathrm{s,tot} \Delta\Psi^{(1)} = 0,
\eeq
\beq
\Psi^{(1)} - \Phi^{(1)} = \frac{F^{(1)}}{F^{(0)}},
\eeq
while, concerning the tensor perturbations, \Eq{eq:h_k} gets modified to~\cite{Zhou:2024doz}
\begin{eqnarray}
\label{eq:ehfR}
    h_{\mathbf{k}}^{\lambda,(2)''}\left(\eta \right)+\frac{1}{F^{(0)}}\left(2\mathcal{H}F^{(0)} +F^{(0)'} \right)h_{\mathbf{k}}^{\lambda,(2)'}\left(\eta \right) +k^2 h_{\mathbf{k}}^{\lambda,(2)}\left(\eta \right)
    =\frac{4}{F^{(0)}} \left(\mathcal{S}_{\mathbf{k}}^{\lambda,(2)} \left(\eta \right)+\sigma^{\lambda,(2)}_{\mathbf{k}}\left( \eta \right)\right) \ ,
\end{eqnarray}
with the source terms $\mathcal{S}_{\mathbf{k}}^{\lambda,(2)} \left(\eta \right)$ and $\sigma^{\lambda,(2)}_{\mathbf{k}}\left( \eta \right)$ reading as 
\begin{eqnarray}
     \mathcal{S}_{\mathbf{k}}^{\lambda,(2)}(\eta)&&=\int\frac{d^3p}{(2\pi)^{3/2}}\varepsilon^{\lambda, lm}(\mathbf{k})p_lp_m\left(\left(1+\frac{4}{3(1+w)}\right)\Phi^{(1)}_{\mathbf{k}-\mathbf{p}}(\eta) \Phi^{(1)}_{\mathbf{p}}(\eta)+2\Psi^{(1)}_{\mathbf{k}-\mathbf{p}}(\eta) \Phi^{(1)}_{\mathbf{p}}(\eta) \right. \nonumber\\
     &&\left.+\frac{8}{3(1+w)\mathcal{H}}\Psi^{(1)'}_{\mathbf{k}-\mathbf{p}}(\eta) \Phi^{(1)}_{\mathbf{p}}(\eta)+\frac{4}{3(1+w)\mathcal{H}^2}\Psi^{(1)'}_{\mathbf{k}-\mathbf{p}}(\eta) \Psi^{(1)'}_{\mathbf{p}}(\eta)  \right. \nonumber\\
     &&\left. -\Psi^{(1)}_{\mathbf{k}-\mathbf{p}}(\eta) \Psi^{(1)}_{\mathbf{p}}(\eta) \right) \ , \label{eq:Ss}\\
  \sigma_{\mathbf{k}}^{\lambda,(2)}(\eta)&&=\int\frac{d^3p}{(2\pi)^{3/2}}\varepsilon^{\lambda, lm}(\mathbf{k})p_lp_m\left(2\Psi^{(1)}_{\mathbf{k}-\mathbf{p}} (\eta) F^{(1)}_{\mathbf{p}}(\eta)+ \left(F^{(0)}-1 \right)\left(3\Psi^{(1)}_{\mathbf{k}-\mathbf{p}} (\eta) \Psi^{(1)}_{\mathbf{p}}(\eta) \right.\right. \nonumber\\
  &&\left.\left.-\Phi^{(1)}_{\mathbf{k}-\mathbf{p}} (\eta) \Phi^{(1)}_{\mathbf{p}}(\eta) \right) \right) \ , \label{eq:Ssigma}
 \end{eqnarray}
where $F$ stands for the derivative of the function $f(R)$ with respect to $R$, i.e. $F\equiv\frac{\mathrm{d}f(R)}{\mathrm{d}R}$. We also note here that in the case of $f(R)$ gravity theories we have the presence of an anisotropic source term of SIGWs denoted as $\sigma_{\mathbf{k}}^{\lambda,(2)}(\eta)$.

At the end, the solution for the $(\times)$ and $(+)$ second-order tensor perturbations will read as 
\begin{eqnarray}
    h_{\mathbf{k}}^{\lambda,(2)}\left(\eta \right)&=&\int\frac{d^3p}{(2\pi)^{3/2}}\varepsilon^{\lambda, lm}(\mathbf{k})p_lp_m I^{(2)}_{hf} \left(u,v,x \right) \zeta_{\mathbf{k}-\mathbf{p}}\zeta_{\mathbf{p}} \ , \label{eq:sol1}
\end{eqnarray}
with the second order kernel function $I^{(2)}_{hf}(u,v,x)$ in Eq.~(\ref{eq:sol1}) satisfying the following equation:
\begin{eqnarray}\label{eq:kernel}
    I^{(2)''}_{hf}\left(u,v,x  \right)+\frac{1}{F^{(0)}}\left(2\mathcal{H}F^{(0)}+F^{(0)'} \right)I^{(2)'}_{hf}\left(u,v,x  \right)
    +k^2I^{(2)}_{hf}\left(u,v,x  \right)=4\left(f^{(2)}_s\left(u,v,x  \right)+f^{(2)}_{\sigma}\left(u,v,x  \right) \right) \ ,
\end{eqnarray}
where
    \begin{align}
        f^{(2)}_s\left(u,v,x  \right)&=\left(1+\frac{4}{3(1+w_{\mathrm{tot}})}\right)T_{\Phi}(ux) T_{\Psi}(vx)+2T_{\Psi}(ux) T_{\Phi}(vx) -T_{\Psi}(ux) T_{\Psi}(vx) \nonumber\\
    &+\frac{8u}{3(1+w_{\mathrm{tot}})\mathcal{H}}\frac{d}{d(ux)}T_{\Psi}(ux) T_{\Phi}(vx)   +\frac{4uv}{3(1+w_{\mathrm{tot}})\mathcal{H}^2}\frac{d}{d(ux)}T_{\Psi}(ux)\frac{d}{d(vx)}T_{\Psi}(vx)           \ ,  \\
    f^{(2)}_{\sigma}\left(u,v,x  \right)&=2T_{\Psi}(ux) T_{s}(vx)+ \left(F^{(0)}-1 \right)\left(3T_{\Psi}(ux) T_{\Psi}(vx) -T_{\Phi}(ux)T_{\Phi}(vx) \right)              \ . 
    \end{align}
\subsubsection{The scalaron contribution}
Additionally, distinctive of the richer phenomenology of an $f(R)$ theory is the existence of an extra massive polarisation mode (scalaron) $\phi_s$, defined as $\phi_s\equiv \frac{\mathrm{d}f(R)}{\mathrm{d}R} = F(R)$, whose equation
on FLRW space-time can be recast as \cite{Katsuragawa:2019uto}
\begin{eqnarray}\label{eq:phis1}
   \phi_{s}^{(1)''}+2\mathcal{H}\phi_{s}^{(1)'}-\left(\Delta+m^2_s  \right)\phi_{s}^{(1)}=-\square^{(1)} \phi_{s}^{(0)} +\frac{\kappa}{3} T^{M,(1)},
\end{eqnarray}
where $T^{M}$ is the trace of the matter stress-energy tensor. Given the wave equation \Eq{eq:phis1}, one can thus regard the scalaron field $\phi_{s}$ as an additional massive scalar mode for the GW polarisation with source term: $\mathcal{S}^{(1)}_{s}=-\square^{(1)} \phi_{s}^{(0)} +\frac{\kappa}{3} T^{M,(1)}$. In the following however, instead of solving \Eq{eq:phis1} for the scalaron mode, which might require numerical investigations, we can express the perturbed scalaron field $\phi_{s}^{(1)}$ in terms of the first-order Bardeen potentials $\phi^{(1)}$ and $\psi^{(1)}$ by writing $\phi_{s}^{(1)}$ from its definition as 
\beq\label{eq:phi_s_1}
\phi_{s,\mathbf{k}}^{(1)}(\eta) = F^{(1)}_{\mathbf{k}}(\eta) = F^{(0)}_R(\eta) R^{(1)}_{\mathbf{k}}(\eta), 
\eeq
where $ R^{(1)}_{\mathbf{k}}$ is the first order scalar curvature (Ricci scalar) given in terms of $\Phi^{(1)}$ and $\Psi^{(1)}$ as follows~\cite{Tsujikawa:2007gd}:
\beq
R^{(1)}_{\mathbf{k}}(\eta) = -\frac{2}{a^2}\Bigl[6(\mathcal{H}^\prime +\mathcal{H}^2)\Phi^{(1)}_{\mathbf{k}}(\eta) + 3\mathcal{H}\left(\frac{\mathrm{d}\Phi^{(1)}_{\mathbf{k}}(\eta)}{\mathrm{d}\eta} + 4\frac{\mathrm{d}\Psi^{(1)}_{\mathbf{k}}(\eta)}{\mathrm{d}\eta}\right) + 3\frac{\mathrm{d}^2\Psi^{(1)}_{\mathbf{k}}(\eta)}{\mathrm{d}\eta^2} - k^2\Phi^{(1)}_{\mathbf{k}}(\eta) + 2k^2\Psi^{(1)}_{\mathbf{k}}(\eta)\Bigr].
\eeq

Relating now on superhorizon scales the first-order Bardeen potentials $\Phi^{(1)}_{\mathbf{k}}$ and $\Psi^{(1)}_{\mathbf{k}}$ with the comoving curvature perturbation $\mathcal{R}_\boldmathsymbol{k}$ as~\cite{Papanikolaou:2021uhe} 
\begin{eqnarray}
\Phi^{(1)}_{\mathbf{k}}(\eta)&=&\left(\frac{3+3w_{\mathrm{tot}}}{5+3w_{\mathrm{tot}}}\right) T_{\Phi}(x)\zeta_{\mathbf{k}} \ \  , \  \  \Psi^{(1)}_{\mathbf{k}}(\eta)=\left(\frac{3+3w_{\mathrm{tot}}}{5+3w_{\mathrm{tot}}}\right) T_{\Psi}(x)\zeta_{\mathbf{k}} \ ,
\end{eqnarray}
one can recast $\phi_{s,\mathbf{k}}^{(1)}(\eta)$ as
\begin{eqnarray}\label{eq:phi_s_1st_order}
\phi_{s,\mathbf{k}}^{(1)}\left(\eta \right)&\equiv&T_{s}(x)\zeta_{\mathbf{k}}=-\frac{2F_R^{(0)}}{a^2}\left(\frac{3+3w_{\mathrm{tot}}}{5+3w_{\mathrm{tot}}}\right)\left(6\left(\mathcal{H}'+\mathcal{H}^2 \right) T_{\Phi}(x)+3\mathcal{H}k\left( \frac{d}{dx}T_{\Phi}(x) +4\frac{d}{dx}T_{\Psi}(x)  \right) \right. \nonumber \\
     &&\left.+3k^2\frac{d^2}{d^2x}T_{\Psi}(x)-k^2T_{\Phi}(x)+2k^2T_{\Psi}(x)   \right)\zeta_{\mathbf{k}},
\end{eqnarray}
where $x=k \eta$.

Finally, from \Eq{eq:phi_s_1st_order} we can infer that the tensor power spectrum associated with the first order scalaron perturbation $\phi_{s,\mathbf{k}}^{(1)}$ will read as 
\beq\label{eq:P_s}
\mathcal{P}^{(1)}_{h,sc}(k) = T^2_s(x)\mathcal{P}_\zeta(k).
\eeq
One then can write the total GW spectral abundance $\Omega_\mathrm{GW}(\eta,k)$ at a time $\eta$ when the modes considered sub-horizon (flat spacetime approximation)~\cite{Maggiore:1999vm}, being the sum of two contributions, namely the GR and the scalaron contributions, and reading as~\cite{Zhou:2024doz}
\beq\label{Omega_GW_2}
\Omega_\mathrm{GW}(\eta,k) = 
\frac{1}{6}\left(\frac{k}{\calH(\eta)}\right)^{2}\left[\frac{\overline{\mathcal{P}^{(2)}
_{h,s}(\eta,k)}}{4} + \mathcal{P}^{(1)}_{h,sc}(\eta,k)\right],
\eeq
where $\mathcal{P}^{(2)}_{h,s}(\eta,k)$ is given by \Eq{eq:P_h_GR} and $ \mathcal{P}^{(1)}_{h,sc}$ by \Eq{eq:P_s}. The bar stands for an oscillation average, i.e the GW envelope.
Finally, considering that the 
radiation energy 
density reads as $\rho_r = 
\frac{\pi^2}{30}g_{*\mathrm{\rho}}T_\mathrm{r}^4$ and that the temperature of 
the primordial plasma $T_\mathrm{r}$ scales as $T_\mathrm{r}\propto 
g^{-1/3}_{*\mathrm{S}}a^{-1}$, one finds that the GW spectral abundance at 
our present epoch reads as~\cite{Espinosa:2018eve}
\beq\label{Omega_GW_RD_0}
\Omega_\mathrm{GW}(\eta_0,k) = 
\Omega^{(0)}_\mathrm{r}\frac{g_{*\mathrm{\rho},\mathrm{*}}}{g_{*\mathrm{\rho},0}}
\left(\frac{g_{*\mathrm{S},\mathrm{0}}}{g_{*\mathrm{S},\mathrm{*}}}\right)^{4/3}
\OmegaGW(\eta_\mathrm{*},k),
\eeq
where $g_{*\mathrm{\rho}}$ and $g_{*\mathrm{S}}$ denote the energy and 
entropy relativistic degrees of freedom. Since the AMDE eras considered here take place before BBN, one can show that $\frac{g_{*\mathrm{\rho},\mathrm{*}}}{g_{*\mathrm{\rho},0}}
\left(\frac{g_{*\mathrm{S},\mathrm{0}}}{g_{*\mathrm{S},\mathrm{*}}}\right)^{4/3}\sim 0.4$ while the radiation abundance today, $\Omega^{(0)}_\mathrm{r} \simeq 10^{-4}$ as measured by Planck~\cite{Planck:2018}. Note that the reference 
conformal time $\eta_\mathrm{*}$ in the case of a sudden transition from the AMDE to the lRD era should be of 
$\mathcal{O}(1)\eta_\mathrm{ra}$~\cite{Inomata:2019ivs}, with $\eta_\mathrm{ra}$ being the time at the onset of the lRD era, i.e. time of the axion decay.

\subsection{Staying within the linear regime}
Before continuing to the derivation of the SIGW signal within the StRVM framework, we need to make sure that we stay within the perturbative regime, where all perturbative quantities are less than one. Before doing so, we should stress that we consider modes contributing to the SIGW signal that re-enter the cosmological horizon during the eMD era driven by the axion field, that is within the range $[k_\mathrm{ra},k_\mathrm{da}]$ where $k_\mathrm{da}$ and $k_\mathrm{ra}$ are the modes crossing the horizon at the beginning and the end of the AMDE era respectively. Thus, the maximum comoving scale (minimum physical scale) considered here is $k_\mathrm{da}$.

However, since the sub-horizon energy density perturbations during a matter-dominated era scale linearly with the scale factor, i.e. $\delta\propto a$, there is a possibility that some scales within the range $[k_\mathrm{ra},k_\mathrm{da}]$ become non-linear, i.e. $\delta_k>1$. To ensure thus that we stay within the linear perturbative regime, we set a non-linear scale $k_\mathrm{NL}$ by requiring that $\delta_{\mathrm{k_\mathrm{NL}}}(\eta_\mathrm{ra}) = 1$. In particular, following the analysis of~\cite{Assadullahi:2009nf,Inomata:2020lmk} one can show that the non-linear ultra-violet (UV) cut-off scale at which $\delta_{k_\mathrm{NL}}(\eta_\mathrm{ra}) = 1$ can be recast as
\begin{equation}
k_\mathrm{NL} \simeq \sqrt{\frac{5}{2}}\mathcal{P}^{-1/4}_\mathcal{R}(k_\mathrm{NL})\mathcal{H}(\eta_\mathrm{ra}).
\end{equation}

Since StRVM can provide an inflationary setup with $n_\mathrm{s} = 0.965$~\cite{gms,dorlis2}, one can assume as a first approximation a scale-invariant curvature power spectrum of amplitude $2.1\times 10^{-9}$ as imposed by Planck~\cite{Planck:2018}, giving rise to $ k_\mathrm{NL} \simeq 470/\eta_\mathrm{ra} \simeq 235 k_\mathrm{ra}$~\cite{Inomata:2019ivs}. However, strictly speaking, one should take into account the tilt of the power spectrum leading to a larger value of $k_\mathrm{NL}$. Accounting therefore for this tilt effect, we find  $k_\mathrm{NL} \simeq 400k_\mathrm{ra}$. Consequently the maximum comoving scale considered here reads as
\beq\label{eq:k_max}
k_\mathrm{max} = \mathrm{min}[k_\mathrm{da},\, 400k_\mathrm{ra}].
\eeq

We need to mention at this stage that, in principle,  one can also account for the emission of non-linear modes with $\delta_{k>k_\mathrm{NL}} > 1$, which can potentially enhance the GW signal~\cite{Jedamzik:2010hq,Eggemeier:2022gyo,Fernandez:2023ddy,Padilla:2024cbq}. The treatment of these modes require however high-cost GR numerical simulations, which lie beyond the scope of the current work. Therefore, by neglecting them we underestimate the GW signal, thus giving a conservative estimate for the GW amplitude.

\subsection{The total scalar induced gravitational wave signal}
Let us study then now the total SIGW signal. Regarding the GR contribution one may infer from \Eq{eq:ehfR}, \Eq{eq:Ss} and \Eq{eq:Ssigma} two contributions, the standard GR one \eqref{eq:Ss} plus an anisotropic source contribution \eqref{eq:Ssigma}. As we show in Appendix \ref{app:geometric_anisotropic_stress}, for the case of an axion driven matter-dominated era, $\Phi^{(1)}_\boldmathsymbol{k} \simeq \Psi^{(1)}_\boldmathsymbol{k}$ leading to a negligible anistropic source term $\sigma^{(2)}_\boldmathsymbol{k}(\eta)$, as it can be seen from \Eq{eq:Ssigma}. One then is met with the usual GR SIGW signal with no anisotropic stress studied already in~\cite{Inomata:2019ivs}. Interestingly enough, due to the sudden transition from the AMDE to the lRD era, one obtains in this case  a resonantly enhanced induced GW signal, peaking around the UV cut-off scale $k_\mathrm{max}$, \Eq{eq:k_max}. This is due to the fact that the time derivative of the Bardeen potential goes very quickly from $\Psi^\prime = 0 $ (since in a MD era $\Psi = \mathrm{constant}$ ) to $\Psi^\prime \neq 0$. This entails a resonantly enhanced production of GWs sourced mainly by the $\mathcal{H}^{-2}\Psi^{\prime 2}$ term in \Eq{eq:Source:def} [See~\cite{Inomata:2019ivs,Domenech:2021ztg} for more details.]. 

At the end, the relevant amplified GR SIGW signal is derived as follows
\beq\label{eq:Omega_GW_resonance}
\Omega^{\mathrm{res}}_\mathrm{GW}(\eta_0,k) = c_g\Omega^{(0)}_\mathrm{r} \Omega^\mathrm{res}_\mathrm{GW}(\eta_\mathrm{lRD},k),
\eeq
where $c_g =\frac{g_{*\mathrm{\rho},\mathrm{*}}}{g_{*\mathrm{\rho},0}}
\left(\frac{g_{*\mathrm{S},\mathrm{0}}}{g_{*\mathrm{S},\mathrm{*}}}\right)^{4/3} \sim 0.4$ and $\eta_\mathrm{lRD}\sim O(\eta_\mathrm{ra})$ stands for a time during the lRD era by which the curvature perturbations decouple from the tensor perturbations, thus one can assume freely propagating GWs. For the case of a scale-invariant curvature power spectrum, $\mathcal{P}_\zeta \simeq 2.1 \times 10^{-9}$, the quanrtity $\Omega^\mathrm{res}_\mathrm{GW}(\eta_\mathrm{lRD},k)$ can be approximately written (after a lengthy but straightforward calculation) as~\cite{Inomata:2019ivs} 
\beq\label{eq:Omega_GW_GR_res}
\Omega^\mathrm{res}_\mathrm{GW}(\eta_\mathrm{lRD},k)\simeq 
\begin{cases}
    3\times 10^{-7}\left(\frac{k}{k_\mathrm{ra}}\right)^3\left(\frac{k_\mathrm{max}}{k_\mathrm{ra}}\right)^5, \; \mathrm{for}\; 150\left(\frac{k_\mathrm{max}}{k_\mathrm{ra}}\right)^{-5/3}\lesssim\frac{k}{k_\mathrm{ra}}\ll 1\\
    10^{-6}\frac{k}{k_\mathrm{ra}}\left(\frac{k_\mathrm{max}}{k_\mathrm{ra}}\right)^5 , \; \mathrm{for}\; 1\ll 1<\frac{k}{k_\mathrm{ra}}\lesssim \left(\frac{k_\mathrm{max}}{k_\mathrm{ra}}\right)^{5/6}\\
    7\times 10^{-7}\left(\frac{k}{k_\mathrm{ra}}\right)^7,  \; \mathrm{for}\; \left(\frac{k_\mathrm{max}}{k_\mathrm{ra}}\right)^{5/6}\lesssim \frac{k}{k_\mathrm{ra}}\lesssim \frac{k_\mathrm{max}}{k_\mathrm{ra}}\\
    0, \; \mathrm{for}\;k_\mathrm{max} \lesssim k\lesssim 2k_\mathrm{max},
\end{cases}
\eeq
where for our setup $k_\mathrm{max}$ is given by \Eq{eq:k_max}.

As mentioned before, the GR induced GW signal studied here is expected to peak at $k_\mathrm{max}$~\cite{Inomata:2019yww}. Therefore, the peak frequency of the GR-induced GW signal, can be computed as follows:
\beq\label{eq:f_GW}
f_\mathrm{GW} = \frac{k}{2\pi a_0} = \frac{k}{2\pi a_{\mathrm{ra}}} \frac{a_{\mathrm{ra}}}{a_\mathrm{eq}} \frac{a_\mathrm{eq}}{a_0}  =
\frac{k}{2\pi a_{\mathrm{ra}}} \left(\frac{\rho_\mathrm{eq}}{\rho_{\mathrm{ra}}}\right)^{1/4}\left(\frac{\rho_\mathrm{0}}{\rho_\mathrm{eq}}\right)^{1/3},
\eeq
where $\rho_\mathrm{eq}=1.096\times 10^{-36}\mathrm{GeV}^4$ and $\rho_0 = 3.6\times 10^{-47}\mathrm{GeV}^4$ are the background energy densities at the matter-radiation equality and today, respectively. Lastly, $\rho_{\mathrm{ra}}$ and $a_{\mathrm{ra}}$ respectively stand for the energy density and the scale factor at the end of the AMDE, namely at the time of the axion decay to radiation.

In order to compute now $\rho_{\mathrm{ra}}$ and $a_{\mathrm{ra}}$, we shall invoke our analysis from the previous section. Specifically, the AMDE starts when the temperature of the Universe is of the order of the axion mass $ T_a \approx m_a$ and finishes when the axion starts to decay to radiation, namely when the Hubble parameter is of the order of the decay rate of the axion's dominant decay channel, that is when $H \approx \Gamma_\mathrm{a} \equiv H_\mathrm{ra} $. Now since the AMDE is by definition a matter-dominated era, $\rho \sim a^{-3}$ so by setting $a_\mathrm{da} \equiv 1 $ we find $a_{\mathrm{ra}} =  (H^ 2_\mathrm{da}/H^ 2_{\mathrm{ra}})^{1/3}$. As was already stated in the previous section, the Hubble parameter at the start of the AMDE is given by

\begin{align}\label{Ha}
H_\mathrm{da} = \Big(\,\frac{\rho_a}{3 M^2_{\rm Pl}}\,\Big)^{1/2} = \Big(\,\frac{\zeta(3)\, m_a \, T_a^3}{3\pi^2 M_{\rm Pl}^2}\,\Big)^{1/2}\, = \Big(\,\frac{\zeta(3)\, m_a^ 4}{3\pi^2 M_{\rm Pl}^2}\,\Big)^{1/2},
\end{align}
With regard now to the decay rate $\Gamma_a$, it can be calculated from \eqref{widthaxion}, which we restate here for the reader's convenience
\bea\label{Ga}
\Gamma_a  \sim \frac{m_a^3}{64\pi \, f_a^2} \sim \frac{\Lambda_1^6}{64\, \pi \, f_a^5} \, \, \ll H_I\,.
\eea
Thus, the peak frequency of our signal can be computed by \Eq{eq:f_GW} with $k = k_\mathrm{max}$ given by \Eq{eq:k_max}. 

Regarding now the scalaron-associated induced GW signal, one can make an analytic prediction for $\Omega^{(sc)}_\mathrm{GW}$ working with the modes that are sub-horizon at the time of the computation of the GW spectrum~\cite{Maggiore:1999vm}. This is the case for our setup, since the scales considered here are such that $k>k_\mathrm{ra}$.
Doing so, one can show that for $k\gg aH$~\cite{Tsujikawa:2007gd}
\beq\label{eq:R_1_sub_horizon}
R^{(1)}_{\mathbf{k}}\left(\eta \right) \simeq -2\frac{k^2}{a^2}\frac{\Phi^{(1)}}{1 + 4\frac{k^2}{a^2}\frac{F^{(0)}_{R}}{F^{(0)}}}.
\eeq
Thus, plugging \Eq{eq:R_1_sub_horizon} in \Eq{eq:phi_s_1} one can show straightforwardly from \Eq{eq:P_s} and \Eq{Omega_GW_2} that during the AMDE era $\Omega^{sc}_\mathrm{GW}(\eta, k)$ reads:
\beq\label{eq:Omega_GW_scalaron}
\Omega^{sc}_\mathrm{GW}(\eta, k) \simeq \frac{24c_g\Omega^{(0)}_\mathrm{r}}{25} \left(\frac{c_2}{\Mp^2}\right)^2  \left(\frac{k}{k_\mathrm{ra}}\right)^6\mathcal{P}_\zeta(k),
\eeq
where we used the fact that (see Appendix \ref{app:geometric_anisotropic_stress}):
\begin{align}\label{condfr}
\frac{k^2}{a^2} \frac{F^{(0)}_{R}}{F^{(0)}}\ll 1\,,
\end{align}
which expresses the condition that one encounters a negligible geometric-in-origin anisotropic stress.

Since the frequency of GWs is defined as $f\equiv\frac{k}{2\pi a_0}$ where $a_0$ is the scale factor today, one can infer from \Eq{eq:Omega_GW_scalaron} that the scalaron associated SIGW signal is characterised by a universal frequency scaling of $f^6$, which characterises not only our  
logarithmically quantum-gravity-corrected gravitational StRVM effective action \eqref{sea4},
but also any other $f(R)$ modified gravity model which satisfy the condition \eqref{condfr}.

\begin{figure}[h!]
\centering
\includegraphics[width=0.80\textwidth]{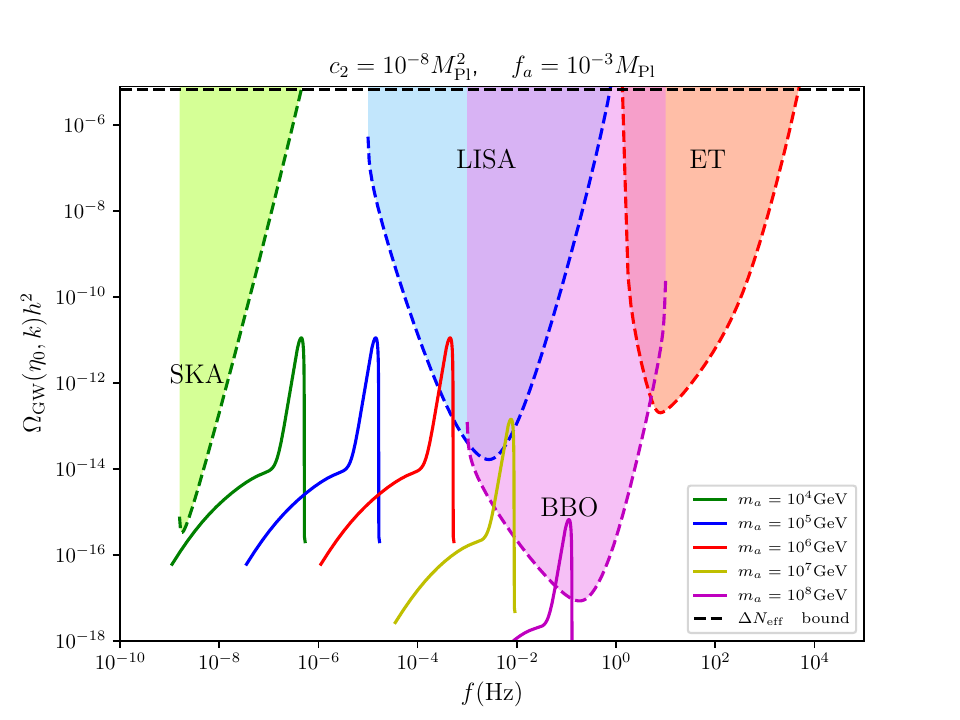}
\caption{\it{ We show the total induced gravitational-wave (GW) signal today as a function of the frequency for $f_a = 10^{-3}\Mp$ and $c_2 = 10^{-8}\Mp^2$ and for different values of the axion mass $m_a$. We superimpose as well the sensitivity curves of future GW detectors, including LISA \cite{LISA:2017pwj, Karnesis:2022vdp}, ET \cite{Maggiore:2019uih}, SKA \cite{Janssen:2014dka}, and BBO \cite{Harry:2006fi}.  }}
\label{fig:Omega_GW_vs_ma}
\end{figure}

Interestingly enough, one can find the minimum value of the quantum correction coefficient $c_2$ for the production of an abundant scalaron-induced GW signal, with a frequency scaling of $f^6$. To this end, we compare the amplitude of the scalaron GW signal \eqref{eq:Omega_GW_scalaron} with that of the GR resonantly enhanced one \eqref{eq:Omega_GW_GR_res} at $k=k_\mathrm{max}$. Since $k_\mathrm{max}$ is the minimum scale between $k_\mathrm{da}$ and $400k_\mathrm{ra}$ we discriminate between two regimes.
\begin{itemize}
\item{$k_\mathrm{da}>400 k_\mathrm{ra}\Rightarrow k_\mathrm{max} =400k_\mathrm{ra}$

In this case, one can show straightforwardly from the aforementioned inequality that
\beq\label{fa_ma_condition_k_max_k_NL}
\left(\frac{f_a}{0.1\Mp}\right)^2\left(\frac{10^9\mathrm{GeV}}{m_a}\right)\geq 0.065,
\eeq
with the scalaron and the GR induced GW amplitudes becoming independent  of the values of the axion mass $m_a$ and coupling $f_a$.
Requiring that $\Omega^{sc}_\mathrm{GW}(\eta_0, 400k_\mathrm{ra}) = \Omega^\mathrm{res}_\mathrm{GW}(\eta_\mathrm{0},400k_\mathrm{ra})$ we arrive at the lower bound:
\beq\label{eq:c_2_constraint_k_max_k_NL}
c_2\geq 2\times 10^{-7}\Mp^2.
\eeq
}
\begin{figure}[h!]
\centering
\includegraphics[width=0.80\textwidth]{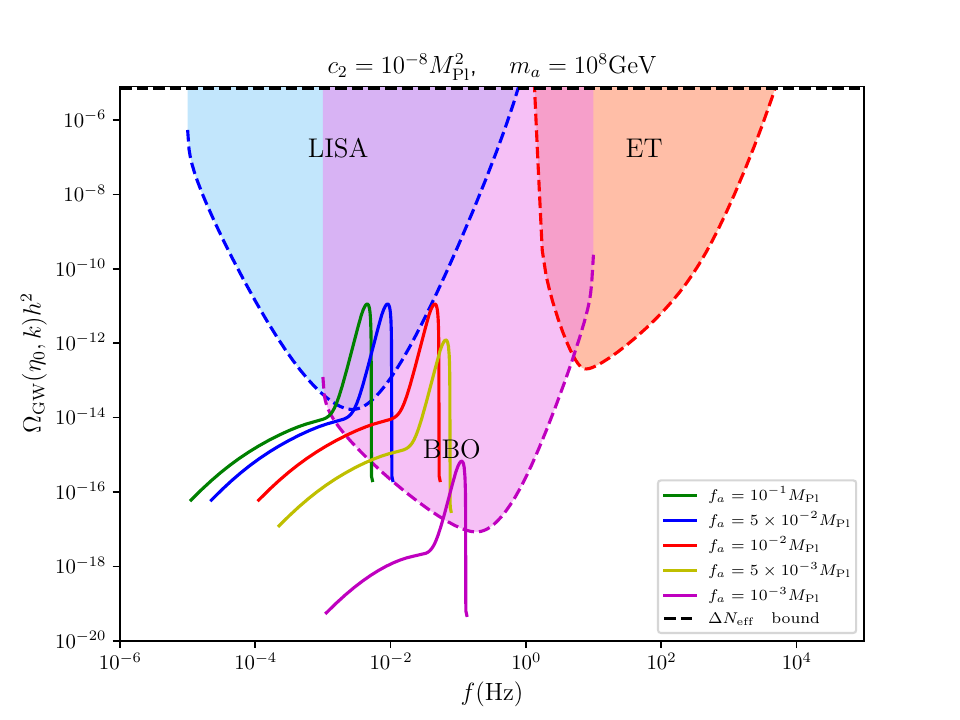}
\caption{\it{ We show the total induced gravitational-wave (GW) signal today as a function of the frequency for $m_a = 10^8\mathrm{GeV}$ and $c_2 = 10^{-8}\Mp^2$ and for different values of the axion coupling $f_a$. We superimpose as well the sensitivity curves of future GW detectors, including LISA \cite{LISA:2017pwj, Karnesis:2022vdp}, ET \cite{Maggiore:2019uih}, SKA \cite{Janssen:2014dka}, and BBO \cite{Harry:2006fi}. }}
\label{fig:Omega_GW_vs_fa}
\end{figure}

\item{
$k_\mathrm{da}<400 k_\mathrm{ra}\Rightarrow k_\mathrm{max} = k_\mathrm{da}$

In this regime, one obtains that
\beq\label{fa_ma_condition_k_max_k_da}
\left(\frac{f_a}{0.1\Mp}\right)^2\left(\frac{10^9\mathrm{GeV}}{m_a}\right)\leq 0.065,
\eeq
with the scalaron and the GR induced GW amplitudes depending now on the values of the axion mass $m_a$ and coupling $f_a$:
\begin{eqnarray}
  \Omega^{sc}_\mathrm{GW}(\eta_0, k_\mathrm{da}) &  = & 7.5\times 10^{4}\left(\frac{c_2}{\Mp^2}\right)^2  \left(\frac{f_a}{0.1\Mp}\right)^4\left(\frac{10^9\mathrm{GeV}}{m_a}\right)^2 \label{eq:Omega_GW_sc_k_max_k_d}\\
\Omega^\mathrm{res}_\mathrm{GW}(\eta_\mathrm{0},k_\mathrm{da})&  = & 2.3\times 10^{-8}\left(\frac{f_a}{0.1\Mp}\right)^{14/3}\left(\frac{10^9\mathrm{GeV}}{m_a}\right)^{7/3} \label{eq:Omega_GW_GR_res_k_max_k_d}. 
\end{eqnarray} 
Requiring, as before, that $\Omega^{sc}_\mathrm{GW}(\eta_0, k_\mathrm{da}) = \Omega^\mathrm{res}_\mathrm{GW}(\eta_\mathrm{0},k_\mathrm{da})$ we obtain:
\beq\label{eq:c_2_constraint_k_max_k_da}
c_2\geq 5.5\times 10^{-7}\Mp^2\left(\frac{f_a}{0.1\Mp}\right)^{2/6}\left(\frac{10^9\mathrm{GeV}}{m_a}\right)^{1/6}.
\eeq

}
\end{itemize}

In \Fig{fig:Omega_GW_vs_ma} and \Fig{fig:Omega_GW_vs_fa} below, we plot the total SIGW spectrum as a function of the frequency by fixing the quantum correction coefficient to $c_2=10^{-8}\Mp^2$ and by varying both $m_a$ and $f_a$. We see that, as we increase the axion mass $m_a$ or decrease the axion coupling $f_a$, the SIGW amplitude is initially constant and then starts to decrease. This is because for small values of $m_a$ or large values of $f_a$ \Eq{fa_ma_condition_k_max_k_NL} is fulfilled, i.e. $k_\mathrm{max}=k_\mathrm{NL}=400k_\mathrm{ra}$, and, as a consequence ({\it cf.}  \eqref{eq:Omega_GW_GR_res} and \eqref{eq:Omega_GW_scalaron}), the GW amplitude is independent of the values of $m_a$ and $f_a$. On the other hand, for large values of $m_a$ or small values of $f_a$, \Eq{fa_ma_condition_k_max_k_da} is fulfilled and the GW amplitude depends on $m_a$ and $f_a$, as follows from \Eq{eq:Omega_GW_sc_k_max_k_d} and \Eq{eq:Omega_GW_GR_res_k_max_k_d}.  Interestingly enough, for relatively high values of the axion coupling $f_a$ above $10^{-3}\Mp$ and axion masses above $1\mathrm{GeV}$, we obtain SIGW signals within the sensitivity curves of GW experiments, namely LISA \cite{LISA:2017pwj, Karnesis:2022vdp}, ET \cite{Maggiore:2019uih}, SKA \cite{Janssen:2014dka}, and BBO \cite{Harry:2006fi}. Thus, these induced GW signals may be detectable in the future serving as a new probe of the StRVM and in principle quantum gravity/string inspired $f(R)$ gravitational theories.

\begin{figure}[h!]
\centering
\includegraphics[width=0.80\textwidth]{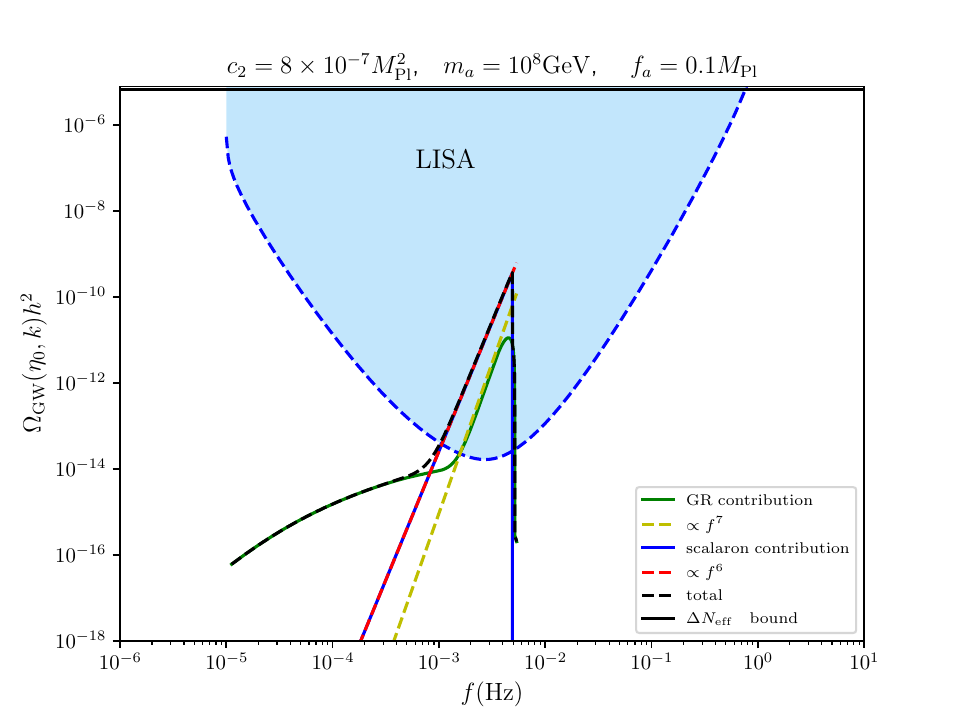}
\caption{\it{ We show in the dashed black curve the total induced gravitational-wave (GW) signal today as a function of the frequency for $m_a = 10^8\mathrm{GeV}$, $f_a = 0.1\Mp$ and $c_2 = 8\times 10^{-7}\Mp^2$. We depict as well in the blue and green solid curves the GR and scalaron contributions to the GW signal respectively while with the yellow dashed and red curves we show their high frequency scaling which go as $f^7$ and $f^6$ respectively. We also superimpose the LISA GW sensitivity curve~\cite{LISA:2017pwj}.}}
\label{fig:Omega_GW_comparison}
\end{figure}

In \Fig{fig:Omega_GW_comparison}, we also show the SIGW signal by fixing the axion mass and the axion coupling to $m_a = 10^{8}~\mathrm{GeV}$ and $f_a = 0.1\Mp$ and by choosing a fiducial value of the $c_2$ quantum correction coefficient, i.e. $c_2 = 8\times 10^{-7}~\Mp^2$ such that the scalaron SIGW signal dominates over the GR one. As we see, we find a characteristic universal $f^6$ scaling, which is  
distinctive not only to the logarithmically quantum corrected gravitational StRVM action \eqref{sea4}, but also to any other $f(R)$ gravity theory with $\frac{k^2}{a^2}\frac{F^{(0)}_{R}}{F^{(0)}}\ll 1$. This feature, therefore, serves as a clear GW prediction of a wide class of $f(R)$ gravity theories, including the string-inspired ones. One can then consider SIGWs as a novel portal to probe the underlying nature of gravity.

\section{Conclusions}\label{sec:concl}

In this work, we have studied scalar induced GWs produced during an early matter-dominated era dominated by heavy axion fields within the context of Chern-Simons running-vacuum cosmology, aka StRVM. In particular, by considering logarithmic quantum corrections of the Einstein-Hilbert gravitational action we derived the scalar induced GW signal treating  the StRVM as an $f(R)$ gravity theory.

Interestingly enough, considering a scale-invariant curvature power spectrum, favored by the Planck CMB observations, as the source of our induced GW signal, we have found that for high values of the logarithmic quantum correction coefficient $c_2$, the SIGW signal associated with the scalaron degree of freedom is the dominant one, giving rise to a distinctive universal $f^6$ frequency scaling. The latter feature  is actually present in any $f(R)$ modified gravity theory with \eqref{condfr},
that is, with negligible geometric anisotropic stress. Par contrast, in the GR case, it is an $f^7$ frequency scaling that characterizes the high-frequency induced GW signal.

Furthermore, for relatively high axion masses above $1~\mathrm{GeV}$ and axion couplings above $10^{-3}\Mp$, we have found a SIGW signal well within the sensitivity curves of GW detectors, namely that of LISA, ET, BBO and SKA, thus being potentially detectable by future GW observatories, and serving as a clear GW signature of a wide class of $f(R)$ gravity theories, not necessarily inspired from string theory. Other $f(R)$ or string-inspired gravitational actions, for which the condition 
\eqref{condfr} may not be valid, will lead, in general, to different characteristic frequency scalings, promoting therefore the portal of SIGWs to a novel probe of the underlying nature of (quantum) gravity.

\acknowledgments
The authors acknowledge participation in the COST Association Actions CA21136 “Addressing observational tensions in cosmology with systematics and fundamental physics (CosmoVerse)” and CA23130 “Bridging high and low energies in search of quantum gravity (BridgeQG)”. CT, TP, SB and ENS acknowledge also participation in the LISA CosWG.
TP acknowledges the support of INFN Sezione di Napoli \textit{iniziativa specifica} QGSKY as well as financial support from the Foundation for Education and European Culture in Greece. The work of N.E.M. is supported in part by the UK Science and Technology Facilities research Council (STFC) and UK Engineering and Physical Sciences Research Council (EPSRC) under the research grants  ST/X000753/1 and EP/V002821/1, respectively.

.\\

\appendix
\section{The geometric anisotropic stress}\label{app:geometric_anisotropic_stress}

In $f(R)$ gravity theories, $\Phi^{(1)}$ and $\Psi^{(1)}$ are connected through the following relation~\cite{Tsujikawa:2007gd,Zhou:2024doz}:
\beq\label{eq:Phi-Psi_geometrical}
\Psi^{(1)}-\Phi^{(1)} = \frac{F^{(1)}}{F^{(0)}}.
\eeq
Writing $F^{(1)}$ as $F^{(1)} = F^{(0)}_R R^{(1)}$ where $R^{(1)}$ is the first order perturbation of the Ricci scalar, one can show that in the sub-horizon regime $R^{(1)}$ can be recast as~\cite{Tsujikawa:2007gd}
 \beq\label{eq:delta_R}
 R^{(1)} \simeq -2 \frac{k^2}{a^2}\frac{\Phi^{(1)}}{1 + 4\frac{k^2}{a^2}\frac{F^{(0)}_{R}}{F^{(0)}}}.
 \eeq
\begin{figure}[ht]
\centering
\includegraphics[width=0.49\textwidth]{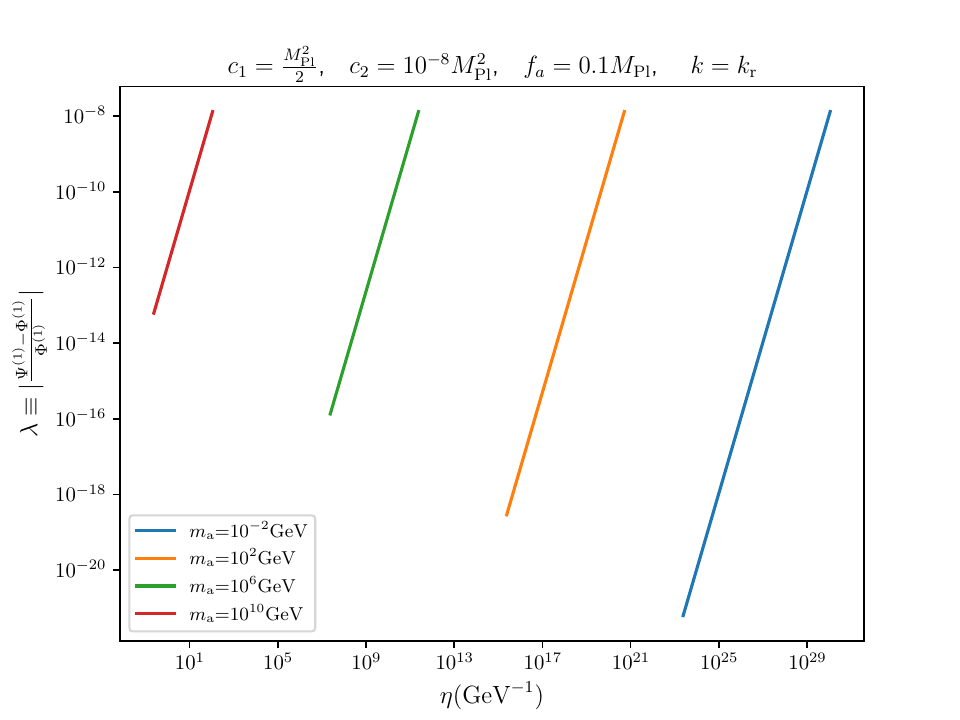}
\includegraphics[width=0.49\textwidth]{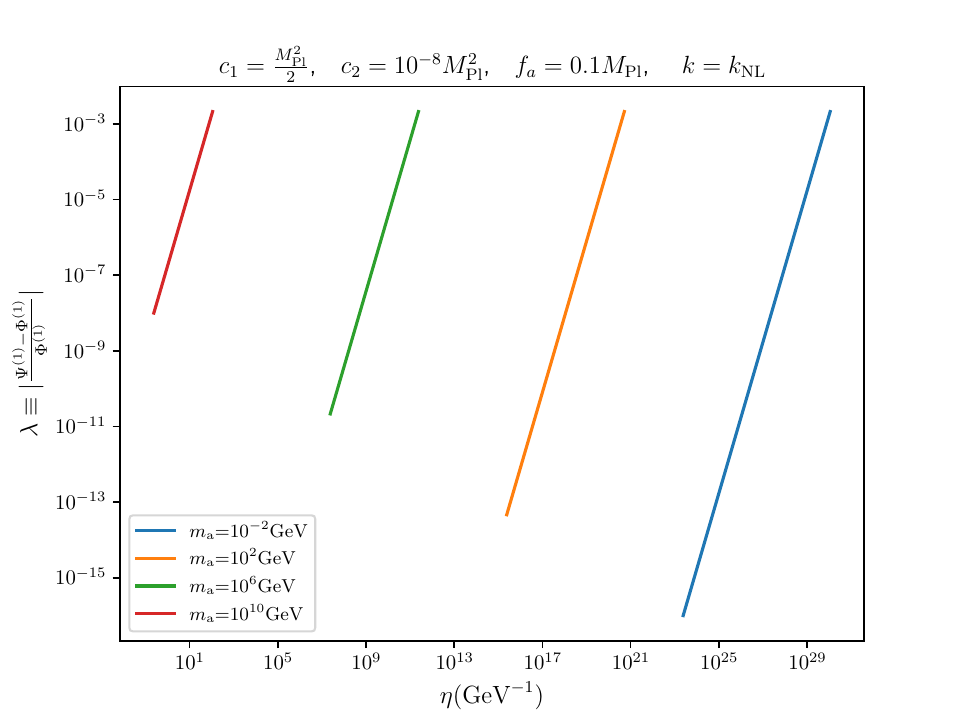}
\caption{\it{\underline{Left Panel}: The dimensionless parameter $\lambda$ as a function of the conformal time for $k=k_\mathrm{r}$ and for different values of the axion mass $m_\mathrm{a}$. 
\underline{Right Panel}: The dimensionless parameter $\lambda$ as a function of the conformal time for $k=k_\mathrm{NL}$ and for different values of the axion mass $m_\mathrm{a}$.}}
\label{fig:lam_vs_ma}
\end{figure}
 At this point, one can naturally define a dimensionless quantity denoted here with $\lambda$ as 
 \beq
 \lambda \equiv \left\vert\frac{\Psi^{(1)} - \Phi^{(1)}}{\Phi^{(1)}}\right\vert,
 \eeq
 which actually quantifies the anisotropic stress of geometrical origin. In the case of $f(R)$ gravity, plugging \Eq{eq:delta_R} into $ F^{(1)} = F^{(0)}_{R}R^{(1)}$  and inserting then $F^{(1)}$ into \Eq{eq:Phi-Psi_geometrical} one finds that 
 \beq\label{eq:gravitional_slip}
 \lambda = \frac{2\frac{k^2}{a^2}\frac{F^{(0)}_{R}}{F^{(0)}}}{1 + 4\frac{k^2}{a^2}\frac{F^{(0)}_{R}}{F^{(0)}}}.
 \eeq

 After a straightforward calculation, we can then show that 
\beq
\lambda = 8.3\times 10^{-2} \frac{c_2}{c_1}\eta^2 k^2.
\eeq

In Fig.~\ref{fig:lam_vs_ma} we plot this quantity for two different values of the axion mass and for two different wave numbers $k$ within the range $[k_\mathrm{d},k_\mathrm{NL}]$. One can see, then, that for $c_2/c_1\ll 1$, we have $\lambda<0.1$, and thus we may safely assume that $\psi^{(1)} \simeq \phi^{(1)}$.

\bibliography{references}

\begin{thebibliography}{91}%
\makeatletter
\providecommand \@ifxundefined [1]{%
 \@ifx{#1\undefined}
}%
\providecommand \@ifnum [1]{%
 \ifnum #1\expandafter \@firstoftwo
 \else \expandafter \@secondoftwo
 \fi
}%
\providecommand \@ifx [1]{%
 \ifx #1\expandafter \@firstoftwo
 \else \expandafter \@secondoftwo
 \fi
}%
\providecommand \natexlab [1]{#1}%
\providecommand \enquote  [1]{``#1''}%
\providecommand \bibnamefont  [1]{#1}%
\providecommand \bibfnamefont [1]{#1}%
\providecommand \citenamefont [1]{#1}%
\providecommand \href@noop [0]{\@secondoftwo}%
\providecommand \href [0]{\begingroup \@sanitize@url \@href}%
\providecommand \@href[1]{\@@startlink{#1}\@@href}%
\providecommand \@@href[1]{\endgroup#1\@@endlink}%
\providecommand \@sanitize@url [0]{\catcode `\\12\catcode `\$12\catcode
  `\&12\catcode `\#12\catcode `\^12\catcode `\_12\catcode `\%12\relax}%
\providecommand \@@startlink[1]{}%
\providecommand \@@endlink[0]{}%
\providecommand \url  [0]{\begingroup\@sanitize@url \@url }%
\providecommand \@url [1]{\endgroup\@href {#1}{\urlprefix }}%
\providecommand \urlprefix  [0]{URL }%
\providecommand \Eprint [0]{\href }%
\providecommand \doibase [0]{https://doi.org/}%
\providecommand \selectlanguage [0]{\@gobble}%
\providecommand \bibinfo  [0]{\@secondoftwo}%
\providecommand \bibfield  [0]{\@secondoftwo}%
\providecommand \translation [1]{[#1]}%
\providecommand \BibitemOpen [0]{}%
\providecommand \bibitemStop [0]{}%
\providecommand \bibitemNoStop [0]{.\EOS\space}%
\providecommand \EOS [0]{\spacefactor3000\relax}%
\providecommand \BibitemShut  [1]{\csname bibitem#1\endcsname}%
\let\auto@bib@innerbib\@empty
\bibitem [{\citenamefont {Sotiriou}\ and\ \citenamefont
  {Faraoni}(2010)}]{Sotiriou:2008rp}%
  \BibitemOpen
  \bibfield  {author} {\bibinfo {author} {\bibfnamefont {T.~P.}\ \bibnamefont
  {Sotiriou}}\ and\ \bibinfo {author} {\bibfnamefont {V.}~\bibnamefont
  {Faraoni}},\ }\bibfield  {title} {\bibinfo {title} {{f(R) Theories Of
  Gravity}},\ }\href {https://doi.org/10.1103/RevModPhys.82.451} {\bibfield
  {journal} {\bibinfo  {journal} {Rev. Mod. Phys.}\ }\textbf {\bibinfo {volume}
  {82}},\ \bibinfo {pages} {451} (\bibinfo {year} {2010})},\ \Eprint
  {https://arxiv.org/abs/0805.1726} {arXiv:0805.1726 [gr-qc]} \BibitemShut
  {NoStop}%
\bibitem [{\citenamefont {Basilakos}\ \emph
  {et~al.}(2020{\natexlab{a}})\citenamefont {Basilakos}, \citenamefont
  {Mavromatos},\ and\ \citenamefont {Sol\`a~Peracaula}}]{bms}%
  \BibitemOpen
  \bibfield  {author} {\bibinfo {author} {\bibfnamefont {S.}~\bibnamefont
  {Basilakos}}, \bibinfo {author} {\bibfnamefont {N.~E.}\ \bibnamefont
  {Mavromatos}},\ and\ \bibinfo {author} {\bibfnamefont {J.}~\bibnamefont
  {Sol\`a~Peracaula}},\ }\bibfield  {title} {\bibinfo {title} {{Gravitational
  and Chiral Anomalies in the Running Vacuum Universe and Matter-Antimatter
  Asymmetry}},\ }\href {https://doi.org/10.1103/PhysRevD.101.045001} {\bibfield
   {journal} {\bibinfo  {journal} {Phys. Rev. D}\ }\textbf {\bibinfo {volume}
  {101}},\ \bibinfo {pages} {045001} (\bibinfo {year} {2020}{\natexlab{a}})},\
  \Eprint {https://arxiv.org/abs/1907.04890} {arXiv:1907.04890 [hep-ph]}
  \BibitemShut {NoStop}%
\bibitem [{\citenamefont {Basilakos}\ \emph
  {et~al.}(2020{\natexlab{b}})\citenamefont {Basilakos}, \citenamefont
  {Mavromatos},\ and\ \citenamefont {Sol\`a~Peracaula}}]{bms2}%
  \BibitemOpen
  \bibfield  {author} {\bibinfo {author} {\bibfnamefont {S.}~\bibnamefont
  {Basilakos}}, \bibinfo {author} {\bibfnamefont {N.~E.}\ \bibnamefont
  {Mavromatos}},\ and\ \bibinfo {author} {\bibfnamefont {J.}~\bibnamefont
  {Sol\`a~Peracaula}},\ }\bibfield  {title} {\bibinfo {title} {{Quantum
  Anomalies in String-Inspired Running Vacuum Universe: Inflation and Axion
  Dark Matter}},\ }\href {https://doi.org/10.1016/j.physletb.2020.135342}
  {\bibfield  {journal} {\bibinfo  {journal} {Phys. Lett. B}\ }\textbf
  {\bibinfo {volume} {803}},\ \bibinfo {pages} {135342} (\bibinfo {year}
  {2020}{\natexlab{b}})},\ \Eprint {https://arxiv.org/abs/2001.03465}
  {arXiv:2001.03465 [gr-qc]} \BibitemShut {NoStop}%
\bibitem [{\citenamefont {Mavromatos}\ and\ \citenamefont
  {Sol\`a~Peracaula}(2021{\natexlab{a}})}]{ms1}%
  \BibitemOpen
  \bibfield  {author} {\bibinfo {author} {\bibfnamefont {N.~E.}\ \bibnamefont
  {Mavromatos}}\ and\ \bibinfo {author} {\bibfnamefont {J.}~\bibnamefont
  {Sol\`a~Peracaula}},\ }\bibfield  {title} {\bibinfo {title}
  {{Stringy-running-vacuum-model inflation: from primordial gravitational waves
  and stiff axion matter to dynamical dark energy}},\ }\href
  {https://doi.org/10.1140/epjs/s11734-021-00197-8} {\bibfield  {journal}
  {\bibinfo  {journal} {Eur. Phys. J. ST}\ }\textbf {\bibinfo {volume} {230}},\
  \bibinfo {pages} {2077} (\bibinfo {year} {2021}{\natexlab{a}})},\ \Eprint
  {https://arxiv.org/abs/2012.07971} {arXiv:2012.07971 [hep-ph]} \BibitemShut
  {NoStop}%
\bibitem [{\citenamefont {Mavromatos}\ and\ \citenamefont
  {Sol\`a~Peracaula}(2021{\natexlab{b}})}]{ms2}%
  \BibitemOpen
  \bibfield  {author} {\bibinfo {author} {\bibfnamefont {N.~E.}\ \bibnamefont
  {Mavromatos}}\ and\ \bibinfo {author} {\bibfnamefont {J.}~\bibnamefont
  {Sol\`a~Peracaula}},\ }\bibfield  {title} {\bibinfo {title} {{Inflationary
  physics and trans-Planckian conjecture in the stringy running vacuum model:
  from the phantom vacuum to the true vacuum}},\ }\href
  {https://doi.org/10.1140/epjp/s13360-021-02149-6} {\bibfield  {journal}
  {\bibinfo  {journal} {Eur. Phys. J. Plus}\ }\textbf {\bibinfo {volume}
  {136}},\ \bibinfo {pages} {1152} (\bibinfo {year} {2021}{\natexlab{b}})},\
  \Eprint {https://arxiv.org/abs/2105.02659} {arXiv:2105.02659 [hep-th]}
  \BibitemShut {NoStop}%
\bibitem [{\citenamefont {G\'omez-Valent}\ \emph {et~al.}(2024)\citenamefont
  {G\'omez-Valent}, \citenamefont {Mavromatos},\ and\ \citenamefont
  {Sol\`a~Peracaula}}]{gms}%
  \BibitemOpen
  \bibfield  {author} {\bibinfo {author} {\bibfnamefont {A.}~\bibnamefont
  {G\'omez-Valent}}, \bibinfo {author} {\bibfnamefont {N.~E.}\ \bibnamefont
  {Mavromatos}},\ and\ \bibinfo {author} {\bibfnamefont {J.}~\bibnamefont
  {Sol\`a~Peracaula}},\ }\bibfield  {title} {\bibinfo {title} {{Stringy running
  vacuum model and current tensions in cosmology}},\ }\href
  {https://doi.org/10.1088/1361-6382/ad0fb8} {\bibfield  {journal} {\bibinfo
  {journal} {Class. Quant. Grav.}\ }\textbf {\bibinfo {volume} {41}},\ \bibinfo
  {pages} {015026} (\bibinfo {year} {2024})},\ \Eprint
  {https://arxiv.org/abs/2305.15774} {arXiv:2305.15774 [gr-qc]} \BibitemShut
  {NoStop}%
\bibitem [{\citenamefont {Asimakis}\ \emph {et~al.}(2022)\citenamefont
  {Asimakis}, \citenamefont {Basilakos}, \citenamefont {Mavromatos},\ and\
  \citenamefont {Saridakis}}]{Asimakis:2021yct}%
  \BibitemOpen
  \bibfield  {author} {\bibinfo {author} {\bibfnamefont {P.}~\bibnamefont
  {Asimakis}}, \bibinfo {author} {\bibfnamefont {S.}~\bibnamefont {Basilakos}},
  \bibinfo {author} {\bibfnamefont {N.~E.}\ \bibnamefont {Mavromatos}},\ and\
  \bibinfo {author} {\bibfnamefont {E.~N.}\ \bibnamefont {Saridakis}},\
  }\bibfield  {title} {\bibinfo {title} {{Big bang nucleosynthesis constraints
  on higher-order modified gravities}},\ }\href
  {https://doi.org/10.1103/PhysRevD.105.084010} {\bibfield  {journal} {\bibinfo
   {journal} {Phys. Rev. D}\ }\textbf {\bibinfo {volume} {105}},\ \bibinfo
  {pages} {084010} (\bibinfo {year} {2022})},\ \Eprint
  {https://arxiv.org/abs/2112.10863} {arXiv:2112.10863 [gr-qc]} \BibitemShut
  {NoStop}%
\bibitem [{\citenamefont {Papanikolaou}\ \emph {et~al.}(2024)\citenamefont
  {Papanikolaou}, \citenamefont {Tzerefos}, \citenamefont {Basilakos},
  \citenamefont {Saridakis},\ and\ \citenamefont
  {Mavromatos}}]{Papanikolaou:2024rlq}%
  \BibitemOpen
  \bibfield  {author} {\bibinfo {author} {\bibfnamefont {T.}~\bibnamefont
  {Papanikolaou}}, \bibinfo {author} {\bibfnamefont {C.}~\bibnamefont
  {Tzerefos}}, \bibinfo {author} {\bibfnamefont {S.}~\bibnamefont {Basilakos}},
  \bibinfo {author} {\bibfnamefont {E.~N.}\ \bibnamefont {Saridakis}},\ and\
  \bibinfo {author} {\bibfnamefont {N.~E.}\ \bibnamefont {Mavromatos}},\
  }\bibfield  {title} {\bibinfo {title} {{Revisiting string-inspired
  running-vacuum models under the lens of light primordial black holes}},\
  }\href {https://doi.org/10.1103/PhysRevD.110.024055} {\bibfield  {journal}
  {\bibinfo  {journal} {Phys. Rev. D}\ }\textbf {\bibinfo {volume} {110}},\
  \bibinfo {pages} {024055} (\bibinfo {year} {2024})},\ \Eprint
  {https://arxiv.org/abs/2402.19373} {arXiv:2402.19373 [gr-qc]} \BibitemShut
  {NoStop}%
\bibitem [{\citenamefont {Inomata}\ \emph
  {et~al.}(2019{\natexlab{a}})\citenamefont {Inomata}, \citenamefont {Kohri},
  \citenamefont {Nakama},\ and\ \citenamefont {Terada}}]{Inomata:2019ivs}%
  \BibitemOpen
  \bibfield  {author} {\bibinfo {author} {\bibfnamefont {K.}~\bibnamefont
  {Inomata}}, \bibinfo {author} {\bibfnamefont {K.}~\bibnamefont {Kohri}},
  \bibinfo {author} {\bibfnamefont {T.}~\bibnamefont {Nakama}},\ and\ \bibinfo
  {author} {\bibfnamefont {T.}~\bibnamefont {Terada}},\ }\bibfield  {title}
  {\bibinfo {title} {{Enhancement of Gravitational Waves Induced by Scalar
  Perturbations due to a Sudden Transition from an Early Matter Era to the
  Radiation Era}},\ }\href {https://doi.org/10.1103/PhysRevD.100.043532}
  {\bibfield  {journal} {\bibinfo  {journal} {Phys. Rev. D}\ }\textbf {\bibinfo
  {volume} {100}},\ \bibinfo {pages} {043532} (\bibinfo {year}
  {2019}{\natexlab{a}})},\ \Eprint {https://arxiv.org/abs/1904.12879}
  {arXiv:1904.12879 [astro-ph.CO]} \BibitemShut {NoStop}%
\bibitem [{\citenamefont {Inomata}\ \emph
  {et~al.}(2019{\natexlab{b}})\citenamefont {Inomata}, \citenamefont {Kohri},
  \citenamefont {Nakama},\ and\ \citenamefont {Terada}}]{Inomata:2019zqy}%
  \BibitemOpen
  \bibfield  {author} {\bibinfo {author} {\bibfnamefont {K.}~\bibnamefont
  {Inomata}}, \bibinfo {author} {\bibfnamefont {K.}~\bibnamefont {Kohri}},
  \bibinfo {author} {\bibfnamefont {T.}~\bibnamefont {Nakama}},\ and\ \bibinfo
  {author} {\bibfnamefont {T.}~\bibnamefont {Terada}},\ }\bibfield  {title}
  {\bibinfo {title} {{Gravitational Waves Induced by Scalar Perturbations
  during a Gradual Transition from an Early Matter Era to the Radiation Era}},\
  }\href {https://doi.org/10.1088/1475-7516/2019/10/071} {\bibfield  {journal}
  {\bibinfo  {journal} {JCAP}\ }\textbf {\bibinfo {volume} {10}},\ \bibinfo
  {pages} {071}},\ \bibinfo {note} {[Erratum: JCAP 08, E01 (2023)]},\ \Eprint
  {https://arxiv.org/abs/1904.12878} {arXiv:1904.12878 [astro-ph.CO]}
  \BibitemShut {NoStop}%
\bibitem [{\citenamefont {Inomata}\ \emph {et~al.}(2020)\citenamefont
  {Inomata}, \citenamefont {Kawasaki}, \citenamefont {Mukaida}, \citenamefont
  {Terada},\ and\ \citenamefont {Yanagida}}]{Inomata:2020lmk}%
  \BibitemOpen
  \bibfield  {author} {\bibinfo {author} {\bibfnamefont {K.}~\bibnamefont
  {Inomata}}, \bibinfo {author} {\bibfnamefont {M.}~\bibnamefont {Kawasaki}},
  \bibinfo {author} {\bibfnamefont {K.}~\bibnamefont {Mukaida}}, \bibinfo
  {author} {\bibfnamefont {T.}~\bibnamefont {Terada}},\ and\ \bibinfo {author}
  {\bibfnamefont {T.~T.}\ \bibnamefont {Yanagida}},\ }\bibfield  {title}
  {\bibinfo {title} {{Gravitational Wave Production right after a Primordial
  Black Hole Evaporation}},\ }\href
  {https://doi.org/10.1103/PhysRevD.101.123533} {\bibfield  {journal} {\bibinfo
   {journal} {Phys. Rev. D}\ }\textbf {\bibinfo {volume} {101}},\ \bibinfo
  {pages} {123533} (\bibinfo {year} {2020})},\ \Eprint
  {https://arxiv.org/abs/2003.10455} {arXiv:2003.10455 [astro-ph.CO]}
  \BibitemShut {NoStop}%
\bibitem [{\citenamefont {Papanikolaou}\ \emph {et~al.}(2021)\citenamefont
  {Papanikolaou}, \citenamefont {Vennin},\ and\ \citenamefont
  {Langlois}}]{Papanikolaou:2020qtd}%
  \BibitemOpen
  \bibfield  {author} {\bibinfo {author} {\bibfnamefont {T.}~\bibnamefont
  {Papanikolaou}}, \bibinfo {author} {\bibfnamefont {V.}~\bibnamefont
  {Vennin}},\ and\ \bibinfo {author} {\bibfnamefont {D.}~\bibnamefont
  {Langlois}},\ }\bibfield  {title} {\bibinfo {title} {{Gravitational waves
  from a universe filled with primordial black holes}},\ }\href
  {https://doi.org/10.1088/1475-7516/2021/03/053} {\bibfield  {journal}
  {\bibinfo  {journal} {JCAP}\ }\textbf {\bibinfo {volume} {03}},\ \bibinfo
  {pages} {053}},\ \Eprint {https://arxiv.org/abs/2010.11573} {arXiv:2010.11573
  [astro-ph.CO]} \BibitemShut {NoStop}%
\bibitem [{\citenamefont {Dom\`enech}\ \emph {et~al.}(2021)\citenamefont
  {Dom\`enech}, \citenamefont {Lin},\ and\ \citenamefont
  {Sasaki}}]{Domenech:2020ssp}%
  \BibitemOpen
  \bibfield  {author} {\bibinfo {author} {\bibfnamefont {G.}~\bibnamefont
  {Dom\`enech}}, \bibinfo {author} {\bibfnamefont {C.}~\bibnamefont {Lin}},\
  and\ \bibinfo {author} {\bibfnamefont {M.}~\bibnamefont {Sasaki}},\
  }\bibfield  {title} {\bibinfo {title} {{Gravitational wave constraints on the
  primordial black hole dominated early universe}},\ }\href
  {https://doi.org/10.1088/1475-7516/2021/11/E01} {\bibfield  {journal}
  {\bibinfo  {journal} {JCAP}\ }\textbf {\bibinfo {volume} {04}},\ \bibinfo
  {pages} {062}},\ \bibinfo {note} {[Erratum: JCAP 11, E01 (2021)]},\ \Eprint
  {https://arxiv.org/abs/2012.08151} {arXiv:2012.08151 [gr-qc]} \BibitemShut
  {NoStop}%
\bibitem [{\citenamefont {Papanikolaou}(2022)}]{Papanikolaou:2022chm}%
  \BibitemOpen
  \bibfield  {author} {\bibinfo {author} {\bibfnamefont {T.}~\bibnamefont
  {Papanikolaou}},\ }\bibfield  {title} {\bibinfo {title} {{Gravitational waves
  induced from primordial black hole fluctuations: the~effect of an extended
  mass function}},\ }\href {https://doi.org/10.1088/1475-7516/2022/10/089}
  {\bibfield  {journal} {\bibinfo  {journal} {JCAP}\ }\textbf {\bibinfo
  {volume} {10}},\ \bibinfo {pages} {089}},\ \Eprint
  {https://arxiv.org/abs/2207.11041} {arXiv:2207.11041 [astro-ph.CO]}
  \BibitemShut {NoStop}%
\bibitem [{\citenamefont {Dom\`enech}\ \emph {et~al.}(2022)\citenamefont
  {Dom\`enech}, \citenamefont {Passaglia},\ and\ \citenamefont
  {Renaux-Petel}}]{Domenech:2021and}%
  \BibitemOpen
  \bibfield  {author} {\bibinfo {author} {\bibfnamefont {G.}~\bibnamefont
  {Dom\`enech}}, \bibinfo {author} {\bibfnamefont {S.}~\bibnamefont
  {Passaglia}},\ and\ \bibinfo {author} {\bibfnamefont {S.}~\bibnamefont
  {Renaux-Petel}},\ }\bibfield  {title} {\bibinfo {title} {{Gravitational waves
  from dark matter isocurvature}},\ }\href
  {https://doi.org/10.1088/1475-7516/2022/03/023} {\bibfield  {journal}
  {\bibinfo  {journal} {JCAP}\ }\textbf {\bibinfo {volume} {03}}\bibfield
  {number} {\bibinfo  {number} { (03)},\ \bibinfo {pages} {023}},\ }\Eprint
  {https://arxiv.org/abs/2112.10163} {arXiv:2112.10163 [astro-ph.CO]}
  \BibitemShut {NoStop}%
\bibitem [{\citenamefont {Dom\`enech}(2021)}]{Domenech:2021ztg}%
  \BibitemOpen
  \bibfield  {author} {\bibinfo {author} {\bibfnamefont {G.}~\bibnamefont
  {Dom\`enech}},\ }\bibfield  {title} {\bibinfo {title} {{Scalar Induced
  Gravitational Waves Review}},\ }\href
  {https://doi.org/10.3390/universe7110398} {\bibfield  {journal} {\bibinfo
  {journal} {Universe}\ }\textbf {\bibinfo {volume} {7}},\ \bibinfo {pages}
  {398} (\bibinfo {year} {2021})},\ \Eprint {https://arxiv.org/abs/2109.01398}
  {arXiv:2109.01398 [gr-qc]} \BibitemShut {NoStop}%
\bibitem [{\citenamefont {Aghanim}\ \emph {et~al.}(2020)\citenamefont {Aghanim}
  \emph {et~al.}}]{Planck:2018}%
  \BibitemOpen
  \bibfield  {author} {\bibinfo {author} {\bibfnamefont {N.}~\bibnamefont
  {Aghanim}} \emph {et~al.} (\bibinfo {collaboration} {Planck}),\ }\bibfield
  {title} {\bibinfo {title} {{Planck 2018 results. VI. Cosmological
  parameters}},\ }\href {https://doi.org/10.1051/0004-6361/201833910}
  {\bibfield  {journal} {\bibinfo  {journal} {Astron. Astrophys.}\ }\textbf
  {\bibinfo {volume} {641}},\ \bibinfo {pages} {A6} (\bibinfo {year} {2020})},\
  \bibinfo {note} {[Erratum: Astron.Astrophys. 652, C4 (2021)]},\ \Eprint
  {https://arxiv.org/abs/1807.06209} {arXiv:1807.06209 [astro-ph.CO]}
  \BibitemShut {NoStop}%
\bibitem [{\citenamefont {Witten}(2000)}]{Witten-wsinst}%
  \BibitemOpen
  \bibfield  {author} {\bibinfo {author} {\bibfnamefont {E.}~\bibnamefont
  {Witten}},\ }\bibfield  {title} {\bibinfo {title} {{World sheet corrections
  via D instantons}},\ }\href {https://doi.org/10.1088/1126-6708/2000/02/030}
  {\bibfield  {journal} {\bibinfo  {journal} {JHEP}\ }\textbf {\bibinfo
  {volume} {02}},\ \bibinfo {pages} {030}},\ \Eprint
  {https://arxiv.org/abs/hep-th/9907041} {arXiv:hep-th/9907041} \BibitemShut
  {NoStop}%
\bibitem [{\citenamefont {Blumenhagen}\ \emph {et~al.}(2005)\citenamefont
  {Blumenhagen}, \citenamefont {Cvetic}, \citenamefont {Langacker},\ and\
  \citenamefont {Shiu}}]{cvetic}%
  \BibitemOpen
  \bibfield  {author} {\bibinfo {author} {\bibfnamefont {R.}~\bibnamefont
  {Blumenhagen}}, \bibinfo {author} {\bibfnamefont {M.}~\bibnamefont {Cvetic}},
  \bibinfo {author} {\bibfnamefont {P.}~\bibnamefont {Langacker}},\ and\
  \bibinfo {author} {\bibfnamefont {G.}~\bibnamefont {Shiu}},\ }\bibfield
  {title} {\bibinfo {title} {{Toward realistic intersecting D-brane models}},\
  }\href {https://doi.org/10.1146/annurev.nucl.55.090704.151541} {\bibfield
  {journal} {\bibinfo  {journal} {Ann. Rev. Nucl. Part. Sci.}\ }\textbf
  {\bibinfo {volume} {55}},\ \bibinfo {pages} {71} (\bibinfo {year} {2005})},\
  \Eprint {https://arxiv.org/abs/hep-th/0502005} {arXiv:hep-th/0502005}
  \BibitemShut {NoStop}%
\bibitem [{\citenamefont {Svrcek}\ and\ \citenamefont {Witten}(2006)}]{svrcek}%
  \BibitemOpen
  \bibfield  {author} {\bibinfo {author} {\bibfnamefont {P.}~\bibnamefont
  {Svrcek}}\ and\ \bibinfo {author} {\bibfnamefont {E.}~\bibnamefont
  {Witten}},\ }\bibfield  {title} {\bibinfo {title} {{Axions In String
  Theory}},\ }\href {https://doi.org/10.1088/1126-6708/2006/06/051} {\bibfield
  {journal} {\bibinfo  {journal} {JHEP}\ }\textbf {\bibinfo {volume} {06}},\
  \bibinfo {pages} {051}},\ \Eprint {https://arxiv.org/abs/hep-th/0605206}
  {arXiv:hep-th/0605206} \BibitemShut {NoStop}%
\bibitem [{\citenamefont {Duncan}\ \emph {et~al.}(1992)\citenamefont {Duncan},
  \citenamefont {Kaloper},\ and\ \citenamefont {Olive}}]{kaloper}%
  \BibitemOpen
  \bibfield  {author} {\bibinfo {author} {\bibfnamefont {M.~J.}\ \bibnamefont
  {Duncan}}, \bibinfo {author} {\bibfnamefont {N.}~\bibnamefont {Kaloper}},\
  and\ \bibinfo {author} {\bibfnamefont {K.~A.}\ \bibnamefont {Olive}},\
  }\bibfield  {title} {\bibinfo {title} {{Axion hair and dynamical torsion from
  anomalies}},\ }\href {https://doi.org/10.1016/0550-3213(92)90052-D}
  {\bibfield  {journal} {\bibinfo  {journal} {Nucl. Phys. B}\ }\textbf
  {\bibinfo {volume} {387}},\ \bibinfo {pages} {215} (\bibinfo {year}
  {1992})}\BibitemShut {NoStop}%
\bibitem [{\citenamefont {Campbell}\ \emph {et~al.}(1991)\citenamefont
  {Campbell}, \citenamefont {Duncan}, \citenamefont {Kaloper},\ and\
  \citenamefont {Olive}}]{kaloper2}%
  \BibitemOpen
  \bibfield  {author} {\bibinfo {author} {\bibfnamefont {B.~A.}\ \bibnamefont
  {Campbell}}, \bibinfo {author} {\bibfnamefont {M.~J.}\ \bibnamefont
  {Duncan}}, \bibinfo {author} {\bibfnamefont {N.}~\bibnamefont {Kaloper}},\
  and\ \bibinfo {author} {\bibfnamefont {K.~A.}\ \bibnamefont {Olive}},\
  }\bibfield  {title} {\bibinfo {title} {{Gravitational dynamics with Lorentz
  Chern-Simons terms}},\ }\href {https://doi.org/10.1016/S0550-3213(05)80045-8}
  {\bibfield  {journal} {\bibinfo  {journal} {Nucl. Phys. B}\ }\textbf
  {\bibinfo {volume} {351}},\ \bibinfo {pages} {778} (\bibinfo {year}
  {1991})}\BibitemShut {NoStop}%
\bibitem [{\citenamefont {Green}\ \emph
  {et~al.}(2012{\natexlab{a}})\citenamefont {Green}, \citenamefont {Schwarz},\
  and\ \citenamefont {Witten}}]{str1}%
  \BibitemOpen
  \bibfield  {author} {\bibinfo {author} {\bibfnamefont {M.~B.}\ \bibnamefont
  {Green}}, \bibinfo {author} {\bibfnamefont {J.~H.}\ \bibnamefont {Schwarz}},\
  and\ \bibinfo {author} {\bibfnamefont {E.}~\bibnamefont {Witten}},\ }\href
  {https://doi.org/10.1017/CBO9781139248563} {\emph {\bibinfo {title}
  {{Superstring Theory Vol. 1}: {25th Anniversary Edition}}}},\ Cambridge
  Monographs on Mathematical Physics\ (\bibinfo  {publisher} {Cambridge
  University Press},\ \bibinfo {year} {2012})\BibitemShut {NoStop}%
\bibitem [{\citenamefont {Green}\ \emph
  {et~al.}(2012{\natexlab{b}})\citenamefont {Green}, \citenamefont {Schwarz},\
  and\ \citenamefont {Witten}}]{str2}%
  \BibitemOpen
  \bibfield  {author} {\bibinfo {author} {\bibfnamefont {M.~B.}\ \bibnamefont
  {Green}}, \bibinfo {author} {\bibfnamefont {J.~H.}\ \bibnamefont {Schwarz}},\
  and\ \bibinfo {author} {\bibfnamefont {E.}~\bibnamefont {Witten}},\ }\href
  {https://doi.org/10.1017/CBO9781139248570} {\emph {\bibinfo {title}
  {{Superstring Theory Vol. 2}: {25th Anniversary Edition}}}},\ Cambridge
  Monographs on Mathematical Physics\ (\bibinfo  {publisher} {Cambridge
  University Press},\ \bibinfo {year} {2012})\BibitemShut {NoStop}%
\bibitem [{\citenamefont {Polchinski}(2007{\natexlab{a}})}]{pol1}%
  \BibitemOpen
  \bibfield  {author} {\bibinfo {author} {\bibfnamefont {J.}~\bibnamefont
  {Polchinski}},\ }\href {https://doi.org/10.1017/CBO9780511816079} {\emph
  {\bibinfo {title} {{String theory. Vol. 1: An introduction to the bosonic
  string}}}},\ Cambridge Monographs on Mathematical Physics\ (\bibinfo
  {publisher} {Cambridge University Press},\ \bibinfo {year}
  {2007})\BibitemShut {NoStop}%
\bibitem [{\citenamefont {Polchinski}(2007{\natexlab{b}})}]{pol2}%
  \BibitemOpen
  \bibfield  {author} {\bibinfo {author} {\bibfnamefont {J.}~\bibnamefont
  {Polchinski}},\ }\href {https://doi.org/10.1017/CBO9780511618123} {\emph
  {\bibinfo {title} {{String theory. Vol. 2: Superstring theory and
  beyond}}}},\ Cambridge Monographs on Mathematical Physics\ (\bibinfo
  {publisher} {Cambridge University Press},\ \bibinfo {year}
  {2007})\BibitemShut {NoStop}%
\bibitem [{\citenamefont {Alexander}\ \emph {et~al.}(2006)\citenamefont
  {Alexander}, \citenamefont {Peskin},\ and\ \citenamefont
  {Sheikh-Jabbari}}]{stephon}%
  \BibitemOpen
  \bibfield  {author} {\bibinfo {author} {\bibfnamefont {S.~H.-S.}\
  \bibnamefont {Alexander}}, \bibinfo {author} {\bibfnamefont {M.~E.}\
  \bibnamefont {Peskin}},\ and\ \bibinfo {author} {\bibfnamefont {M.~M.}\
  \bibnamefont {Sheikh-Jabbari}},\ }\bibfield  {title} {\bibinfo {title}
  {{Leptogenesis from gravity waves in models of inflation}},\ }\href
  {https://doi.org/10.1103/PhysRevLett.96.081301} {\bibfield  {journal}
  {\bibinfo  {journal} {Phys. Rev. Lett.}\ }\textbf {\bibinfo {volume} {96}},\
  \bibinfo {pages} {081301} (\bibinfo {year} {2006})},\ \Eprint
  {https://arxiv.org/abs/hep-th/0403069} {arXiv:hep-th/0403069} \BibitemShut
  {NoStop}%
\bibitem [{\citenamefont {Lyth}\ \emph {et~al.}(2005)\citenamefont {Lyth},
  \citenamefont {Quimbay},\ and\ \citenamefont {Rodriguez}}]{Lyth}%
  \BibitemOpen
  \bibfield  {author} {\bibinfo {author} {\bibfnamefont {D.~H.}\ \bibnamefont
  {Lyth}}, \bibinfo {author} {\bibfnamefont {C.}~\bibnamefont {Quimbay}},\ and\
  \bibinfo {author} {\bibfnamefont {Y.}~\bibnamefont {Rodriguez}},\ }\bibfield
  {title} {\bibinfo {title} {{Leptogenesis and tensor polarisation from a
  gravitational Chern-Simons term}},\ }\href
  {https://doi.org/10.1088/1126-6708/2005/03/016} {\bibfield  {journal}
  {\bibinfo  {journal} {JHEP}\ }\textbf {\bibinfo {volume} {03}},\ \bibinfo
  {pages} {016}},\ \Eprint {https://arxiv.org/abs/hep-th/0501153}
  {arXiv:hep-th/0501153} \BibitemShut {NoStop}%
\bibitem [{\citenamefont {Dorlis}\ \emph
  {et~al.}(2024{\natexlab{a}})\citenamefont {Dorlis}, \citenamefont
  {Mavromatos},\ and\ \citenamefont {Vlachos}}]{Dorlis:2024yqw}%
  \BibitemOpen
  \bibfield  {author} {\bibinfo {author} {\bibfnamefont {P.}~\bibnamefont
  {Dorlis}}, \bibinfo {author} {\bibfnamefont {N.~E.}\ \bibnamefont
  {Mavromatos}},\ and\ \bibinfo {author} {\bibfnamefont {S.-N.}\ \bibnamefont
  {Vlachos}},\ }\bibfield  {title} {\bibinfo {title} {{Condensate-induced
  inflation from primordial gravitational waves in string-inspired Chern-Simons
  gravity}},\ }\href {https://doi.org/10.1103/PhysRevD.110.063512} {\bibfield
  {journal} {\bibinfo  {journal} {Phys. Rev. D}\ }\textbf {\bibinfo {volume}
  {110}},\ \bibinfo {pages} {063512} (\bibinfo {year} {2024}{\natexlab{a}})},\
  \Eprint {https://arxiv.org/abs/2403.09005} {arXiv:2403.09005 [gr-qc]}
  \BibitemShut {NoStop}%
\bibitem [{\citenamefont {Jackiw}\ and\ \citenamefont {Pi}(2003)}]{Jackiw}%
  \BibitemOpen
  \bibfield  {author} {\bibinfo {author} {\bibfnamefont {R.}~\bibnamefont
  {Jackiw}}\ and\ \bibinfo {author} {\bibfnamefont {S.~Y.}\ \bibnamefont
  {Pi}},\ }\bibfield  {title} {\bibinfo {title} {{Chern-Simons modification of
  general relativity}},\ }\href {https://doi.org/10.1103/PhysRevD.68.104012}
  {\bibfield  {journal} {\bibinfo  {journal} {Phys. Rev. D}\ }\textbf {\bibinfo
  {volume} {68}},\ \bibinfo {pages} {104012} (\bibinfo {year} {2003})},\
  \Eprint {https://arxiv.org/abs/gr-qc/0308071} {arXiv:gr-qc/0308071}
  \BibitemShut {NoStop}%
\bibitem [{\citenamefont {Alexander}\ and\ \citenamefont
  {Yunes}(2009)}]{Alexander:2009tp}%
  \BibitemOpen
  \bibfield  {author} {\bibinfo {author} {\bibfnamefont {S.}~\bibnamefont
  {Alexander}}\ and\ \bibinfo {author} {\bibfnamefont {N.}~\bibnamefont
  {Yunes}},\ }\bibfield  {title} {\bibinfo {title} {{Chern-Simons Modified
  General Relativity}},\ }\href {https://doi.org/10.1016/j.physrep.2009.07.002}
  {\bibfield  {journal} {\bibinfo  {journal} {Phys. Rept.}\ }\textbf {\bibinfo
  {volume} {480}},\ \bibinfo {pages} {1} (\bibinfo {year} {2009})},\ \Eprint
  {https://arxiv.org/abs/0907.2562} {arXiv:0907.2562 [hep-th]} \BibitemShut
  {NoStop}%
\bibitem [{\citenamefont {McAllister}\ \emph {et~al.}(2010)\citenamefont
  {McAllister}, \citenamefont {Silverstein},\ and\ \citenamefont
  {Westphal}}]{silver}%
  \BibitemOpen
  \bibfield  {author} {\bibinfo {author} {\bibfnamefont {L.}~\bibnamefont
  {McAllister}}, \bibinfo {author} {\bibfnamefont {E.}~\bibnamefont
  {Silverstein}},\ and\ \bibinfo {author} {\bibfnamefont {A.}~\bibnamefont
  {Westphal}},\ }\bibfield  {title} {\bibinfo {title} {{Gravity Waves and
  Linear Inflation from Axion Monodromy}},\ }\href
  {https://doi.org/10.1103/PhysRevD.82.046003} {\bibfield  {journal} {\bibinfo
  {journal} {Phys. Rev. D}\ }\textbf {\bibinfo {volume} {82}},\ \bibinfo
  {pages} {046003} (\bibinfo {year} {2010})},\ \Eprint
  {https://arxiv.org/abs/0808.0706} {arXiv:0808.0706 [hep-th]} \BibitemShut
  {NoStop}%
\bibitem [{\citenamefont {Sola~Peracaula}(2022)}]{rvm1}%
  \BibitemOpen
  \bibfield  {author} {\bibinfo {author} {\bibfnamefont {J.}~\bibnamefont
  {Sola~Peracaula}},\ }\bibfield  {title} {\bibinfo {title} {{The cosmological
  constant problem and running vacuum in the expanding universe}},\ }\href
  {https://doi.org/10.1098/rsta.2021.0182} {\bibfield  {journal} {\bibinfo
  {journal} {Phil. Trans. Roy. Soc. Lond. A}\ }\textbf {\bibinfo {volume}
  {380}},\ \bibinfo {pages} {20210182} (\bibinfo {year} {2022})},\ \Eprint
  {https://arxiv.org/abs/2203.13757} {arXiv:2203.13757 [gr-qc]} \BibitemShut
  {NoStop}%
\bibitem [{\citenamefont {Sol\`a}\ and\ \citenamefont
  {G\'omez-Valent}(2015)}]{rvm2}%
  \BibitemOpen
  \bibfield  {author} {\bibinfo {author} {\bibfnamefont {J.}~\bibnamefont
  {Sol\`a}}\ and\ \bibinfo {author} {\bibfnamefont {A.}~\bibnamefont
  {G\'omez-Valent}},\ }\bibfield  {title} {\bibinfo {title} {{The
  $\bar{\Lambda}{\rm CDM}$ cosmology: From inflation to dark energy through
  running \ensuremath{\Lambda}}},\ }\href
  {https://doi.org/10.1142/S0218271815410035} {\bibfield  {journal} {\bibinfo
  {journal} {Int. J. Mod. Phys. D}\ }\textbf {\bibinfo {volume} {24}},\
  \bibinfo {pages} {1541003} (\bibinfo {year} {2015})},\ \Eprint
  {https://arxiv.org/abs/1501.03832} {arXiv:1501.03832 [gr-qc]} \BibitemShut
  {NoStop}%
\bibitem [{\citenamefont {Sol\`a~Peracaula}\ and\ \citenamefont
  {Yu}(2020{\natexlab{a}})}]{rvm3}%
  \BibitemOpen
  \bibfield  {author} {\bibinfo {author} {\bibfnamefont {J.}~\bibnamefont
  {Sol\`a~Peracaula}}\ and\ \bibinfo {author} {\bibfnamefont {H.}~\bibnamefont
  {Yu}},\ }\bibfield  {title} {\bibinfo {title} {{Particle and entropy
  production in the Running Vacuum Universe}},\ }\href
  {https://doi.org/10.1007/s10714-020-2657-4} {\bibfield  {journal} {\bibinfo
  {journal} {Gen. Rel. Grav.}\ }\textbf {\bibinfo {volume} {52}},\ \bibinfo
  {pages} {17} (\bibinfo {year} {2020}{\natexlab{a}})},\ \Eprint
  {https://arxiv.org/abs/1910.01638} {arXiv:1910.01638 [gr-qc]} \BibitemShut
  {NoStop}%
\bibitem [{\citenamefont {Moreno-Pulido}\ and\ \citenamefont
  {Sola}(2020)}]{rvmqft1}%
  \BibitemOpen
  \bibfield  {author} {\bibinfo {author} {\bibfnamefont {C.}~\bibnamefont
  {Moreno-Pulido}}\ and\ \bibinfo {author} {\bibfnamefont {J.}~\bibnamefont
  {Sola}},\ }\bibfield  {title} {\bibinfo {title} {{Running vacuum in quantum
  field theory in curved spacetime: renormalizing $\rho_{vac}$ without $\sim
  m^4$ terms}},\ }\href {https://doi.org/10.1140/epjc/s10052-020-8238-6}
  {\bibfield  {journal} {\bibinfo  {journal} {Eur. Phys. J. C}\ }\textbf
  {\bibinfo {volume} {80}},\ \bibinfo {pages} {692} (\bibinfo {year} {2020})},\
  \Eprint {https://arxiv.org/abs/2005.03164} {arXiv:2005.03164 [gr-qc]}
  \BibitemShut {NoStop}%
\bibitem [{\citenamefont {Moreno-Pulido}\ and\ \citenamefont
  {Sola~Peracaula}(2022{\natexlab{a}})}]{rvmqft2}%
  \BibitemOpen
  \bibfield  {author} {\bibinfo {author} {\bibfnamefont {C.}~\bibnamefont
  {Moreno-Pulido}}\ and\ \bibinfo {author} {\bibfnamefont {J.}~\bibnamefont
  {Sola~Peracaula}},\ }\bibfield  {title} {\bibinfo {title} {{Renormalizing the
  vacuum energy in cosmological spacetime: implications for the cosmological
  constant problem}},\ }\href {https://doi.org/10.1140/epjc/s10052-022-10484-w}
  {\bibfield  {journal} {\bibinfo  {journal} {Eur. Phys. J. C}\ }\textbf
  {\bibinfo {volume} {82}},\ \bibinfo {pages} {551} (\bibinfo {year}
  {2022}{\natexlab{a}})},\ \Eprint {https://arxiv.org/abs/2201.05827}
  {arXiv:2201.05827 [gr-qc]} \BibitemShut {NoStop}%
\bibitem [{\citenamefont {Moreno-Pulido}\ and\ \citenamefont
  {Sola~Peracaula}(2022{\natexlab{b}})}]{rvmqft3}%
  \BibitemOpen
  \bibfield  {author} {\bibinfo {author} {\bibfnamefont {C.}~\bibnamefont
  {Moreno-Pulido}}\ and\ \bibinfo {author} {\bibfnamefont {J.}~\bibnamefont
  {Sola~Peracaula}},\ }\bibfield  {title} {\bibinfo {title} {{Equation of state
  of the running vacuum}},\ }\href
  {https://doi.org/10.1140/epjc/s10052-022-11117-y} {\bibfield  {journal}
  {\bibinfo  {journal} {Eur. Phys. J. C}\ }\textbf {\bibinfo {volume} {82}},\
  \bibinfo {pages} {1137} (\bibinfo {year} {2022}{\natexlab{b}})},\ \Eprint
  {https://arxiv.org/abs/2207.07111} {arXiv:2207.07111 [gr-qc]} \BibitemShut
  {NoStop}%
\bibitem [{\citenamefont {Moreno-Pulido}\ \emph {et~al.}(2023)\citenamefont
  {Moreno-Pulido}, \citenamefont {Sola~Peracaula},\ and\ \citenamefont
  {Cheraghchi}}]{rvmqft4}%
  \BibitemOpen
  \bibfield  {author} {\bibinfo {author} {\bibfnamefont {C.}~\bibnamefont
  {Moreno-Pulido}}, \bibinfo {author} {\bibfnamefont {J.}~\bibnamefont
  {Sola~Peracaula}},\ and\ \bibinfo {author} {\bibfnamefont {S.}~\bibnamefont
  {Cheraghchi}},\ }\bibfield  {title} {\bibinfo {title} {{Running vacuum in QFT
  in FLRW spacetime: the dynamics of $\rho _{\textrm{vac}}(H)$ from the
  quantized matter fields}},\ }\href
  {https://doi.org/10.1140/epjc/s10052-023-11772-9} {\bibfield  {journal}
  {\bibinfo  {journal} {Eur. Phys. J. C}\ }\textbf {\bibinfo {volume} {83}},\
  \bibinfo {pages} {637} (\bibinfo {year} {2023})},\ \Eprint
  {https://arxiv.org/abs/2301.05205} {arXiv:2301.05205 [gr-qc]} \BibitemShut
  {NoStop}%
\bibitem [{\citenamefont {Mavromatos}(2023)}]{Mavromatos:2022xdo}%
  \BibitemOpen
  \bibfield  {author} {\bibinfo {author} {\bibfnamefont {N.~E.}\ \bibnamefont
  {Mavromatos}},\ }\bibfield  {title} {\bibinfo {title} {{Lorentz Symmetry
  Violation in String-Inspired Effective Modified Gravity Theories}},\ }\href
  {https://doi.org/10.1007/978-3-031-31520-6_1} {\bibfield  {journal} {\bibinfo
   {journal} {Lect. Notes Phys.}\ }\textbf {\bibinfo {volume} {1017}},\
  \bibinfo {pages} {3} (\bibinfo {year} {2023})},\ \Eprint
  {https://arxiv.org/abs/2205.07044} {arXiv:2205.07044 [hep-th]} \BibitemShut
  {NoStop}%
\bibitem [{\citenamefont {Tong}(2005)}]{Tong:2005un}%
  \BibitemOpen
  \bibfield  {author} {\bibinfo {author} {\bibfnamefont {D.}~\bibnamefont
  {Tong}},\ }\bibfield  {title} {\bibinfo {title} {{TASI lectures on solitons:
  Instantons, monopoles, vortices and kinks}},\ }in\ \href@noop {} {\emph
  {\bibinfo {booktitle} {{Theoretical Advanced Study Institute in Elementary
  Particle Physics}: {Many Dimensions of String Theory}}}}\ (\bibinfo {year}
  {2005})\ \Eprint {https://arxiv.org/abs/hep-th/0509216}
  {arXiv:hep-th/0509216} \BibitemShut {NoStop}%
\bibitem [{\citenamefont {Eguchi}\ \emph {et~al.}(1980)\citenamefont {Eguchi},
  \citenamefont {Gilkey},\ and\ \citenamefont {Hanson}}]{Eguchi:1980jx}%
  \BibitemOpen
  \bibfield  {author} {\bibinfo {author} {\bibfnamefont {T.}~\bibnamefont
  {Eguchi}}, \bibinfo {author} {\bibfnamefont {P.~B.}\ \bibnamefont {Gilkey}},\
  and\ \bibinfo {author} {\bibfnamefont {A.~J.}\ \bibnamefont {Hanson}},\
  }\bibfield  {title} {\bibinfo {title} {{Gravitation, Gauge Theories and
  Differential Geometry}},\ }\href
  {https://doi.org/10.1016/0370-1573(80)90130-1} {\bibfield  {journal}
  {\bibinfo  {journal} {Phys. Rept.}\ }\textbf {\bibinfo {volume} {66}},\
  \bibinfo {pages} {213} (\bibinfo {year} {1980})}\BibitemShut {NoStop}%
\bibitem [{\citenamefont {Akrami}\ \emph {et~al.}(2020)\citenamefont {Akrami}
  \emph {et~al.}}]{Planck}%
  \BibitemOpen
  \bibfield  {author} {\bibinfo {author} {\bibfnamefont {Y.}~\bibnamefont
  {Akrami}} \emph {et~al.} (\bibinfo {collaboration} {Planck}),\ }\bibfield
  {title} {\bibinfo {title} {{Planck 2018 results. X. Constraints on
  inflation}},\ }\href {https://doi.org/10.1051/0004-6361/201833887} {\bibfield
   {journal} {\bibinfo  {journal} {Astron. Astrophys.}\ }\textbf {\bibinfo
  {volume} {641}},\ \bibinfo {pages} {A10} (\bibinfo {year} {2020})},\ \Eprint
  {https://arxiv.org/abs/1807.06211} {arXiv:1807.06211 [astro-ph.CO]}
  \BibitemShut {NoStop}%
\bibitem [{\citenamefont {Bedroya}\ \emph {et~al.}(2020)\citenamefont
  {Bedroya}, \citenamefont {Brandenberger}, \citenamefont {Loverde},\ and\
  \citenamefont {Vafa}}]{TCH1}%
  \BibitemOpen
  \bibfield  {author} {\bibinfo {author} {\bibfnamefont {A.}~\bibnamefont
  {Bedroya}}, \bibinfo {author} {\bibfnamefont {R.}~\bibnamefont
  {Brandenberger}}, \bibinfo {author} {\bibfnamefont {M.}~\bibnamefont
  {Loverde}},\ and\ \bibinfo {author} {\bibfnamefont {C.}~\bibnamefont
  {Vafa}},\ }\bibfield  {title} {\bibinfo {title} {{Trans-Planckian Censorship
  and Inflationary Cosmology}},\ }\href
  {https://doi.org/10.1103/PhysRevD.101.103502} {\bibfield  {journal} {\bibinfo
   {journal} {Phys. Rev. D}\ }\textbf {\bibinfo {volume} {101}},\ \bibinfo
  {pages} {103502} (\bibinfo {year} {2020})},\ \Eprint
  {https://arxiv.org/abs/1909.11106} {arXiv:1909.11106 [hep-th]} \BibitemShut
  {NoStop}%
\bibitem [{\citenamefont {Bedroya}\ and\ \citenamefont {Vafa}(2020)}]{TCH2}%
  \BibitemOpen
  \bibfield  {author} {\bibinfo {author} {\bibfnamefont {A.}~\bibnamefont
  {Bedroya}}\ and\ \bibinfo {author} {\bibfnamefont {C.}~\bibnamefont {Vafa}},\
  }\bibfield  {title} {\bibinfo {title} {{Trans-Planckian Censorship and the
  Swampland}},\ }\href {https://doi.org/10.1007/JHEP09(2020)123} {\bibfield
  {journal} {\bibinfo  {journal} {JHEP}\ }\textbf {\bibinfo {volume} {09}},\
  \bibinfo {pages} {123}},\ \Eprint {https://arxiv.org/abs/1909.11063}
  {arXiv:1909.11063 [hep-th]} \BibitemShut {NoStop}%
\bibitem [{\citenamefont {Mavromatos}\ \emph {et~al.}(2022)\citenamefont
  {Mavromatos}, \citenamefont {Spanos},\ and\ \citenamefont {Stamou}}]{stamou}%
  \BibitemOpen
  \bibfield  {author} {\bibinfo {author} {\bibfnamefont {N.~E.}\ \bibnamefont
  {Mavromatos}}, \bibinfo {author} {\bibfnamefont {V.~C.}\ \bibnamefont
  {Spanos}},\ and\ \bibinfo {author} {\bibfnamefont {I.~D.}\ \bibnamefont
  {Stamou}},\ }\bibfield  {title} {\bibinfo {title} {{Primordial black holes
  and gravitational waves in multiaxion-Chern-Simons inflation}},\ }\href
  {https://doi.org/10.1103/PhysRevD.106.063532} {\bibfield  {journal} {\bibinfo
   {journal} {Phys. Rev. D}\ }\textbf {\bibinfo {volume} {106}},\ \bibinfo
  {pages} {063532} (\bibinfo {year} {2022})},\ \Eprint
  {https://arxiv.org/abs/2206.07963} {arXiv:2206.07963 [hep-th]} \BibitemShut
  {NoStop}%
\bibitem [{\citenamefont {Lima}\ \emph {et~al.}(2013)\citenamefont {Lima},
  \citenamefont {Basilakos},\ and\ \citenamefont {Sola}}]{Lima1}%
  \BibitemOpen
  \bibfield  {author} {\bibinfo {author} {\bibfnamefont {J.~A.~S.}\
  \bibnamefont {Lima}}, \bibinfo {author} {\bibfnamefont {S.}~\bibnamefont
  {Basilakos}},\ and\ \bibinfo {author} {\bibfnamefont {J.}~\bibnamefont
  {Sola}},\ }\bibfield  {title} {\bibinfo {title} {{Expansion History with
  Decaying Vacuum: A Complete Cosmological Scenario}},\ }\href
  {https://doi.org/10.1093/mnras/stt220} {\bibfield  {journal} {\bibinfo
  {journal} {Mon. Not. Roy. Astron. Soc.}\ }\textbf {\bibinfo {volume} {431}},\
  \bibinfo {pages} {923} (\bibinfo {year} {2013})},\ \Eprint
  {https://arxiv.org/abs/1209.2802} {arXiv:1209.2802 [gr-qc]} \BibitemShut
  {NoStop}%
\bibitem [{\citenamefont {Perico}\ \emph {et~al.}(2013)\citenamefont {Perico},
  \citenamefont {Lima}, \citenamefont {Basilakos},\ and\ \citenamefont
  {Sola}}]{Lima2}%
  \BibitemOpen
  \bibfield  {author} {\bibinfo {author} {\bibfnamefont {E.~L.~D.}\
  \bibnamefont {Perico}}, \bibinfo {author} {\bibfnamefont {J.~A.~S.}\
  \bibnamefont {Lima}}, \bibinfo {author} {\bibfnamefont {S.}~\bibnamefont
  {Basilakos}},\ and\ \bibinfo {author} {\bibfnamefont {J.}~\bibnamefont
  {Sola}},\ }\bibfield  {title} {\bibinfo {title} {{Complete Cosmic History
  with a dynamical $\Lambda=\Lambda(H)$ term}},\ }\href
  {https://doi.org/10.1103/PhysRevD.88.063531} {\bibfield  {journal} {\bibinfo
  {journal} {Phys. Rev. D}\ }\textbf {\bibinfo {volume} {88}},\ \bibinfo
  {pages} {063531} (\bibinfo {year} {2013})},\ \Eprint
  {https://arxiv.org/abs/1306.0591} {arXiv:1306.0591 [astro-ph.CO]}
  \BibitemShut {NoStop}%
\bibitem [{\citenamefont {Kolb}\ and\ \citenamefont {S.}(2019)}]{kolb}%
  \BibitemOpen
  \bibfield  {author} {\bibinfo {author} {\bibfnamefont {E.~W.}\ \bibnamefont
  {Kolb}}\ and\ \bibinfo {author} {\bibfnamefont {T.~M.}\ \bibnamefont {S.}},\
  }\href {https://doi.org/10.1201/9780429492860} {\emph {\bibinfo {title} {{The
  Early Universe}}}},\ Vol.~\bibinfo {volume} {69}\ (\bibinfo  {publisher}
  {Taylor and Francis},\ \bibinfo {year} {2019})\BibitemShut {NoStop}%
\bibitem [{\citenamefont {Lima}\ \emph {et~al.}(2016)\citenamefont {Lima},
  \citenamefont {Basilakos},\ and\ \citenamefont {Sol\`a}}]{thermal1}%
  \BibitemOpen
  \bibfield  {author} {\bibinfo {author} {\bibfnamefont {J.~A.~S.}\
  \bibnamefont {Lima}}, \bibinfo {author} {\bibfnamefont {S.}~\bibnamefont
  {Basilakos}},\ and\ \bibinfo {author} {\bibfnamefont {J.}~\bibnamefont
  {Sol\`a}},\ }\bibfield  {title} {\bibinfo {title} {{Thermodynamical aspects
  of running vacuum models}},\ }\href
  {https://doi.org/10.1140/epjc/s10052-016-4060-6} {\bibfield  {journal}
  {\bibinfo  {journal} {Eur. Phys. J. C}\ }\textbf {\bibinfo {volume} {76}},\
  \bibinfo {pages} {228} (\bibinfo {year} {2016})},\ \Eprint
  {https://arxiv.org/abs/1509.00163} {arXiv:1509.00163 [gr-qc]} \BibitemShut
  {NoStop}%
\bibitem [{\citenamefont {Lima}\ \emph {et~al.}(2015)\citenamefont {Lima},
  \citenamefont {Basilakos},\ and\ \citenamefont {Sol\`a}}]{thermal2}%
  \BibitemOpen
  \bibfield  {author} {\bibinfo {author} {\bibfnamefont {J.~A.~S.}\
  \bibnamefont {Lima}}, \bibinfo {author} {\bibfnamefont {S.}~\bibnamefont
  {Basilakos}},\ and\ \bibinfo {author} {\bibfnamefont {J.}~\bibnamefont
  {Sol\`a}},\ }\bibfield  {title} {\bibinfo {title} {{Nonsingular Decaying
  Vacuum Cosmology and Entropy Production}},\ }\href
  {https://doi.org/10.1007/s10714-015-1888-2} {\bibfield  {journal} {\bibinfo
  {journal} {Gen. Rel. Grav.}\ }\textbf {\bibinfo {volume} {47}},\ \bibinfo
  {pages} {40} (\bibinfo {year} {2015})},\ \Eprint
  {https://arxiv.org/abs/1412.5196} {arXiv:1412.5196 [gr-qc]} \BibitemShut
  {NoStop}%
\bibitem [{\citenamefont {Sol\`a~Peracaula}\ and\ \citenamefont
  {Yu}(2020{\natexlab{b}})}]{thermal3}%
  \BibitemOpen
  \bibfield  {author} {\bibinfo {author} {\bibfnamefont {J.}~\bibnamefont
  {Sol\`a~Peracaula}}\ and\ \bibinfo {author} {\bibfnamefont {H.}~\bibnamefont
  {Yu}},\ }\bibfield  {title} {\bibinfo {title} {{Particle and entropy
  production in the Running Vacuum Universe}},\ }\href
  {https://doi.org/10.1007/s10714-020-2657-4} {\bibfield  {journal} {\bibinfo
  {journal} {Gen. Rel. Grav.}\ }\textbf {\bibinfo {volume} {52}},\ \bibinfo
  {pages} {17} (\bibinfo {year} {2020}{\natexlab{b}})},\ \Eprint
  {https://arxiv.org/abs/1910.01638} {arXiv:1910.01638 [gr-qc]} \BibitemShut
  {NoStop}%
\bibitem [{\citenamefont {Dorlis}\ \emph
  {et~al.}(2024{\natexlab{b}})\citenamefont {Dorlis}, \citenamefont
  {Mavromatos},\ and\ \citenamefont {Vlachos}}]{dorlis2}%
  \BibitemOpen
  \bibfield  {author} {\bibinfo {author} {\bibfnamefont {P.}~\bibnamefont
  {Dorlis}}, \bibinfo {author} {\bibfnamefont {N.~E.}\ \bibnamefont
  {Mavromatos}},\ and\ \bibinfo {author} {\bibfnamefont {S.-N.}\ \bibnamefont
  {Vlachos}},\ }\bibfield  {title} {\bibinfo {title} {{Quantum-Ordering
  Ambiguities in Weak Chern-Simons 4D Gravity and Metastability of the
  Condensate-Induced Inflation}},\ }\href@noop {} {\  (\bibinfo {year}
  {2024}{\natexlab{b}})},\ \Eprint {https://arxiv.org/abs/2411.12519}
  {arXiv:2411.12519 [gr-qc]} \BibitemShut {NoStop}%
\bibitem [{\citenamefont {Carr}\ \emph {et~al.}(2018)\citenamefont {Carr},
  \citenamefont {Dimopoulos}, \citenamefont {Owen},\ and\ \citenamefont
  {Tenkanen}}]{carr}%
  \BibitemOpen
  \bibfield  {author} {\bibinfo {author} {\bibfnamefont {B.}~\bibnamefont
  {Carr}}, \bibinfo {author} {\bibfnamefont {K.}~\bibnamefont {Dimopoulos}},
  \bibinfo {author} {\bibfnamefont {C.}~\bibnamefont {Owen}},\ and\ \bibinfo
  {author} {\bibfnamefont {T.}~\bibnamefont {Tenkanen}},\ }\bibfield  {title}
  {\bibinfo {title} {{Primordial Black Hole Formation During Slow Reheating
  After Inflation}},\ }\href {https://doi.org/10.1103/PhysRevD.97.123535}
  {\bibfield  {journal} {\bibinfo  {journal} {Phys. Rev. D}\ }\textbf {\bibinfo
  {volume} {97}},\ \bibinfo {pages} {123535} (\bibinfo {year} {2018})},\
  \Eprint {https://arxiv.org/abs/1804.08639} {arXiv:1804.08639 [astro-ph.CO]}
  \BibitemShut {NoStop}%
\bibitem [{\citenamefont {Blumenhagen}\ and\ \citenamefont
  {Plauschinn}(2014)}]{blumen}%
  \BibitemOpen
  \bibfield  {author} {\bibinfo {author} {\bibfnamefont {R.}~\bibnamefont
  {Blumenhagen}}\ and\ \bibinfo {author} {\bibfnamefont {E.}~\bibnamefont
  {Plauschinn}},\ }\bibfield  {title} {\bibinfo {title} {{Towards Universal
  Axion Inflation and Reheating in String Theory}},\ }\href
  {https://doi.org/10.1016/j.physletb.2014.08.007} {\bibfield  {journal}
  {\bibinfo  {journal} {Phys. Lett. B}\ }\textbf {\bibinfo {volume} {736}},\
  \bibinfo {pages} {482} (\bibinfo {year} {2014})},\ \Eprint
  {https://arxiv.org/abs/1404.3542} {arXiv:1404.3542 [hep-th]} \BibitemShut
  {NoStop}%
\bibitem [{\citenamefont {Halverson}\ \emph {et~al.}(2019)\citenamefont
  {Halverson}, \citenamefont {Long}, \citenamefont {Nelson},\ and\
  \citenamefont {Salinas}}]{Halverson}%
  \BibitemOpen
  \bibfield  {author} {\bibinfo {author} {\bibfnamefont {J.}~\bibnamefont
  {Halverson}}, \bibinfo {author} {\bibfnamefont {C.}~\bibnamefont {Long}},
  \bibinfo {author} {\bibfnamefont {B.}~\bibnamefont {Nelson}},\ and\ \bibinfo
  {author} {\bibfnamefont {G.}~\bibnamefont {Salinas}},\ }\bibfield  {title}
  {\bibinfo {title} {{Axion reheating in the string landscape}},\ }\href
  {https://doi.org/10.1103/PhysRevD.99.086014} {\bibfield  {journal} {\bibinfo
  {journal} {Phys. Rev. D}\ }\textbf {\bibinfo {volume} {99}},\ \bibinfo
  {pages} {086014} (\bibinfo {year} {2019})},\ \Eprint
  {https://arxiv.org/abs/1903.04495} {arXiv:1903.04495 [hep-th]} \BibitemShut
  {NoStop}%
\bibitem [{\citenamefont {Blumenhagen}\ \emph {et~al.}(2007)\citenamefont
  {Blumenhagen}, \citenamefont {Cvetic},\ and\ \citenamefont
  {Weigand}}]{cvetic2}%
  \BibitemOpen
  \bibfield  {author} {\bibinfo {author} {\bibfnamefont {R.}~\bibnamefont
  {Blumenhagen}}, \bibinfo {author} {\bibfnamefont {M.}~\bibnamefont
  {Cvetic}},\ and\ \bibinfo {author} {\bibfnamefont {T.}~\bibnamefont
  {Weigand}},\ }\bibfield  {title} {\bibinfo {title} {{Spacetime instanton
  corrections in 4D string vacua: The Seesaw mechanism for D-Brane models}},\
  }\href {https://doi.org/10.1016/j.nuclphysb.2007.02.016} {\bibfield
  {journal} {\bibinfo  {journal} {Nucl. Phys. B}\ }\textbf {\bibinfo {volume}
  {771}},\ \bibinfo {pages} {113} (\bibinfo {year} {2007})},\ \Eprint
  {https://arxiv.org/abs/hep-th/0609191} {arXiv:hep-th/0609191} \BibitemShut
  {NoStop}%
\bibitem [{\citenamefont {Blumenhagen}\ \emph {et~al.}(2009)\citenamefont
  {Blumenhagen}, \citenamefont {Cvetic}, \citenamefont {Kachru},\ and\
  \citenamefont {Weigand}}]{Blumenhagen:2009qh}%
  \BibitemOpen
  \bibfield  {author} {\bibinfo {author} {\bibfnamefont {R.}~\bibnamefont
  {Blumenhagen}}, \bibinfo {author} {\bibfnamefont {M.}~\bibnamefont {Cvetic}},
  \bibinfo {author} {\bibfnamefont {S.}~\bibnamefont {Kachru}},\ and\ \bibinfo
  {author} {\bibfnamefont {T.}~\bibnamefont {Weigand}},\ }\bibfield  {title}
  {\bibinfo {title} {{D-Brane Instantons in Type II Orientifolds}},\ }\href
  {https://doi.org/10.1146/annurev.nucl.010909.083113} {\bibfield  {journal}
  {\bibinfo  {journal} {Ann. Rev. Nucl. Part. Sci.}\ }\textbf {\bibinfo
  {volume} {59}},\ \bibinfo {pages} {269} (\bibinfo {year} {2009})},\ \Eprint
  {https://arxiv.org/abs/0902.3251} {arXiv:0902.3251 [hep-th]} \BibitemShut
  {NoStop}%
\bibitem [{\citenamefont {Polchinski}(1994)}]{Polchinski:1994fq}%
  \BibitemOpen
  \bibfield  {author} {\bibinfo {author} {\bibfnamefont {J.}~\bibnamefont
  {Polchinski}},\ }\bibfield  {title} {\bibinfo {title} {{Combinatorics of
  boundaries in string theory}},\ }\href
  {https://doi.org/10.1103/PhysRevD.50.R6041} {\bibfield  {journal} {\bibinfo
  {journal} {Phys. Rev. D}\ }\textbf {\bibinfo {volume} {50}},\ \bibinfo
  {pages} {R6041} (\bibinfo {year} {1994})},\ \Eprint
  {https://arxiv.org/abs/hep-th/9407031} {arXiv:hep-th/9407031} \BibitemShut
  {NoStop}%
\bibitem [{\citenamefont {Ellis}\ \emph {et~al.}(2019)\citenamefont {Ellis},
  \citenamefont {Garcia}, \citenamefont {Nagata}, \citenamefont {Nanopoulos},\
  and\ \citenamefont {Olive}}]{Ellis:2018moe}%
  \BibitemOpen
  \bibfield  {author} {\bibinfo {author} {\bibfnamefont {J.}~\bibnamefont
  {Ellis}}, \bibinfo {author} {\bibfnamefont {M.~A.~G.}\ \bibnamefont
  {Garcia}}, \bibinfo {author} {\bibfnamefont {N.}~\bibnamefont {Nagata}},
  \bibinfo {author} {\bibfnamefont {D.~V.}\ \bibnamefont {Nanopoulos}},\ and\
  \bibinfo {author} {\bibfnamefont {K.~A.}\ \bibnamefont {Olive}},\ }\bibfield
  {title} {\bibinfo {title} {{Symmetry Breaking and Reheating after Inflation
  in No-Scale Flipped SU(5)}},\ }\href
  {https://doi.org/10.1088/1475-7516/2019/04/009} {\bibfield  {journal}
  {\bibinfo  {journal} {JCAP}\ }\textbf {\bibinfo {volume} {04}},\ \bibinfo
  {pages} {009}},\ \Eprint {https://arxiv.org/abs/1812.08184} {arXiv:1812.08184
  [hep-ph]} \BibitemShut {NoStop}%
\bibitem [{\citenamefont {Basilakos}\ \emph {et~al.}(2024)\citenamefont
  {Basilakos}, \citenamefont {Nanopoulos}, \citenamefont {Papanikolaou},
  \citenamefont {Saridakis},\ and\ \citenamefont
  {Tzerefos}}]{Basilakos:2023jvp}%
  \BibitemOpen
  \bibfield  {author} {\bibinfo {author} {\bibfnamefont {S.}~\bibnamefont
  {Basilakos}}, \bibinfo {author} {\bibfnamefont {D.~V.}\ \bibnamefont
  {Nanopoulos}}, \bibinfo {author} {\bibfnamefont {T.}~\bibnamefont
  {Papanikolaou}}, \bibinfo {author} {\bibfnamefont {E.~N.}\ \bibnamefont
  {Saridakis}},\ and\ \bibinfo {author} {\bibfnamefont {C.}~\bibnamefont
  {Tzerefos}},\ }\bibfield  {title} {\bibinfo {title} {{Induced gravitational
  waves from flipped SU(5) superstring theory at nHz}},\ }\href
  {https://doi.org/10.1016/j.physletb.2024.138446} {\bibfield  {journal}
  {\bibinfo  {journal} {Phys. Lett. B}\ }\textbf {\bibinfo {volume} {849}},\
  \bibinfo {pages} {138446} (\bibinfo {year} {2024})},\ \Eprint
  {https://arxiv.org/abs/2309.15820} {arXiv:2309.15820 [astro-ph.CO]}
  \BibitemShut {NoStop}%
\bibitem [{\citenamefont {Conlon}\ and\ \citenamefont
  {Marsh}(2013)}]{Conlon:2013isa}%
  \BibitemOpen
  \bibfield  {author} {\bibinfo {author} {\bibfnamefont {J.~P.}\ \bibnamefont
  {Conlon}}\ and\ \bibinfo {author} {\bibfnamefont {M.~C.~D.}\ \bibnamefont
  {Marsh}},\ }\bibfield  {title} {\bibinfo {title} {{The Cosmophenomenology of
  Axionic Dark Radiation}},\ }\href {https://doi.org/10.1007/JHEP10(2013)214}
  {\bibfield  {journal} {\bibinfo  {journal} {JHEP}\ }\textbf {\bibinfo
  {volume} {10}},\ \bibinfo {pages} {214}},\ \Eprint
  {https://arxiv.org/abs/1304.1804} {arXiv:1304.1804 [hep-ph]} \BibitemShut
  {NoStop}%
\bibitem [{\citenamefont {Marsh}(2016)}]{Marsh:2015xka}%
  \BibitemOpen
  \bibfield  {author} {\bibinfo {author} {\bibfnamefont {D.~J.~E.}\
  \bibnamefont {Marsh}},\ }\bibfield  {title} {\bibinfo {title} {{Axion
  Cosmology}},\ }\href {https://doi.org/10.1016/j.physrep.2016.06.005}
  {\bibfield  {journal} {\bibinfo  {journal} {Phys. Rept.}\ }\textbf {\bibinfo
  {volume} {643}},\ \bibinfo {pages} {1} (\bibinfo {year} {2016})},\ \Eprint
  {https://arxiv.org/abs/1510.07633} {arXiv:1510.07633 [astro-ph.CO]}
  \BibitemShut {NoStop}%
\bibitem [{\citenamefont {Matarrese}\ \emph {et~al.}(1993)\citenamefont
  {Matarrese}, \citenamefont {Pantano},\ and\ \citenamefont
  {Saez}}]{Matarrese:1992rp}%
  \BibitemOpen
  \bibfield  {author} {\bibinfo {author} {\bibfnamefont {S.}~\bibnamefont
  {Matarrese}}, \bibinfo {author} {\bibfnamefont {O.}~\bibnamefont {Pantano}},\
  and\ \bibinfo {author} {\bibfnamefont {D.}~\bibnamefont {Saez}},\ }\bibfield
  {title} {\bibinfo {title} {{A General relativistic approach to the nonlinear
  evolution of collisionless matter}},\ }\href
  {https://doi.org/10.1103/PhysRevD.47.1311} {\bibfield  {journal} {\bibinfo
  {journal} {Phys. Rev. D}\ }\textbf {\bibinfo {volume} {47}},\ \bibinfo
  {pages} {1311} (\bibinfo {year} {1993})}\BibitemShut {NoStop}%
\bibitem [{\citenamefont {Matarrese}\ \emph {et~al.}(1994)\citenamefont
  {Matarrese}, \citenamefont {Pantano},\ and\ \citenamefont
  {Saez}}]{Matarrese:1993zf}%
  \BibitemOpen
  \bibfield  {author} {\bibinfo {author} {\bibfnamefont {S.}~\bibnamefont
  {Matarrese}}, \bibinfo {author} {\bibfnamefont {O.}~\bibnamefont {Pantano}},\
  and\ \bibinfo {author} {\bibfnamefont {D.}~\bibnamefont {Saez}},\ }\bibfield
  {title} {\bibinfo {title} {{General relativistic dynamics of irrotational
  dust: Cosmological implications}},\ }\href
  {https://doi.org/10.1103/PhysRevLett.72.320} {\bibfield  {journal} {\bibinfo
  {journal} {Phys. Rev. Lett.}\ }\textbf {\bibinfo {volume} {72}},\ \bibinfo
  {pages} {320} (\bibinfo {year} {1994})},\ \Eprint
  {https://arxiv.org/abs/astro-ph/9310036} {arXiv:astro-ph/9310036}
  \BibitemShut {NoStop}%
\bibitem [{\citenamefont {Matarrese}\ \emph {et~al.}(1998)\citenamefont
  {Matarrese}, \citenamefont {Mollerach},\ and\ \citenamefont
  {Bruni}}]{Matarrese:1997ay}%
  \BibitemOpen
  \bibfield  {author} {\bibinfo {author} {\bibfnamefont {S.}~\bibnamefont
  {Matarrese}}, \bibinfo {author} {\bibfnamefont {S.}~\bibnamefont
  {Mollerach}},\ and\ \bibinfo {author} {\bibfnamefont {M.}~\bibnamefont
  {Bruni}},\ }\bibfield  {title} {\bibinfo {title} {{Second order perturbations
  of the Einstein-de Sitter universe}},\ }\href
  {https://doi.org/10.1103/PhysRevD.58.043504} {\bibfield  {journal} {\bibinfo
  {journal} {Phys. Rev. D}\ }\textbf {\bibinfo {volume} {58}},\ \bibinfo
  {pages} {043504} (\bibinfo {year} {1998})},\ \Eprint
  {https://arxiv.org/abs/astro-ph/9707278} {arXiv:astro-ph/9707278}
  \BibitemShut {NoStop}%
\bibitem [{\citenamefont {Mollerach}\ \emph {et~al.}(2004)\citenamefont
  {Mollerach}, \citenamefont {Harari},\ and\ \citenamefont
  {Matarrese}}]{Mollerach:2003nq}%
  \BibitemOpen
  \bibfield  {author} {\bibinfo {author} {\bibfnamefont {S.}~\bibnamefont
  {Mollerach}}, \bibinfo {author} {\bibfnamefont {D.}~\bibnamefont {Harari}},\
  and\ \bibinfo {author} {\bibfnamefont {S.}~\bibnamefont {Matarrese}},\
  }\bibfield  {title} {\bibinfo {title} {{CMB polarization from secondary
  vector and tensor modes}},\ }\href
  {https://doi.org/10.1103/PhysRevD.69.063002} {\bibfield  {journal} {\bibinfo
  {journal} {Phys. Rev. D}\ }\textbf {\bibinfo {volume} {69}},\ \bibinfo
  {pages} {063002} (\bibinfo {year} {2004})},\ \Eprint
  {https://arxiv.org/abs/astro-ph/0310711} {arXiv:astro-ph/0310711}
  \BibitemShut {NoStop}%
\bibitem [{\citenamefont {Hwang}\ \emph {et~al.}(2017)\citenamefont {Hwang},
  \citenamefont {Jeong},\ and\ \citenamefont {Noh}}]{Hwang:2017oxa}%
  \BibitemOpen
  \bibfield  {author} {\bibinfo {author} {\bibfnamefont {J.-C.}\ \bibnamefont
  {Hwang}}, \bibinfo {author} {\bibfnamefont {D.}~\bibnamefont {Jeong}},\ and\
  \bibinfo {author} {\bibfnamefont {H.}~\bibnamefont {Noh}},\ }\bibfield
  {title} {\bibinfo {title} {{Gauge dependence of gravitational waves generated
  from scalar perturbations}},\ }\href
  {https://doi.org/10.3847/1538-4357/aa74be} {\bibfield  {journal} {\bibinfo
  {journal} {Astrophys. J.}\ }\textbf {\bibinfo {volume} {842}},\ \bibinfo
  {pages} {46} (\bibinfo {year} {2017})},\ \Eprint
  {https://arxiv.org/abs/1704.03500} {arXiv:1704.03500 [astro-ph.CO]}
  \BibitemShut {NoStop}%
\bibitem [{\citenamefont {Tomikawa}\ and\ \citenamefont
  {Kobayashi}(2020)}]{Tomikawa:2019tvi}%
  \BibitemOpen
  \bibfield  {author} {\bibinfo {author} {\bibfnamefont {K.}~\bibnamefont
  {Tomikawa}}\ and\ \bibinfo {author} {\bibfnamefont {T.}~\bibnamefont
  {Kobayashi}},\ }\bibfield  {title} {\bibinfo {title} {{Gauge dependence of
  gravitational waves generated at second order from scalar perturbations}},\
  }\href {https://doi.org/10.1103/PhysRevD.101.083529} {\bibfield  {journal}
  {\bibinfo  {journal} {Phys. Rev. D}\ }\textbf {\bibinfo {volume} {101}},\
  \bibinfo {pages} {083529} (\bibinfo {year} {2020})},\ \Eprint
  {https://arxiv.org/abs/1910.01880} {arXiv:1910.01880 [gr-qc]} \BibitemShut
  {NoStop}%
\bibitem [{\citenamefont {De~Luca}\ \emph {et~al.}(2020)\citenamefont
  {De~Luca}, \citenamefont {Franciolini}, \citenamefont {Kehagias},\ and\
  \citenamefont {Riotto}}]{DeLuca:2019ufz}%
  \BibitemOpen
  \bibfield  {author} {\bibinfo {author} {\bibfnamefont {V.}~\bibnamefont
  {De~Luca}}, \bibinfo {author} {\bibfnamefont {G.}~\bibnamefont
  {Franciolini}}, \bibinfo {author} {\bibfnamefont {A.}~\bibnamefont
  {Kehagias}},\ and\ \bibinfo {author} {\bibfnamefont {A.}~\bibnamefont
  {Riotto}},\ }\bibfield  {title} {\bibinfo {title} {{On the Gauge Invariance
  of Cosmological Gravitational Waves}},\ }\href
  {https://doi.org/10.1088/1475-7516/2020/03/014} {\bibfield  {journal}
  {\bibinfo  {journal} {JCAP}\ }\textbf {\bibinfo {volume} {03}},\ \bibinfo
  {pages} {014}},\ \Eprint {https://arxiv.org/abs/1911.09689} {arXiv:1911.09689
  [gr-qc]} \BibitemShut {NoStop}%
\bibitem [{\citenamefont {Inomata}\ and\ \citenamefont
  {Terada}(2020)}]{Inomata:2019yww}%
  \BibitemOpen
  \bibfield  {author} {\bibinfo {author} {\bibfnamefont {K.}~\bibnamefont
  {Inomata}}\ and\ \bibinfo {author} {\bibfnamefont {T.}~\bibnamefont
  {Terada}},\ }\bibfield  {title} {\bibinfo {title} {{Gauge Independence of
  Induced Gravitational Waves}},\ }\href
  {https://doi.org/10.1103/PhysRevD.101.023523} {\bibfield  {journal} {\bibinfo
   {journal} {Phys. Rev. D}\ }\textbf {\bibinfo {volume} {101}},\ \bibinfo
  {pages} {023523} (\bibinfo {year} {2020})},\ \Eprint
  {https://arxiv.org/abs/1912.00785} {arXiv:1912.00785 [gr-qc]} \BibitemShut
  {NoStop}%
\bibitem [{\citenamefont {Dom\`enech}\ and\ \citenamefont
  {Sasaki}(2021)}]{Domenech:2020xin}%
  \BibitemOpen
  \bibfield  {author} {\bibinfo {author} {\bibfnamefont {G.}~\bibnamefont
  {Dom\`enech}}\ and\ \bibinfo {author} {\bibfnamefont {M.}~\bibnamefont
  {Sasaki}},\ }\bibfield  {title} {\bibinfo {title} {{Approximate gauge
  independence of the induced gravitational wave spectrum}},\ }\href
  {https://doi.org/10.1103/PhysRevD.103.063531} {\bibfield  {journal} {\bibinfo
   {journal} {Phys. Rev. D}\ }\textbf {\bibinfo {volume} {103}},\ \bibinfo
  {pages} {063531} (\bibinfo {year} {2021})},\ \Eprint
  {https://arxiv.org/abs/2012.14016} {arXiv:2012.14016 [gr-qc]} \BibitemShut
  {NoStop}%
\bibitem [{\citenamefont {Espinosa}\ \emph {et~al.}(2018)\citenamefont
  {Espinosa}, \citenamefont {Racco},\ and\ \citenamefont
  {Riotto}}]{Espinosa:2018eve}%
  \BibitemOpen
  \bibfield  {author} {\bibinfo {author} {\bibfnamefont {J.~R.}\ \bibnamefont
  {Espinosa}}, \bibinfo {author} {\bibfnamefont {D.}~\bibnamefont {Racco}},\
  and\ \bibinfo {author} {\bibfnamefont {A.}~\bibnamefont {Riotto}},\
  }\bibfield  {title} {\bibinfo {title} {{A Cosmological Signature of the SM
  Higgs Instability: Gravitational Waves}},\ }\href
  {https://doi.org/10.1088/1475-7516/2018/09/012} {\bibfield  {journal}
  {\bibinfo  {journal} {JCAP}\ }\textbf {\bibinfo {volume} {1809}}\bibfield
  {number} {\bibinfo  {number} { (09)},\ \bibinfo {pages} {012}},\ }\Eprint
  {https://arxiv.org/abs/1804.07732} {arXiv:1804.07732 [hep-ph]} \BibitemShut
  {NoStop}%
\bibitem [{\citenamefont {Baumann}\ \emph {et~al.}(2007)\citenamefont
  {Baumann}, \citenamefont {Steinhardt}, \citenamefont {Takahashi},\ and\
  \citenamefont {Ichiki}}]{Baumann:2007zm}%
  \BibitemOpen
  \bibfield  {author} {\bibinfo {author} {\bibfnamefont {D.}~\bibnamefont
  {Baumann}}, \bibinfo {author} {\bibfnamefont {P.~J.}\ \bibnamefont
  {Steinhardt}}, \bibinfo {author} {\bibfnamefont {K.}~\bibnamefont
  {Takahashi}},\ and\ \bibinfo {author} {\bibfnamefont {K.}~\bibnamefont
  {Ichiki}},\ }\bibfield  {title} {\bibinfo {title} {{Gravitational Wave
  Spectrum Induced by Primordial Scalar Perturbations}},\ }\href
  {https://doi.org/10.1103/PhysRevD.76.084019} {\bibfield  {journal} {\bibinfo
  {journal} {Phys. Rev.}\ }\textbf {\bibinfo {volume} {D76}},\ \bibinfo {pages}
  {084019} (\bibinfo {year} {2007})},\ \Eprint
  {https://arxiv.org/abs/hep-th/0703290} {arXiv:hep-th/0703290 [hep-th]}
  \BibitemShut {NoStop}%
\bibitem [{\citenamefont {Ananda}\ \emph {et~al.}(2007)\citenamefont {Ananda},
  \citenamefont {Clarkson},\ and\ \citenamefont {Wands}}]{Ananda:2006af}%
  \BibitemOpen
  \bibfield  {author} {\bibinfo {author} {\bibfnamefont {K.~N.}\ \bibnamefont
  {Ananda}}, \bibinfo {author} {\bibfnamefont {C.}~\bibnamefont {Clarkson}},\
  and\ \bibinfo {author} {\bibfnamefont {D.}~\bibnamefont {Wands}},\ }\bibfield
   {title} {\bibinfo {title} {{The Cosmological gravitational wave background
  from primordial density perturbations}},\ }\href
  {https://doi.org/10.1103/PhysRevD.75.123518} {\bibfield  {journal} {\bibinfo
  {journal} {Phys. Rev.}\ }\textbf {\bibinfo {volume} {D75}},\ \bibinfo {pages}
  {123518} (\bibinfo {year} {2007})},\ \Eprint
  {https://arxiv.org/abs/gr-qc/0612013} {arXiv:gr-qc/0612013 [gr-qc]}
  \BibitemShut {NoStop}%
\bibitem [{\citenamefont {Kohri}\ and\ \citenamefont
  {Terada}(2018)}]{Kohri:2018awv}%
  \BibitemOpen
  \bibfield  {author} {\bibinfo {author} {\bibfnamefont {K.}~\bibnamefont
  {Kohri}}\ and\ \bibinfo {author} {\bibfnamefont {T.}~\bibnamefont {Terada}},\
  }\bibfield  {title} {\bibinfo {title} {{Semianalytic calculation of
  gravitational wave spectrum nonlinearly induced from primordial curvature
  perturbations}},\ }\href {https://doi.org/10.1103/PhysRevD.97.123532}
  {\bibfield  {journal} {\bibinfo  {journal} {Phys. Rev.}\ }\textbf {\bibinfo
  {volume} {D97}},\ \bibinfo {pages} {123532} (\bibinfo {year} {2018})},\
  \Eprint {https://arxiv.org/abs/1804.08577} {arXiv:1804.08577 [gr-qc]}
  \BibitemShut {NoStop}%
\bibitem [{\citenamefont {Zhou}\ \emph {et~al.}(2024)\citenamefont {Zhou},
  \citenamefont {Kuang}, \citenamefont {Wu}, \citenamefont {Chen},
  \citenamefont {L\"u},\ and\ \citenamefont {Chang}}]{Zhou:2024doz}%
  \BibitemOpen
  \bibfield  {author} {\bibinfo {author} {\bibfnamefont {J.-Z.}\ \bibnamefont
  {Zhou}}, \bibinfo {author} {\bibfnamefont {Y.-T.}\ \bibnamefont {Kuang}},
  \bibinfo {author} {\bibfnamefont {D.}~\bibnamefont {Wu}}, \bibinfo {author}
  {\bibfnamefont {F.-Y.}\ \bibnamefont {Chen}}, \bibinfo {author}
  {\bibfnamefont {H.}~\bibnamefont {L\"u}},\ and\ \bibinfo {author}
  {\bibfnamefont {Z.}~\bibnamefont {Chang}},\ }\bibfield  {title} {\bibinfo
  {title} {{Scalar induced gravitational waves in f(R) gravity}},\ }\href@noop
  {} {\  (\bibinfo {year} {2024})},\ \Eprint {https://arxiv.org/abs/2409.07702}
  {arXiv:2409.07702 [gr-qc]} \BibitemShut {NoStop}%
\bibitem [{\citenamefont {Tsujikawa}(2007)}]{Tsujikawa:2007gd}%
  \BibitemOpen
  \bibfield  {author} {\bibinfo {author} {\bibfnamefont {S.}~\bibnamefont
  {Tsujikawa}},\ }\bibfield  {title} {\bibinfo {title} {{Matter density
  perturbations and effective gravitational constant in modified gravity models
  of dark energy}},\ }\href {https://doi.org/10.1103/PhysRevD.76.023514}
  {\bibfield  {journal} {\bibinfo  {journal} {Phys. Rev. D}\ }\textbf {\bibinfo
  {volume} {76}},\ \bibinfo {pages} {023514} (\bibinfo {year} {2007})},\
  \Eprint {https://arxiv.org/abs/0705.1032} {arXiv:0705.1032 [astro-ph]}
  \BibitemShut {NoStop}%
\bibitem [{\citenamefont {Katsuragawa}\ \emph {et~al.}(2019)\citenamefont
  {Katsuragawa}, \citenamefont {Nakamura}, \citenamefont {Ikeda},\ and\
  \citenamefont {Capozziello}}]{Katsuragawa:2019uto}%
  \BibitemOpen
  \bibfield  {author} {\bibinfo {author} {\bibfnamefont {T.}~\bibnamefont
  {Katsuragawa}}, \bibinfo {author} {\bibfnamefont {T.}~\bibnamefont
  {Nakamura}}, \bibinfo {author} {\bibfnamefont {T.}~\bibnamefont {Ikeda}},\
  and\ \bibinfo {author} {\bibfnamefont {S.}~\bibnamefont {Capozziello}},\
  }\bibfield  {title} {\bibinfo {title} {{Gravitational Waves in $F(R)$
  Gravity: Scalar Waves and the Chameleon Mechanism}},\ }\href
  {https://doi.org/10.1103/PhysRevD.99.124050} {\bibfield  {journal} {\bibinfo
  {journal} {Phys. Rev. D}\ }\textbf {\bibinfo {volume} {99}},\ \bibinfo
  {pages} {124050} (\bibinfo {year} {2019})},\ \Eprint
  {https://arxiv.org/abs/1902.02494} {arXiv:1902.02494 [gr-qc]} \BibitemShut
  {NoStop}%
\bibitem [{\citenamefont {Papanikolaou}\ \emph {et~al.}(2022)\citenamefont
  {Papanikolaou}, \citenamefont {Tzerefos}, \citenamefont {Basilakos},\ and\
  \citenamefont {Saridakis}}]{Papanikolaou:2021uhe}%
  \BibitemOpen
  \bibfield  {author} {\bibinfo {author} {\bibfnamefont {T.}~\bibnamefont
  {Papanikolaou}}, \bibinfo {author} {\bibfnamefont {C.}~\bibnamefont
  {Tzerefos}}, \bibinfo {author} {\bibfnamefont {S.}~\bibnamefont
  {Basilakos}},\ and\ \bibinfo {author} {\bibfnamefont {E.~N.}\ \bibnamefont
  {Saridakis}},\ }\bibfield  {title} {\bibinfo {title} {{Scalar induced
  gravitational waves from primordial black hole Poisson fluctuations in f(R)
  gravity}},\ }\href {https://doi.org/10.1088/1475-7516/2022/10/013} {\bibfield
   {journal} {\bibinfo  {journal} {JCAP}\ }\textbf {\bibinfo {volume} {10}},\
  \bibinfo {pages} {013}},\ \Eprint {https://arxiv.org/abs/2112.15059}
  {arXiv:2112.15059 [astro-ph.CO]} \BibitemShut {NoStop}%
\bibitem [{\citenamefont {Maggiore}(2000)}]{Maggiore:1999vm}%
  \BibitemOpen
  \bibfield  {author} {\bibinfo {author} {\bibfnamefont {M.}~\bibnamefont
  {Maggiore}},\ }\bibfield  {title} {\bibinfo {title} {{Gravitational wave
  experiments and early universe cosmology}},\ }\href
  {https://doi.org/10.1016/S0370-1573(99)00102-7} {\bibfield  {journal}
  {\bibinfo  {journal} {Phys. Rept.}\ }\textbf {\bibinfo {volume} {331}},\
  \bibinfo {pages} {283} (\bibinfo {year} {2000})},\ \Eprint
  {https://arxiv.org/abs/gr-qc/9909001} {arXiv:gr-qc/9909001} \BibitemShut
  {NoStop}%
\bibitem [{\citenamefont {Assadullahi}\ and\ \citenamefont
  {Wands}(2009)}]{Assadullahi:2009nf}%
  \BibitemOpen
  \bibfield  {author} {\bibinfo {author} {\bibfnamefont {H.}~\bibnamefont
  {Assadullahi}}\ and\ \bibinfo {author} {\bibfnamefont {D.}~\bibnamefont
  {Wands}},\ }\bibfield  {title} {\bibinfo {title} {{Gravitational waves from
  an early matter era}},\ }\href {https://doi.org/10.1103/PhysRevD.79.083511}
  {\bibfield  {journal} {\bibinfo  {journal} {Phys. Rev. D}\ }\textbf {\bibinfo
  {volume} {79}},\ \bibinfo {pages} {083511} (\bibinfo {year} {2009})},\
  \Eprint {https://arxiv.org/abs/0901.0989} {arXiv:0901.0989 [astro-ph.CO]}
  \BibitemShut {NoStop}%
\bibitem [{\citenamefont {Jedamzik}\ \emph {et~al.}(2010)\citenamefont
  {Jedamzik}, \citenamefont {Lemoine},\ and\ \citenamefont
  {Martin}}]{Jedamzik:2010hq}%
  \BibitemOpen
  \bibfield  {author} {\bibinfo {author} {\bibfnamefont {K.}~\bibnamefont
  {Jedamzik}}, \bibinfo {author} {\bibfnamefont {M.}~\bibnamefont {Lemoine}},\
  and\ \bibinfo {author} {\bibfnamefont {J.}~\bibnamefont {Martin}},\
  }\bibfield  {title} {\bibinfo {title} {{Generation of gravitational waves
  during early structure formation between cosmic inflation and reheating}},\
  }\href {https://doi.org/10.1088/1475-7516/2010/04/021} {\bibfield  {journal}
  {\bibinfo  {journal} {JCAP}\ }\textbf {\bibinfo {volume} {04}},\ \bibinfo
  {pages} {021}},\ \Eprint {https://arxiv.org/abs/1002.3278} {arXiv:1002.3278
  [astro-ph.CO]} \BibitemShut {NoStop}%
\bibitem [{\citenamefont {Eggemeier}\ \emph {et~al.}(2023)\citenamefont
  {Eggemeier}, \citenamefont {Niemeyer}, \citenamefont {Jedamzik},\ and\
  \citenamefont {Easther}}]{Eggemeier:2022gyo}%
  \BibitemOpen
  \bibfield  {author} {\bibinfo {author} {\bibfnamefont {B.}~\bibnamefont
  {Eggemeier}}, \bibinfo {author} {\bibfnamefont {J.~C.}\ \bibnamefont
  {Niemeyer}}, \bibinfo {author} {\bibfnamefont {K.}~\bibnamefont {Jedamzik}},\
  and\ \bibinfo {author} {\bibfnamefont {R.}~\bibnamefont {Easther}},\
  }\bibfield  {title} {\bibinfo {title} {{Stochastic gravitational waves from
  postinflationary structure formation}},\ }\href
  {https://doi.org/10.1103/PhysRevD.107.043503} {\bibfield  {journal} {\bibinfo
   {journal} {Phys. Rev. D}\ }\textbf {\bibinfo {volume} {107}},\ \bibinfo
  {pages} {043503} (\bibinfo {year} {2023})},\ \Eprint
  {https://arxiv.org/abs/2212.00425} {arXiv:2212.00425 [astro-ph.CO]}
  \BibitemShut {NoStop}%
\bibitem [{\citenamefont {Fernandez}\ \emph {et~al.}(2024)\citenamefont
  {Fernandez}, \citenamefont {Foster}, \citenamefont {Lillard},\ and\
  \citenamefont {Shelton}}]{Fernandez:2023ddy}%
  \BibitemOpen
  \bibfield  {author} {\bibinfo {author} {\bibfnamefont {N.}~\bibnamefont
  {Fernandez}}, \bibinfo {author} {\bibfnamefont {J.~W.}\ \bibnamefont
  {Foster}}, \bibinfo {author} {\bibfnamefont {B.}~\bibnamefont {Lillard}},\
  and\ \bibinfo {author} {\bibfnamefont {J.}~\bibnamefont {Shelton}},\
  }\bibfield  {title} {\bibinfo {title} {{Stochastic Gravitational Waves from
  Early Structure Formation}},\ }\href
  {https://doi.org/10.1103/PhysRevLett.133.111002} {\bibfield  {journal}
  {\bibinfo  {journal} {Phys. Rev. Lett.}\ }\textbf {\bibinfo {volume} {133}},\
  \bibinfo {pages} {111002} (\bibinfo {year} {2024})},\ \Eprint
  {https://arxiv.org/abs/2312.12499} {arXiv:2312.12499 [astro-ph.CO]}
  \BibitemShut {NoStop}%
\bibitem [{\citenamefont {Padilla}\ \emph {et~al.}(2024)\citenamefont
  {Padilla}, \citenamefont {Hidalgo}, \citenamefont {Malik},\ and\
  \citenamefont {Mulryne}}]{Padilla:2024cbq}%
  \BibitemOpen
  \bibfield  {author} {\bibinfo {author} {\bibfnamefont {L.~E.}\ \bibnamefont
  {Padilla}}, \bibinfo {author} {\bibfnamefont {J.~C.}\ \bibnamefont
  {Hidalgo}}, \bibinfo {author} {\bibfnamefont {K.~A.}\ \bibnamefont {Malik}},\
  and\ \bibinfo {author} {\bibfnamefont {D.}~\bibnamefont {Mulryne}},\
  }\bibfield  {title} {\bibinfo {title} {{Detecting the Stochastic
  Gravitational Wave Background from Primordial Black Holes in Slow-reheating
  Scenarios}},\ }\href@noop {} {\  (\bibinfo {year} {2024})},\ \Eprint
  {https://arxiv.org/abs/2405.19271} {arXiv:2405.19271 [astro-ph.CO]}
  \BibitemShut {NoStop}%
\bibitem [{\citenamefont {Amaro-Seoane}\ \emph {et~al.}(2017)\citenamefont
  {Amaro-Seoane} \emph {et~al.}}]{LISA:2017pwj}%
  \BibitemOpen
  \bibfield  {author} {\bibinfo {author} {\bibfnamefont {P.}~\bibnamefont
  {Amaro-Seoane}} \emph {et~al.} (\bibinfo {collaboration} {LISA}),\ }\bibfield
   {title} {\bibinfo {title} {{Laser Interferometer Space Antenna}},\
  }\href@noop {} {\  (\bibinfo {year} {2017})},\ \Eprint
  {https://arxiv.org/abs/1702.00786} {arXiv:1702.00786 [astro-ph.IM]}
  \BibitemShut {NoStop}%
\bibitem [{\citenamefont {Karnesis}\ \emph {et~al.}(2024)\citenamefont
  {Karnesis} \emph {et~al.}}]{Karnesis:2022vdp}%
  \BibitemOpen
  \bibfield  {author} {\bibinfo {author} {\bibfnamefont {N.}~\bibnamefont
  {Karnesis}} \emph {et~al.},\ }\bibfield  {title} {\bibinfo {title} {{The
  Laser Interferometer Space Antenna mission in Greece White Paper}},\ }\href
  {https://doi.org/10.1142/S0218271824500275} {\bibfield  {journal} {\bibinfo
  {journal} {Int. J. Mod. Phys. D}\ }\textbf {\bibinfo {volume} {33}},\
  \bibinfo {pages} {2450027} (\bibinfo {year} {2024})},\ \Eprint
  {https://arxiv.org/abs/2209.04358} {arXiv:2209.04358 [gr-qc]} \BibitemShut
  {NoStop}%
\bibitem [{\citenamefont {Maggiore}\ \emph {et~al.}(2020)\citenamefont
  {Maggiore} \emph {et~al.}}]{Maggiore:2019uih}%
  \BibitemOpen
  \bibfield  {author} {\bibinfo {author} {\bibfnamefont {M.}~\bibnamefont
  {Maggiore}} \emph {et~al.},\ }\bibfield  {title} {\bibinfo {title} {{Science
  Case for the Einstein Telescope}},\ }\href
  {https://doi.org/10.1088/1475-7516/2020/03/050} {\bibfield  {journal}
  {\bibinfo  {journal} {JCAP}\ }\textbf {\bibinfo {volume} {03}},\ \bibinfo
  {pages} {050}},\ \Eprint {https://arxiv.org/abs/1912.02622} {arXiv:1912.02622
  [astro-ph.CO]} \BibitemShut {NoStop}%
\bibitem [{\citenamefont {Janssen}\ \emph {et~al.}(2015)\citenamefont {Janssen}
  \emph {et~al.}}]{Janssen:2014dka}%
  \BibitemOpen
  \bibfield  {author} {\bibinfo {author} {\bibfnamefont {G.}~\bibnamefont
  {Janssen}} \emph {et~al.},\ }\bibfield  {title} {\bibinfo {title}
  {{Gravitational wave astronomy with the SKA}},\ }\href
  {https://doi.org/10.22323/1.215.0037} {\bibfield  {journal} {\bibinfo
  {journal} {PoS}\ }\textbf {\bibinfo {volume} {AASKA14}},\ \bibinfo {pages}
  {037} (\bibinfo {year} {2015})},\ \Eprint {https://arxiv.org/abs/1501.00127}
  {arXiv:1501.00127 [astro-ph.IM]} \BibitemShut {NoStop}%
\bibitem [{\citenamefont {Harry}\ \emph {et~al.}(2006)\citenamefont {Harry},
  \citenamefont {Fritschel}, \citenamefont {Shaddock}, \citenamefont
  {Folkner},\ and\ \citenamefont {Phinney}}]{Harry:2006fi}%
  \BibitemOpen
  \bibfield  {author} {\bibinfo {author} {\bibfnamefont {G.~M.}\ \bibnamefont
  {Harry}}, \bibinfo {author} {\bibfnamefont {P.}~\bibnamefont {Fritschel}},
  \bibinfo {author} {\bibfnamefont {D.~A.}\ \bibnamefont {Shaddock}}, \bibinfo
  {author} {\bibfnamefont {W.}~\bibnamefont {Folkner}},\ and\ \bibinfo {author}
  {\bibfnamefont {E.~S.}\ \bibnamefont {Phinney}},\ }\bibfield  {title}
  {\bibinfo {title} {{Laser interferometry for the big bang observer}},\ }\href
  {https://doi.org/10.1088/0264-9381/23/15/008} {\bibfield  {journal} {\bibinfo
   {journal} {Class. Quant. Grav.}\ }\textbf {\bibinfo {volume} {23}},\
  \bibinfo {pages} {4887} (\bibinfo {year} {2006})},\ \bibinfo {note}
  {[Erratum: Class.Quant.Grav. 23, 7361 (2006)]}\BibitemShut {NoStop}%
\end{thebibliography}%

\end{document}